\documentclass[twocolumn]{aastex7}
\usepackage{newtxtext,newtxmath}
\usepackage{graphbox,graphicx}
\usepackage[flushleft]{threeparttable}
\usepackage[T1]{fontenc}
\usepackage{graphicx}	
\usepackage{amsmath}	
\usepackage{siunitx}    
\usepackage{nicefrac}
\usepackage{tabularx}
\usepackage{booktabs}
\usepackage{rotating}
\usepackage{xcolor}

\begin{document}

\title{A Comprehensive Study of Morphology and Kinematics in Extended Nebulae Around UV Luminous Quasars at $z\approx1$}

\author[0000-0002-2662-9363]{Zhuoqi (Will) Liu}
\affiliation{Department of Astronomy, University of Michigan, 1085 S. University, Ann Arbor, MI 48109, USA}
\email{zql@umich.edu}

\author[0000-0001-9487-8583]{Sean D. Johnson}
\affiliation{Department of Astronomy, University of Michigan, 1085 S. University, Ann Arbor, MI 48109, USA}
\email{}

\author[0000-0002-5564-9873]{Eric F. Bell}
\affiliation{Department of Astronomy, University of Michigan, 1085 S. University, Ann Arbor, MI 48109, USA}
\email{}

\author[0000-0002-2941-646X]{Zhijie Qu}
\affiliation{Department of Astronomy \& Astrophysics, The University of Chicago, 5640 S. Ellis Avenue, Chicago, IL 60637, USA}
\email{}

\author[0000-0002-2470-5756]{Benoît Epinat}
\affiliation{Aix-Marseille Univ., CNRS, CNES, LAM, 38 Rue Frédéric Joliot Curie, 13338 Marseille, France}
\affiliation{French-Chilean Laboratory for Astronomy, IRL 3386, CNRS and Universidad de Concepción, Departamento de Astronomía, Barrio Universitario s/n, Concepción, Chile}
\email{}

\author[0000-0001-8813-4182]{Hsiao-Wen Chen}
\affiliation{Department of Astronomy \& Astrophysics, The University of Chicago, Chicago, IL 60637, USA}
\email{}

\author[0000-0002-9946-4731]{Marc Rafelski}
\affiliation{Space Telescope Science Institute, 3700 San Martin Drive, Baltimore, MD 21218, USA}
\affiliation{Department of Physics and Astronomy, Johns Hopkins University, Baltimore, MD 21218, USA}
\email{}

\author[0000-0002-0311-2812]{Jennifer~I-Hsiu Li}
\affiliation{Center for AstroPhysical Surveys, National Center for Supercomputing Applications, University of Illinois Urbana-Champaign, Urbana, IL, 61801, USA}
\affiliation{Michigan Institute for Data Science, University of Michigan, Ann Arbor, MI, 48109, USA}
\affiliation{Department of Astronomy, University of Michigan, Ann Arbor, MI, 48109, USA}
\email{}

\author[0000-0001-7396-3578]{Alexander Beckett}
\affiliation{Aix-Marseille Univ, CNRS, CNES, LAM (Laboratoire d’Astrophysique de Marseille), Marseille, France}
\email{}

\author[0000-0002-8739-3163]{Mandy C. Chen}
\affiliation{The Observatories of the Carnegie Institution for Science, 813 Santa Barbara Street, Pasadena, CA 91101, USA}
\affiliation{Cahill Center for Astronomy and Astrophysics, California Institute of Technology, Pasadena, CA 91125, USA}
\email{}

\author[0009-0000-0797-7365]{Sayak Dutta}
\affiliation{Inter-University Centre for Astronomy \& Astrophysics, Post Bag 04, Pune, India, 411007}
\email{}

\author[0009-0003-8927-2140]{David DePalma}
\affiliation{MIT--Kavli Institute for Astrophysics and Space Research, 77 Massachusetts Avenue, Cambridge, Massachusetts 02139}
\email{}

\author[0000-0002-8459-5413]{Gwen C. Rudie}
\affiliation{The Observatories of the Carnegie Institution for Science, 813 Santa Barbara Street, Pasadena, CA 91101, USA}
\email{}

\author[0000-0002-0668-5560]{Joop Schaye}
\affiliation{Leiden Observatory, Leiden University, PO Box 9513, NL-2300 RA Leiden, The Netherlands}
\email{}

\author{Patrick Petitjean}
\affiliation{Institut d’Astrophysique de Paris, 98bis Boulevard Arago 75014 Paris - France}
\email{}

\author[0000-0001-5804-1428]{Sebastiano Cantalupo}
\affiliation{Department of Physics, University of Milan Bicocca, Piazza della Scienza 3, I-20126 Milano, Italy}
\email{}

\author[0000-0001-6846-9399]{Elise Fuller}
\affiliation{Department of Astronomy, University of Michigan, 1085 S. University, Ann Arbor, MI 48109, USA}
\affiliation{Department of Astrophysical and Planetary Sciences, University of Colorado Boulder, 2000 Colorado Ave, Boulder, CO 80309, USA}
\email{}

\author[0000-0002-0417-1494]{Wolfram Kollatschny}
\affiliation{Institut für Astrophysik und Geophysik, Universität Göttingen, Friedrich-Hund Platz 1, D-37077 Göttingen, Germany}
\email{}

\author[0000-0003-0389-0902]{Sebastián López}
\affiliation{Departamento de Astronomía, Universidad de Chile, Camino El Observatorio 1515, Las Condes, Santiago, Chile}
\email{}

\author[0000-0002-9141-9792]{Nishant Mishra}
\affiliation{Department of Astronomy, University of Michigan, 1085 S. University, Ann Arbor, MI 48109, USA}
\affiliation{The Observatories of the Carnegie Institution for Science, 813 Santa Barbara Street, Pasadena, CA 91101, USA}
\email{}

\author[0000-0003-3938-8762]{Sowgat Muzahid}
\affiliation{Inter-University Centre for Astronomy and Astrophysics (IUCAA), Post Bag 4, Ganeshkhind, Pune 411 007, India}
\email{}

\author{Andrea Travascio}
\affiliation{Department of Physics, University of Milan Bicocca, Piazza della Scienza 3, I-20126 Milano, Italy}
\email{}

\author[0000-0002-8739-3163]{Fakhri S. Zahedy}
\altaffiliation{Deceased}
\affiliation{Department of Physics, University of North Texas, Denton, TX 76201, USA}
\email{}




\begin{abstract}
Gas flows between galaxies and the circumgalactic medium (CGM) play a central role in galaxy evolution and can become observable as giant nebulae when illuminated by the quasars. We present an ensemble study of nebulae around 30 UV-luminous quasars at $z\approx0.4{-}1.4$ from the CUBS and MUSEQuBES surveys, 27 of which are detected in extended [O\,II] and/or [O\,III] emission. Based on a joint analysis of nebular morphology and surrounding galaxy environments, we introduce three morpho-kinematic classifications. We identify eleven irregular, large-scale ($>\!50$ kpc) systems, many of which are likely interaction-related; twelve compact host-galaxy-scale nebula, likely tracing CGM/ISM gas; and four systems with complex morphologies of uncertain origin. We introduce a quantitative measure of the spatial and kinematic association between nebulae and quasar-host group galaxies, finding a statistically significant association for ten nebulae, most of which are irregular, large-scale nebulae, consistent with qualitative analysis. Radio jets are detected in six systems, with no strong correlation found between radio activity and nebular emission. The $[\rm O\,II]$ nebulae are more asymmetric than their Ly$\alpha$ counterparts at $z>2$, but bear more similarity to H\,I gas observed in 21\,cm around local elliptical galaxies. Blueshifted–redshifted patterns, likely tracing gas rotation, are observed in roughly 30\% of the systems, though disturbed kinematics suggest that feedback may also be important. These results show that giant quasar nebulae are not a uniform class of objects, but instead arise through multiple pathways shaped by host-galaxy gas, galaxy interactions, group environments, and quasar activity, with the most striking cases associated with galaxy interactions.
\end{abstract}


\keywords{quasars: supermassive black holes -- galaxies: groups -- intergalactic medium}

\section{Introduction} \label{sec:intro}
Gas exchange between galaxies and their surrounding environments plays a fundamental role in regulating galaxy growth. Inflows from the circumgalactic and intergalactic medium (CGM and IGM) replenish galactic gas reservoirs, sustaining star formation and black hole growth \citep[e.g.,][]{1997ApJ...477..765C, 2013ApJ...768...74T}. Conversely, chemically enriched material can be redistributed from the interstellar medium (ISM) into the CGM and IGM through tidal interactions \citep[e.g.,][]{2016MNRAS.461.2630M}, ram-pressure stripping \citep[e.g.,][]{2006ApJ...647..910H}, and outflows, driven by mechanisms such as active galactic nuclei (AGN) and stellar feedback \citep[for a review, see][]{2012ARA&A..50..455F, 2018Galax...6..114Z}. Accordingly, characterizing the density, extent, physical state, and kinematics of circumgalactic gas on a galaxy-by-galaxy basis is central to understanding the baryon cycle and, more broadly, galaxy formation and evolution. This remains a long-term priority identified by the 2020 Decadal Survey \citep{2021pdaa.book.....N}.

In the local Universe, extended gas reservoirs can be traced directly through H\,I 21 cm emission \citep[for a review, see][]{2012ARA&A..50..491P}. However, detecting such diffuse 21 cm emission remains infeasible beyond the local Universe, except through stacking large samples of galaxies \citep[e.g.,][]{2022ApJ...937..103C, 2023ApJ...947...16A, 2025ApJS..279...19C, 2025ApJ...993L..18D}. As a result, constraints on the CGM in the distant Universe have largely come from absorption features seen in the spectra of UV-bright background sources passing through the halos of foreground galaxies \citep[for a review, see][]{2017ARA&A..55..389T, 2026enap....4..370C}. This technique typically samples only a single sightline per system, limiting our ability to recover the morphology and spatially resolved kinematics of gas flows, except in rare cases \citep[e.g.,][]{2014MNRAS.438.1435C, 2016MNRAS.458.2423Z, 2018Natur.554..493L, 2024MNRAS.528.1895D, 2024CmPhy...7..286B}. Alternatively, wide-field integral-field spectrographs (IFS) such as the Multi-Unit Spectroscopic Explorer (MUSE; \citealt{2010SPIE.7735E..08B}) allow giant nebulae to be mapped directly, providing spatially resolved constraints on the morphology and kinematics of circumgalactic gas. Recent studies have detected CGM emission around galaxies in Ly$\alpha$, $\rm [O\,II]$, $\rm Mg\,II$, and $\rm Si\,II$ by stacking deep MUSE observations over large galaxy samples \citep[e.g.,][]{2023MNRAS.522..535D, 2023Natur.624...53G, 2024A&A...688A..37G, 2024A&A...691A.255K}. However, these methods are inherently limited to ensemble-averaged properties and do not capture the diversity of individual circumgalactic environments. Although extended nebulae have been detected in a small number of individual systems, such detections generally require exceptionally deep observations (tens to hundreds of hours) and/or particularly favorable targets, and therefore remain limited to rare cases \citep[e.g.,][]{2016A&A...587A..98W, 2019ApJ...878L..33C, 2021MNRAS.507.4294Z, 2022A&A...663A..11L, 2024SciA...10P8629Z}.

Quasars provide a powerful alternative probe. Their intense radiation fields photoionize surrounding gas over large distances, boosting recombination and collisionally excited emission and rendering otherwise diffuse material directly observable. At $z>2$, systematic IFS surveys have shown that giant H\,I Ly$\alpha$ nebulae extending beyond $100\,\mathrm{kpc}$ are ubiquitous around quasars \citep[e.g.,][]{2014Natur.506...63C, 2016ApJ...831...39B, 2019ApJS..245...23C, 2020ApJ...894....3O, 2021MNRAS.503.3044F, 2021MNRAS.502..494M, 2025NatAs...9..577T, 2026A&A...707A.380G}. At $z<1.5$, the capabilities of IFS have led to discoveries of giant nebulae around quasars emitting in $\rm [O\,II]$, $\rm H\beta$, and $\rm [O\,III]$ \citep[e.g.,][]{2018ApJ...869L...1J, 2022ApJ...940L..40J, 2024ApJ...966..218J, 2021MNRAS.505.5497H, 2024MNRAS.527.5429L, 2024ApJ...971..134Z, 2023MNRAS.518.2354C, 2024ApJ...962...98C, 2024A&A...683A.205E, 2025ApJ...984..140L}. Giant quasar nebulae therefore offer a unique opportunity to study the morphology and kinematics of circumgalactic gas on an object-by-object basis. At the same time, the observed gas properties may arise not only from quasar illumination, but also from the physical conditions associated with rapid, radiatively efficient accretion, including enhanced gas inflow, galaxy interactions, mergers, or AGN-driven outflows.

The Cosmic Ultraviolet Baryon Surveys (CUBS; \citealt{2020MNRAS.497..498C}) and the MUSE Quasar Blind Emitters Survey (MUSEQuBES; \citealt{2020MNRAS.496.1013M} and \citealt{2024MNRAS.528.3745D}) investigate the CGM and IGM around galaxies at $z\approx 0.1{-}1.4$, using high-signal-to-noise-ratio COS absorption spectra of UV-luminous quasars. Deep MUSE observations in these fields simultaneously enable comprehensive galaxy redshift surveys and the serendipitous detection of extended nebular emission. Recent work has demonstrated that large ionized circumgalactic nebulae are frequently detected around these quasars \citep{2024ApJ...966..218J}.

Several nebulae in the CUBS+MUSEQuBES sample have been examined in detail, revealing a diversity of origins and physical processes, after careful morphology and kinematics characterization. \citet{2022ApJ...940L..40J} found a nebula showcasing a $> 100 \, \rm kpc$ long filament, likely tracing gas accretion directly onto the quasar host galaxy. \citet{2018ApJ...869L...1J} and \citet{2024MNRAS.527.5429L} discovered extended nebulae shaped by interactions with group galaxies via ram pressure- and tidal-stripping, typically located in massive galaxy groups. \citet{2025ApJ...984..140L} reported a system showing a blueshifted-redshifted velocity pattern and multi-component emission features, suggestive of a combination of gas rotation and AGN feedback. Moreover, \citet{2023MNRAS.518.2354C, 2024ApJ...962...98C} used velocity structure functions (VSFs) of spatially resolved quasar nebulae to place the first empirical constraints on multiscale gas motions, showing that one system is consistent with a Kolmogorov-like cascade and that the ensemble of extended nebulae is broadly consistent with predominantly subsonic turbulence. Although several individual case studies exist, the sample has not yet been uniformly examined with respect to the galactic environment and the origin of the nebulae.

In this paper, we present a systematic study of the nebulae in 30 quasar fields from the CUBS+MUSEQuBES survey. In Section \ref{sec:OD}, we describe the observations and data reduction. In Section \ref{sec:DCN}, we detail the generation of surface brightness maps and the kinematic fitting procedure. In Section \ref{sec:CGE}, we outline the characterization of the galactic environment for each quasar field. In Section \ref{sec:NE}, we examine the implications of nebular morphology and kinematics for individual sources and present our classification scheme. In Section \ref{sec:D}, we investigate the potential origins of the nebulae based on their morphology and kinematics, and explore correlations between galaxies and nebulae. Finally, we summarize our results and make concluding remarks in Section \ref{sec:SC}.

Throughout the paper, we adopt a flat $\Lambda$ cosmology with $\Omega_{\rm m}=0.3$, $\Omega_{\rm \Lambda}=0.7$, and $H_{0} = 70 \, \rm km \, s^{-1} Mpc^{-1}$. All magnitudes are given in the AB system \citep{1983ApJ...266..713O}, unless otherwise stated. All distances are given in proper kpc. 

\section{Observations and Data} \label{sec:OD} 
The CUBS \citep{2020MNRAS.497..498C} and MUSEQuBES \citep{2020MNRAS.496.1013M, 2024MNRAS.528.3745D} surveys together contain a total of 31 UV luminous quasars at $z<1.5$. The CUBS quasars were selected to be NUV bright ($m_{\rm NUV}< 17$) and to lie at $z\approx0.8{-}1.4$. The MUSEQuBES quasars were selected primarily based on the availability of high-S/N FUV spectra (\(\mathrm{S/N}_{\rm G130M}>11\), \(\mathrm{S/N}_{\rm G160M}>5\)) from the Hubble Space Telescope (HST). As a result, the selected quasars primarily lie at \(z \approx 0.4\text{--}1.4\) with \(m_{\rm FUV}<18\). Each quasar field was observed with MUSE on the Very Large Telescope (VLT) as part of CUBS survey (PI: H.-W. Chen, PID: 0104.A-0147) or MUSEQuBES survey (PI: J. Schaye, PID: 094.A-0131(B) \& 096.A-0222(A)). The CUBS observations, conducted in MUSE wide-field mode with ground-layer adaptive optics \citep{2016SPIE.9909E..2SK, 2018SPIE10703E..02M}, have integration times ranging from 1.4 to 3.4 hr and seeing conditions of $\rm FWHM=0.5{-}0.7 \, arcsec$. In contrast, the MUSEQuBES observations were also carried out in the MUSE wide-field mode but under natural seeing conditions ($\rm FWHM=0.5{-}1.0 \, arcsec$) with integration times of 2 to 10 hours. One cube is heavily contaminated by sky lines at the wavelengths corresponding to key emission lines at the quasar redshift, rendering it unusable for this analysis and leaving 30 usable MUSE datacubes (14 from CUBS and 16 from MUSEQuBES). The names and redshifts of these quasar fields are listed in Table \ref{table:nebulae}. Further details on the observations for each quasar field are provided in \citet{2024ApJ...966..218J}. We reduced the MUSE data using three independent pipelines: CubEx \citep{2019MNRAS.483.5188C}, the MUSE GTO team pipeline \citep{2014ASPC..485..451W}, and the ESO reduction pipeline \citep{2012SPIE.8451E..0BW}. All three pipelines produced consistent results, though with some differences in illumination corrections and night-sky-subtraction \citep[for more details, see][]{2020MNRAS.491.2057L, 2024ApJ...966..218J}. All data presented in this work are based on the ESO reduction pipeline. For simplicity, we converted the air wavelengths provided by the three pipelines to vacuum wavelengths.

To achieve higher sensitivity and spatial resolution for galaxy surveys in the quasar fields, we obtained HST imaging for all quasar fields. These images were taken with the Advanced Camera for Surveys (ACS) in the F814W filter (PI: L. Straka, PID: 14660; PI: J. S. Mulchaey, PID: 13024; PI: N. Lehner, PID: 14269; PI: A. Beckett, PID: 17815) with exposure times ranging from 1200 to 3000 seconds. We retrieved the reduced images from the Barbara A. Mikulski Archive for Space Telescopes (MAST) when available. For fields observed most recently (the CUBS fields; Program: DISCS; PI: A. Beckett, PID: 17815), we performed a largely standard HST ACS reduction, incorporating procedures described in \citet{2022ApJ...924...14P} and \citet{2025ApJ...992..155B}. Briefly, we inspected the individual exposures and rejected frames with failed guide-star lock, masked large-scale artifacts, and corrected for background offsets between the two chips. The cleaned exposures were then combined using the \texttt{TWEAKREG} and \texttt{ASTRODRIZZLE} routines \citep{2012drzp.book.....G, 2021drzp.book....2H}. For the HST ACS observations of J0119$-$2010, the dataset is currently incomplete due to guiding failures, thus we adopt the STScI-processed drizzled image produced from the two available exposures. 

To ensure uniform astrometry across datasets, we aligned the MUSE datacubes and HST images to the Gaia reference frame using objects detected in both Gaia \citep{2023A&A...674A...1G} and Legacy Surveys \citep{2019AJ....157..168D} with the \texttt{Astrometry} package \citep{2022zndo...6462441W}. This procedure yields an astrometric accuracy of approximately $0.10''$ and $0.08''$, based on the standard deviation of the residual offsets of sources in Legacy Survey versus MUSE and HST, respectively. For each field, astrometric alignment uses between 30 and 100 matched objects, depending on source density.

\section{Detection and Characterization of the Nebulae} \label{sec:DCN}
We adopted the same methodology as \citet{2024ApJ...966..218J} to recover the quasar nebulae in the CUBS+MUSEQuBES survey, including quasar light subtraction, continuum subtraction, and optimal extraction.

As a first step, we modeled and subtracted the unresolved quasar light including both the continuum and the line contributions from the narrow- and broad-line regions, following the approach of \citet{2018ApJ...869L...1J} and \citet{2021MNRAS.505.5497H}. This method does not assume a fixed PSF shape, but instead uses the spectral information in the IFU datacube and the fact that quasar and galaxy spectral energy distributions differ. In particular, the wavelength dependence of the seeing causes the contaminating spectrum from the quasar to appear relatively shallow close to the quasar centroid and steeper at larger radii. To capture this effect, the quasar contribution was modeled as a linear combination of two non-negative matrix factorization components derived from quasar-dominated central spaxels: one component that approximates the artificially flattened quasar contamination closer to the centroid, and another component that approximates the artificially steeper quasar contamination farther from the centroid. For each contaminated spaxel, the observed spectrum was fit with these quasar components together with galaxy eigenspectra representing the host/galaxy contribution \citep[see also][]{2017ApJ...850...40R}. We then subtracted only the best-fit quasar component from the datacube, removing the nuclear continuum and broad/narrow emission-line contamination while avoiding over-subtraction of extended galaxy and nebular emission. Earlier studies have shown that this method successfully eliminates the spatially unresolved nuclear continuum as well as both broad- and narrow-line emission components. 

To remove continuum from galaxies or stars, we performed continuum subtraction over the entire datacube in a local wavelength range around the $\rm [O\,II]$ and $\rm [O\,III]$ emission lines, as described in \citet{2024MNRAS.527.5429L}. We first fit the continuum on the blue and red sides of these lines, excluding the spectral region within $\pm 500{-}1000 \ \mathrm{km \ s^{-1}}$ of the expected observed [O\,II] and [O\,III] wavelengths at the quasar redshift. The exact masking range is adjusted as needed to minimize night-sky contamination while retaining adequate wavelength coverage for the continuum fit. We then conducted 3D segmentation, which is widely applied in the detection of giant Ly$\alpha$ nebulae at higher redshift \citep{2016ApJ...831...39B, 2019A&A...631A..18A, 2021ApJ...923..252S}. 

To perform the 3D segmentation, we smoothed the datacube in both spatial and spectral dimensions with a Gaussian kernel with standard deviation of $1.5$ pixel and constructed a corresponding $\rm  S/N$ datacube. We note that this smoothing level effectively reduces read noise without reducing detectability of narrow features because it is comparable to the PSF and LSF. For each wavelength slice, we identified spaxels with $\rm S/N > 1.5{-}3.0$, grouped adjacent spaxels above this threshold, and required at least 10 connected spaxels to define a detection. Most nebulae are extracted using a threshold of ${\rm S/N} > 3.0$, though we modified the threshold in some cases to account for local noise levels due to the presence of imperfect sky-line subtraction. \textcolor{black}{We note that the estimated uncertainty does not fully characterize the effects of imperfect sky-line subtraction. Residuals from strong sky lines can be spatially correlated across neighboring spaxels, so although the estimated uncertainty reflects the average noise level, it does not capture the correlated residual structures associated with individual sky lines.} Experimentation with different ${\rm S/N}$ thresholds demonstrates that threshold choice does not substantively change our results. We used the detection with the largest connected area to define the spatial segmentation and located the wavelength slice where this area peaked. Starting from this slice, we extended the segmentation spectrally in both increasing and decreasing directions, identifying the contiguous wavelength range that satisfied the $\rm S/N$ threshold for each spaxel and stopping when the condition failed. After masking the detected region, we repeated the procedure for the next largest structure until no additional detections remained. Finally, we integrated the unsmoothed flux in each spaxel over the spectrally defined interval to generate surface brightness maps. For pixels below the $\rm S/N$ threshold, we adopted a background level based on three spectral pixels centered at the wavelength where the mean S/N per pixel of the nebula is the highest. We note that the spectral range used prior to 3D segmentation was individually selected and adjusted for each field to minimize sky-line contamination and optimize nebular emission detection. Additionally, imperfections in the continuum subtraction may lead the 3D segmentation to incorrectly identify residuals from stars or galaxies at different redshifts within the continuum-subtracted datacubes. In such cases, we manually removed the segments after inspecting their images and spectra. 

The resulting $\rm [O\,II]$ and $\rm [O\,III]$ surface brightness maps for these nebulae are shown in the second and third columns of Figures \ref{fig:L_1}-\ref{fig:N_1}. \textcolor{black}{To better illustrate the spatial extent of the nebulae, we overlay contour lines at the surface brightness levels labeled in each panel, in units of $10^{-17}\,\rm erg\,cm^{-2}\,s^{-1}\,arcsec^{-2}$. These contours are intended solely to visualize the extent of the nebulae and are independent of the segmentation threshold adopted to define a detection. In each panel, the contour corresponds to the $3\sigma$ detection limit, averaged over an area of 1 arcsec$^2$ using the variance measured in the corresponding image. The [O\,II] contour level is identical to that adopted by \citet{2024ApJ...966..218J}.} To better visualize the nebula around the quasar, we present HST ACS+F814W images in the first column of Figures \ref{fig:L_1}-\ref{fig:N_1}, overlaid with $\rm [O\,II]$ and $\rm [O\,III]$ emission contours respectively. 

To measure the kinematics of each nebula, we jointly fit Gaussian line profiles to the quasar- and continuum- subtracted $\rm [O \, II]$ and $\rm [O \, III]$ datacubes. We first convolved the data spatially with a Gaussian kernel with standard deviation of $0.3''$, chosen to match the typical seeing disk. We then used the segmentation map from the optimal extraction to define the spatial fitting region, while adopting a broad spectral window that includes both the full line emission and sufficient continuum. Although we constructed the segmentation map using both spatial and spectral smoothing, we performed the line fitting without spectral smoothing. For each spaxel, we began with a single kinematic component and jointly modeled the $\rm [O\,II]$ doublet with two Gaussians and $\rm [O\,III]$ with one Gaussian, with shared redshift and velocity dispersion parameters across the lines. Prior to the final fit, the observed line width is modeled as the quadrature sum of the velocity dispersion and the wavelength-dependent instrumental line spread function (LSF) dispersion. In most cases, the kinematic parameters shared by $\rm [O\,II]$ and $\rm [O\,III]$ result in good fits to the data. In spaxels where the line profiles of $\rm [O\,II]$ and $\rm [O\,III]$ are inconsistent, we allowed the $\rm [O\,III]$ component to be fit at a different velocity than the $\rm [O\,II]$ component. 

Most regions in most nebulae are well fit by a single kinematic component, with fewer than 5\% of spaxels requiring multiple components (normally with shared redshift and velocity dispersion parameters). However, in 3C\,57, J2135$-$5316, and J0119$-$2010, a substantial number of spaxels require two or even three kinematic components to achieve a good fit. The typical velocity offsets between components range from $100$ to $1500\ \mathrm{km\ s^{-1}}$, while the velocity dispersion of individual components spans $30{-}400\ \mathrm{km\ s^{-1}}$. As a result, some components are strongly blended, whereas others are well separated in velocity space. The threshold of whether adopting higher number of components is indicated by the resulting $\chi^2$ and the Bayesian Information Criterion (BIC). When the addition of extra components resulted in a lower BIC, we selected the model with the larger number of components \citep[for details on this approach, see][]{2007MNRAS.377L..74L}. Complex kinematic substructure is easier to detect in spaxels with higher S/N. Consequently, line asymmetries near a nebular outskirts may be missed because of the lower S/N in those regions. Therefore, we calculated a total S/N for the flux summed along the spectral channels defined by the segmentation map. To ensure highly robust kinematic maps, we only display fitted kinematic results for spaxels with total $\rm{S/N}>7$. When both $\rm [O\,II]$ and $\rm [O\,III]$ are available, we calculated the total S/N for each line separately and then combine the two values in quadrature. In all other cases, we computed the S/N using only the $\rm [O\,II]$ line. For nebulae including HE\,0226$-$4110 and TXS\,0206$-$048, we adjusted to higher values due to the presence of strong sky line. Finally, we visually inspected all spaxels in each nebula to check if the best-fit model sufficiently reproduces the data. 

\begin{table*}
\caption{Summary of the nebulae around the CUBS and MUSEQuBES quasars, ordered by classification and redshift.}
\label{table:nebulae}
\begin{tabularx}{\linewidth}{lcrrrrrrrl}
\hline
\hline
Quasar & $z$$^a$ & $N_{\rm gal}$$^b$ & $N_{\rm enc}$$^c$ & $95\%\, \mathrm{interval\ of}\ v_{50} \, ^{d}$ & median $\sigma$$^e$    & $\rm A_{OII}$$^f$ & size$^g$ & CKAF (p-value)$^h$ & Note$^i$\\
       &         &                   &                   & ($\rm km \, s^{-1}$)                           & ($\rm km \, s^{-1}$)   &                   &  (kpc)   &                    &        \\
\hline
Irregular, Large-scale \\
\hline
HE\,0226$-$4110$^*$     & 0.4936  &  10  &  2  & [ $-365$, $363$] & 139& 0.88 & 84  & 55  (0.23) &       \\
PKS\,0405$-$123$^{*,j}$ & 0.5731  &  37  & 12  & [$-1042$, $193$] & 99 & 1.70 &129  & 842 (<0.01)& BR    \\
HE\,0238$-$1904$^{*,k}$ & 0.6282  &  34  &  4  & [ $-156$, $610$] & 102& 1.28 &103  & 600 (<0.01)& BR, MC \\
PKS\,0552$-$640$^{*,l}$ & 0.6824  &  11  &  3  & [ $-494$, $177$] & 107& 0.88 &153  & 160 (0.04) &       \\
J0454$-$6116$^{\#, l}$  & 0.7864  &  18  &  2  & [  $-85$, $614$] & 141& 0.83 &102  & 102 (0.39) &       \\
J0119$-$2010$^\#$       & 0.8160  &  8   &  1  & [$-361$,  $985$] & 140& 0.71 & 96  & 21  (0.44) & MC    \\
HE\,0246$-$4101$^\#$    & 0.8840  &  7   &  4  & [ $-134$, $180$] & 109& 1.23 & 74  & 276 (<0.01)&       \\
PKS\,0355$-$483$^\#$    & 1.0128  &  25  &  5  & [$ -182$, $577$] & 136& 1.18 & 50  & 213 (<0.01)&       \\
HE\,0439$-$5254$^*$     & 1.0530  &  7   &  1  & [  $-42$, $484$] & 215& 0.78 & 47  & 56  (0.02) &       \\
TXS\,0206$-$048$^{*,j}$ & 1.1317  &  32  &  5  & [$ -442$, $598$] & 186& 1.08 &200  & 86  (0.91) &       \\
Q1354$+$048$^*$         & 1.2335  &  17  &  2  & [$  -32$, $642$] & 90 & 1.42 &126  & 118 (0.02) &       \\
\hline
Host-Galaxy-Scale \\
\hline
HE\,0435$-$5304$^*$     & 0.4279  &  2   &  0  & [  $-98$, $ 69$] &  80& 0.90 & 55  & 0   (0.16) & BR    \\
3C\,57$^{*,k}$          & 0.6718  &  5   &  1  & [$ -202$, $296$] & 148& 0.67 & 71  & 56  (0.04) & BR, MC\\
J0110$-$1648$^\#$       & 0.7822  &  2   &  0  & [ $ -80$, $329$] &  81& 0.93 & 29  & 0   (0.50) & BR    \\
HE\,0112$-$4145$^\#$    & 1.0238  &  9   &  1  & [$ -233$, $316$] & 165& 0.54 & 38  & 47  (0.04) & BR    \\
J0154$-$0712$^\#$       & 1.2957  &  7   &  1  & [$ -413$, $275$] & 136& 1.80 & 63  & 15  (0.14) & BR    \\
Q1435$-$0134$^*$        & 1.3117  &  31  &  0  & [$ -122$, $706$] & 248& 1.55 & 63  & 12  (0.90) & BR    \\
PG\,1522$+$101$^*$      & 1.3302  &  11  &  0  & [$ -324$, $917$] & 112& 1.21 & 50  & 0   (0.90) & BR   \\
\hline
J0028$-$3305$^\#$       & 0.8915  &  6   &  0  & [ $ -75$, $167$] & 93 & 1.48 & 42  & 0   (0.55) &     \\
HE\,0419$-$5657$^\#$    & 0.9481  &  3   &  1  & [ $-335$, $299$] & 140& 1.41 & 35  & 47  (0.01) &     \\
Q0107$-$025$^{*, m}$    & 0.9545  &  7   &  0  & [ $ -3 $, $295$] & 110& 1.35 & 28  & 0   (0.62) &     \\
HE\,1003$+$0149$^*$     & 1.0807  &  3   &  0  & [$ -418$, $845$] & 220& 0.45 & 53  & 0   (0.42) &     \\
HE\,0331$-$4112$^\#$    & 1.1153  &  7   &  1  & [$-1058$, $524$] & 200& 0.88 & 32  & 3   (0.37) &     \\
\hline
\multicolumn{10}{l}{Complex Morphology and Kinematics}\\ 
\hline
J2135$-$5316$^{\#, l}$  & 0.8123  &  2   &  0  & [$ -344$, $159$] & 101& 0.57 & 83  & 5   (0.17) & MC   \\
Q0107$-$0235$^{*, m}$   & 0.9574  &  22  &  0  & [$ -581$, $424$] & 125& 0.84 & 90  & 34  (0.61) &       \\
PKS\,2242$-$498$^\#$    & 1.0011  &  8   &  0  & [$ -303$, $414$] & 141& 1.04 & 71  & 0   (0.90) & BR   \\
PKS\,0232$-$04$^*$      & 1.4450  &  6   &  0  & [$ -535$, $340$] & 162& 0.84 &116  & 0   (0.94) & BR   \\
\hline
No Nebulae \\
\hline
HE\,0153$-$4520$^*$     & 0.4532  &  0   &  0  &     X            &   X&   X  &  7  & X   &     \\
HE\,2305$-$5315$^\#$    & 1.0733  &  4   &  0  &     X            &   X&   X  & 15  & X   &     \\
HE\,2336$-$5540$^\#$    & 1.3531  &  16  &  0  &     X            &   X&   X  &  7  & X   &     \\
\hline
\multicolumn{10}{l}{$^{*}$ and $^{\#}$ denote quasar fields from the MUSEQuBES and CUBS surveys, respectively.}\\ 
\multicolumn{10}{l}{$^a$ Quasar systemic redshift, see \citet{2024ApJ...966..218J}.}\\ 
\multicolumn{10}{l}{$^b$ Number of group galaxies excluding the quasar host galaxy in the MUSE $60'' \times 60''$ FoV with $|\Delta v| < \rm 3000 \, km \, s^{-1}$ and iterative sigma-clipping.}\\ 
\multicolumn{10}{l}{$^c$ Number of group galaxies projected within the spatial extent of the main nebular structure, excluding the quasar host galaxy.}\\ 
\multicolumn{10}{l}{$^d$ The 95\% interval (2.5th{-}97.5th percentiles) of the nebular $v_{50}$ distribution for each pixel.}\\ 
\multicolumn{10}{l}{$^e$ Median value of the nebular $\sigma$ map.}\\ 
\multicolumn{10}{l}{$^f$ Shape asymmetry using [O\,II] segmentation map.}\\ 
\multicolumn{10}{l}{$^g$ Nebula size (kpc) in [O\,II], estimated using the same methodology as \cite{2024ApJ...966..218J}.}\\ 
\multicolumn{10}{l}{$^h$ Nebular CKAF values, quantify the overall spatial and kinematic coincidence between each nebula and galaxies in the quasar host group.}\\
\multicolumn{10}{l}{The corresponding p-values are also listed and are defined as the fraction of 10,000 simulations in which the CKAF exceeds the observed value.}\\
\multicolumn{10}{l}{$^i$ Additional nebular features are listed; BR denotes blueshifted–redshifted and MC indicates multi-component emission.}\\ 
\multicolumn{10}{l}{$^j$ For more details on these nebulae and their environment, see \cite{2018ApJ...869L...1J, 2022ApJ...940L..40J}.}\\ 
\multicolumn{10}{l}{$^k$ For more details on these nebulae and their environment, see \cite{2023ApJ...943L..25Z}, \cite{2024MNRAS.527.5429L}, and \cite{2025ApJ...984..140L}.}\\ 
\multicolumn{10}{l}{$^l$ For more details on these nebulae and their environments, see \cite{2023MNRAS.518.2354C, 2024ApJ...962...98C}.}\\ 
\multicolumn{10}{l}{$^m$ Q0107$-$025 is referred to as Q0107$-$025B, and Q0107$-$0235 as Q0107$-$025A, in \citet{2023MNRAS.521.1113B}.}\\
\end{tabularx}
\end{table*}

\begin{figure*}
    \centering
    \includegraphics[scale=0.6]{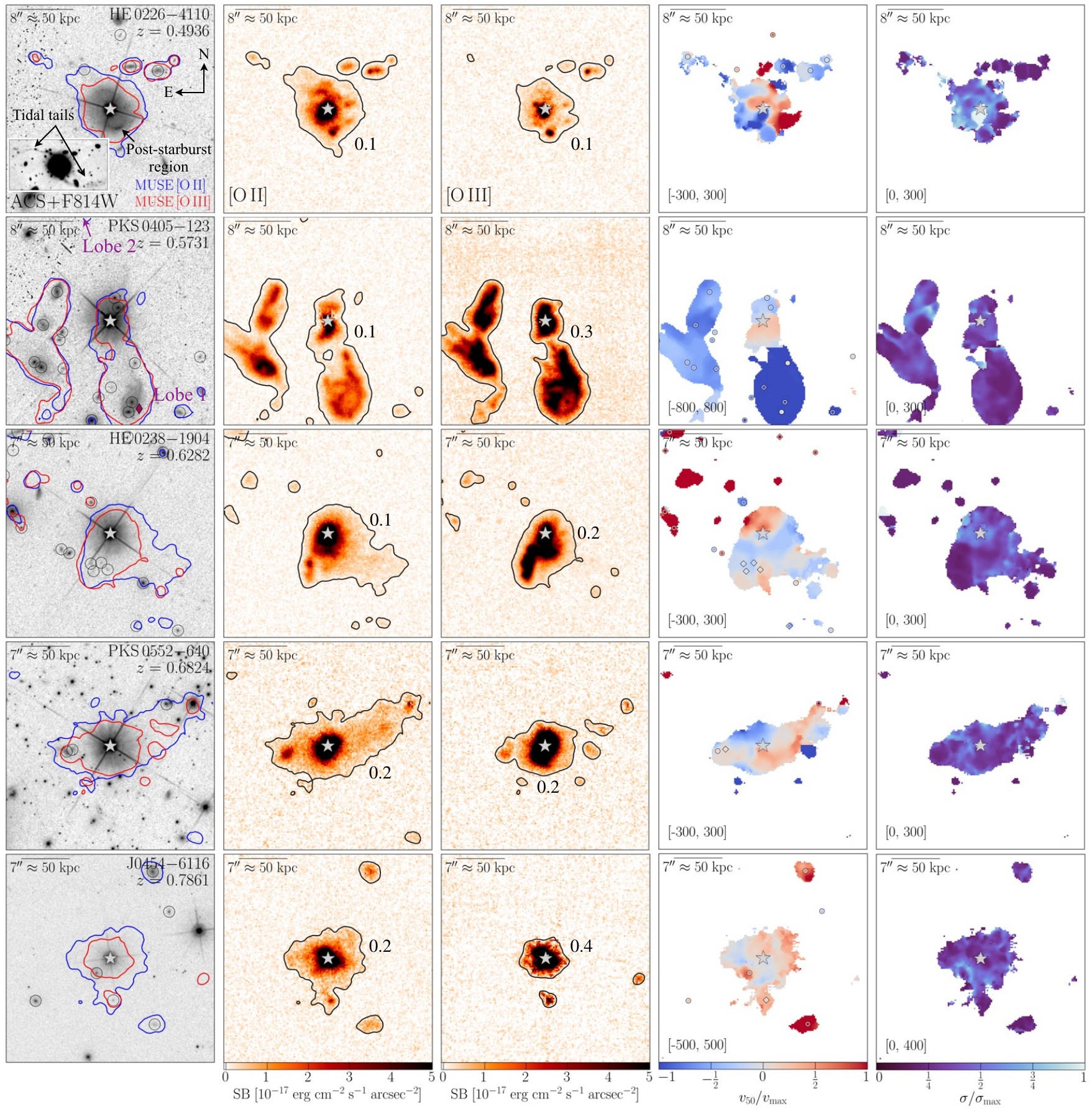}
    \caption{Visualization of irregular, large-scale nebulae (L) around quasars in $30'' \times 30''$ cutout. First column: HST ACS+F814W, with group galaxies marked by black circles. Quasar names and redshifts are indicated in the top right of each panel. A scale bar corresponding to 50 proper kpc at the redshift of the quasar is shown in the top left of each panel. Radio lobes within the FoV are marked with purple thin diamonds, while lobes outside the FoV are indicated by purple arrows. Second and Third columns: narrow-band $\rm [O \, II]$ and $\rm [O \, III]$ surface brightness maps generated from optimal extraction. These maps are overlaid with $\rm [O\,II]$ and $\rm [O\,III]$ surface brightness contours, with contour levels labeled in each panel in units of $10^{-17}\,\rm erg\,cm^{-2}\,s^{-1}\,arcsec^{-2}$. The $\rm [O\,II]$ contour level is identical to that used by \citet{2024ApJ...966..218J}. $\rm [O \, II]$ and $\rm [O\,III]$ contours are also overlaid on the HST image in blue and red respectively. Fourth and Fifth columns: Maps of the nebular $v_{50}$ and $\sigma$ relative to the quasar systemic velocity. Group galaxies are marked with circles if absorption lines are detected and diamonds if only emission lines are present. Their LOS velocities are color-coded within the circles/diamonds using the same colorbar as the nebula. The extrema of the colorbar are labeled in the bottom left of each panel, with the corresponding colormap shown in the last row. The inset panel in the HST image of HE\,0226$-$4110 is a white light image generated from the MUSE datacube.}
    \label{fig:L_1}
\end{figure*}

\begin{figure*}
    \centering
    \addtocounter{figure}{-1} 
    \renewcommand{\thefigure}{\arabic{figure}} 
    \includegraphics[scale=0.6]{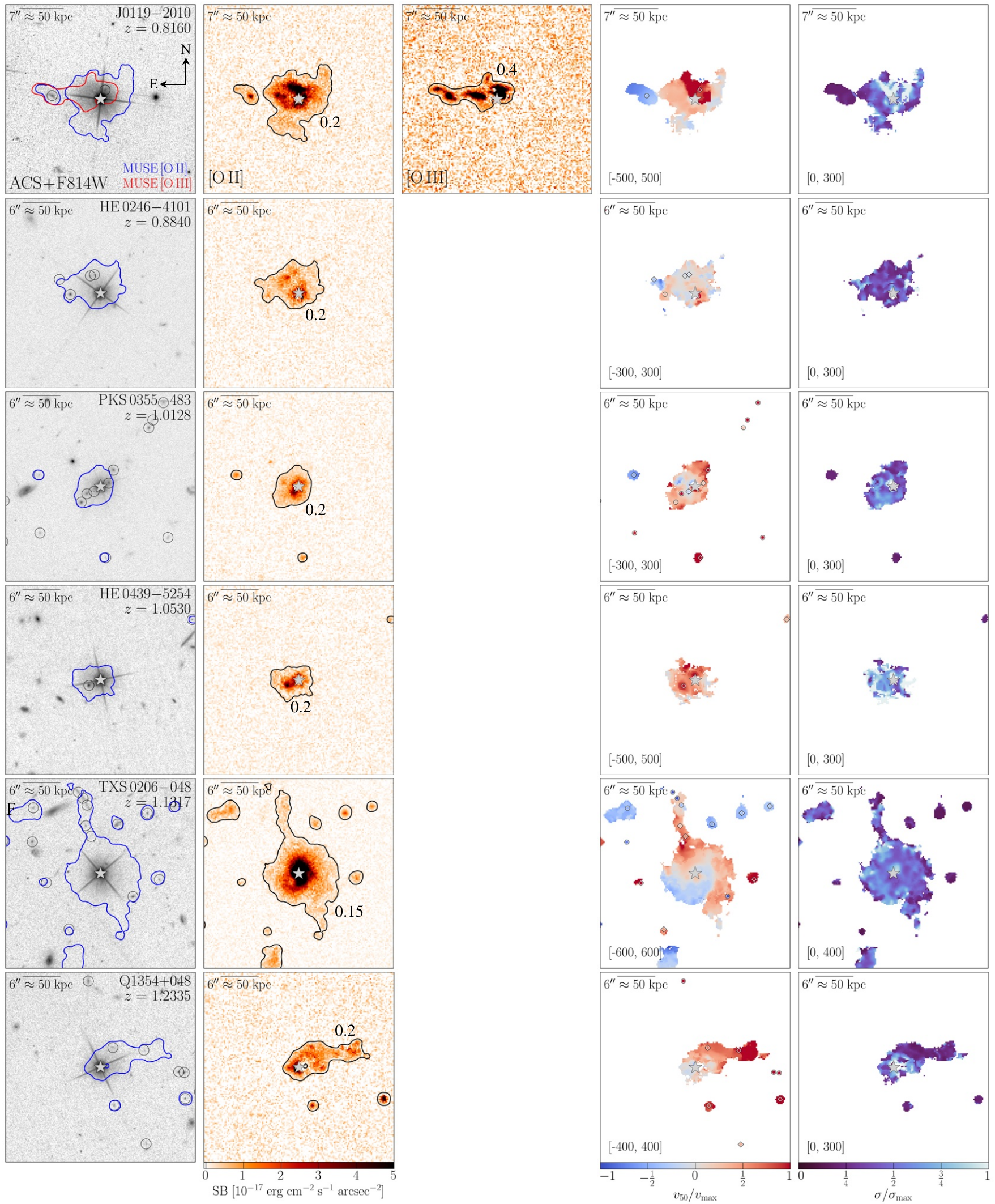}
    \caption{Continued for irregular, large-scale nebulae (L).}
    \label{fig:L_2}
    \renewcommand{\thefigure}{\arabic{figure}} 
\end{figure*}

To characterize the nebular kinematics, we produced LOS velocity ($v_{50}$) relative to the quasar systemic velocity and velocity dispersion ($\sigma$) maps, which visualize the line center and width. The quasar systemic velocity is derived from emission-line fitting \citep[see][]{2024ApJ...966..218J}. These quantities are derived from the cumulative flux distribution of the fitted Gaussian line profiles. Specifically, $v_{50}$ is defined as the velocity accumulating 50\% of the total model flux, while $\sigma$ is computed from the velocity interval between the 10\% and 90\% cumulative flux levels (also known as $\rm W_{80}$), divided by 2.563. By definition, $v_{50}$ and $\sigma$ will respectively equal to the line centroid and standard deviation for nebulae that are well characterized by a single Gaussian kinematic component. We adopt \(\sigma\) rather than \(\rm W_{80}\) as our primary kinematic measure to enable direct comparison with other studies, particularly 21 cm studies that commonly report velocity dispersion. To retain comparability with the widely used \(\rm W_{80}\), we define \(\sigma\) as a fixed multiplicative scaling of \(\rm W_{80}\), allowing straightforward conversion between the two measures. Whenever available, we used the $\rm [O\,III]$ emission line to derive these kinematic maps, since the $\rm [O\,II]$ doublet introduces additional complexity in the line profile. For spaxels without detectable $\rm [O\,III]$, we instead relied on $\rm [O\,II]$. In spaxels where both lines are detected, the measured $v_{50}$ and $\sigma$ values are generally consistent between $\rm [O\,II]$ and $\rm [O\,III]$. We show the resulting $v_{50}$ and $\sigma$ maps in fourth and fifth columns in Figures \ref{fig:L_1}-\ref{fig:N_1}. Additionally, we summarize the nebular kinematics in Table \ref{table:nebulae}, where the velocities are given by the 2.5th–97.5th percentile range of the spaxel distribution in the spatially resolved $v_{50}$ map, and the velocity dispersion is given by the median value of the spatially resolved $\sigma$ map.

\section{Characterization of the Galactic Environment} \label{sec:CGE}
Previous case studies have shown that giant quasar nebulae can originate from ISM gas of nearby galaxies. This gas produces extended nebular emission when it is stripped through tidal interactions or ram-pressure effects and subsequently illuminated by the quasar \citep{2018ApJ...869L...1J, 2021MNRAS.505.5497H, 2024MNRAS.527.5429L}. These findings highlight the role of the surrounding galaxies in supplying material to the quasar environment. It is therefore essential to characterize the galactic environment of quasar fields.

To characterize the galactic environments of the quasar fields, we conducted galaxy surveys for each field in the CUBS+MUSEQuBES sample, following the methodology described in \citet{2021MNRAS.505.5497H} and \citet{2024MNRAS.527.5429L}. We first identified candidate galaxies as continuum sources in the background-subtracted MUSE white-light and ACS$+$F814W images, using detection thresholds of $1.25\sigma$ and $1\sigma$, respectively. We then combined the two source catalogs and visually inspected the resulting masks to remove artifacts. These deliberately marginal thresholds were adopted to minimize the chance of missing faint, but real sources. As a result, many detected objects are too faint for reliable redshift measurements. We extracted the galaxy spectra from the MUSE datacube at the continuum centroid using an aperture of $\rm radius=0.6''$. We then fit each object with linear combinations of SDSS galaxy eigenspectra \citep{2012AJ....144..144B} to determine its redshift. For each fit, we evaluated models over a redshift grid spanning $z=0{-}2.0$ with $\Delta z=0.0001$ and adopted the solution that minimized the global $\chi^2$. We visually inspected every best-fit spectrum to confirm the robustness of the measurement. When spectra showed both emission and absorption features, we masked the strong emission lines and measured the redshift from stellar absorption features. This step mitigates contamination from nebular emission that may overlap with the galaxies in projection but is not physically associated with them. From previous applications of this method, we estimate a typical redshift uncertainty with a standard deviation of approximately $ 20\,\mathrm{km\,s^{-1}}$ \citep{2021MNRAS.505.5497H}. We note that this value applies under conditions of sufficient S/N, and the uncertainty is expected to increase at lower S/N. Based on the background sky noise and source counts in the ACS$+$F814W image, we estimate that the imaging catalog is complete to $m_{\rm F814W} \approx 26\text{--}27$, with the precise completeness limit depending on source angular size.

We first defined candidate group member galaxies across the MUSE FoV ($\approx 300-500\, \rm kpc$) as those within LOS velocities $|\Delta v| < \rm 3000 \, km \, s^{-1}$ relative to the quasar systemic velocity. We then robustly identified group members using an iterative Gaussian maximum-likelihood sigma-clipping procedure, following the method adopted by \citet{2024ApJ...965..143L}. At each iteration, we estimated the group mean velocity and velocity dispersion from the candidate velocity distribution, removed galaxies lying outside $3\sigma$ from the current mean velocity, and repeated the procedure until the membership converges. The final member sample therefore represents a kinematically coherent galaxy structure, likely associated with the quasar. For each field, we report the total number of candidate group galaxies (many of them lie outside the cutout images shown in Figures \ref{fig:L_1}-\ref{fig:N_1}) and the number of candidate group galaxies projected within the spatial extent of the nebula (considering only the main structure and excluding individual galaxies enclosed within their own emission) in Table \ref{table:nebulae}. Compared to previous studies, we identify seven more galaxies in the field of PKS\,0405$-$123 than \citet{2018ApJ...869L...1J}, and five more galaxies in the field of TXS\,0206$-$048 than \citet{2022ApJ...940L..40J}, due to our broader velocity criterion for group candidacy. In Figures \ref{fig:L_1}-\ref{fig:N_1}, we also mark the centroid of group galaxies in the HST ACS+F814W images for each quasar field. Additionally, we show their LOS velocity relative to the quasar systemic velocity in the $v_{50}$ maps and we mark galaxies with absorption-line redshifts as circles and those based solely on emission lines as diamonds. We note that some galaxies are identified via emission lines although they are not captured by the optimal extraction and do not show emission in surface brightness maps because emission from individual galaxies often does not meet the spatial extension or S/N threshold in the segmentation process (see Section \ref{sec:DCN}).

\section{The Nebular Morphology and Kinematics} \label{sec:NE}
Giant quasar nebulae display a wide variety of morphologies and kinematic signatures. Some nebular emission extends over $100\, \rm kpc$, while other nebular structures remain confined within $\approx10 \, \rm kpc$ around the quasar, exhibiting significant diversity in both morphology and kinematics. For instance, \citet{2022ApJ...940L..40J} presented a nebula with a striking $\approx 100 \, \rm kpc$ filament. \citet{2025ApJ...984..140L} identified a nebula exhibiting blueshifted-redshifted pattern similar to rotating disks often seen in H\,I 21 cm. Additionally, \citet{2018ApJ...869L...1J}, \citet{2021MNRAS.505.5497H}, and \cite{2024MNRAS.527.5429L} reported large-scale nebulae with size of $>3000 \rm \, kpc^2$ and narrow line widths. In this section, we define four classes that characterize the morphology and kinematics of each nebula in the CUBS+MUSEQuBES sample. These categories provide a framework for exploring the potential shared physical origins within each group.

\begin{enumerate}
    \item L: Irregular, large-scale ($\approx 50{-}200 \, \rm kpc$) morphology with disordered kinematics, often encompassing multiple group galaxies. Such properties suggest that the emitting gas may trace group-scale circumgalactic or intragroup gas. Possible physical origins include tidal debris from galaxy interactions, ram-pressure-stripped gas, filamentary accretion, and/or cool gas embedded in the group environment. Please see Paragraph 2 of Section \ref{TBO} for a more detailed discussion of their possible origins.
    
    \item S: Host-galaxy–scale ($\approx 30{-}70 \, \rm kpc$), centered around the quasar. These systems generally do not encompass group galaxies and therefore appear more closely connected to the quasar host galaxy. Their kinematics are diverse, ranging from ordered blueshifted--redshifted velocity patterns to more disturbed structures. These features suggest that they may originate from gas associated with the quasar host, potentially tracing rotating or dynamically disturbed gas. Please see Paragraph 5 of Section \ref{TBO} for a more detailed discussion of their possible origins.
    
    \item A: Complex morphology and kinematics, exhibiting a combination of the kinematic and morphological characteristics described above, but with additional ambiguous features. These characteristics do not provide clear evidence for a unique classification, making it difficult to assign them unambiguously to either the L or S class. Please see Paragraph 6 of Section \ref{TBO} for a more detailed discussion of their possible origins.
    
    \item N: No extended nebular emission is detected around the quasars.
\end{enumerate}

\begin{figure*}
    \centering
    \includegraphics[scale=0.6]{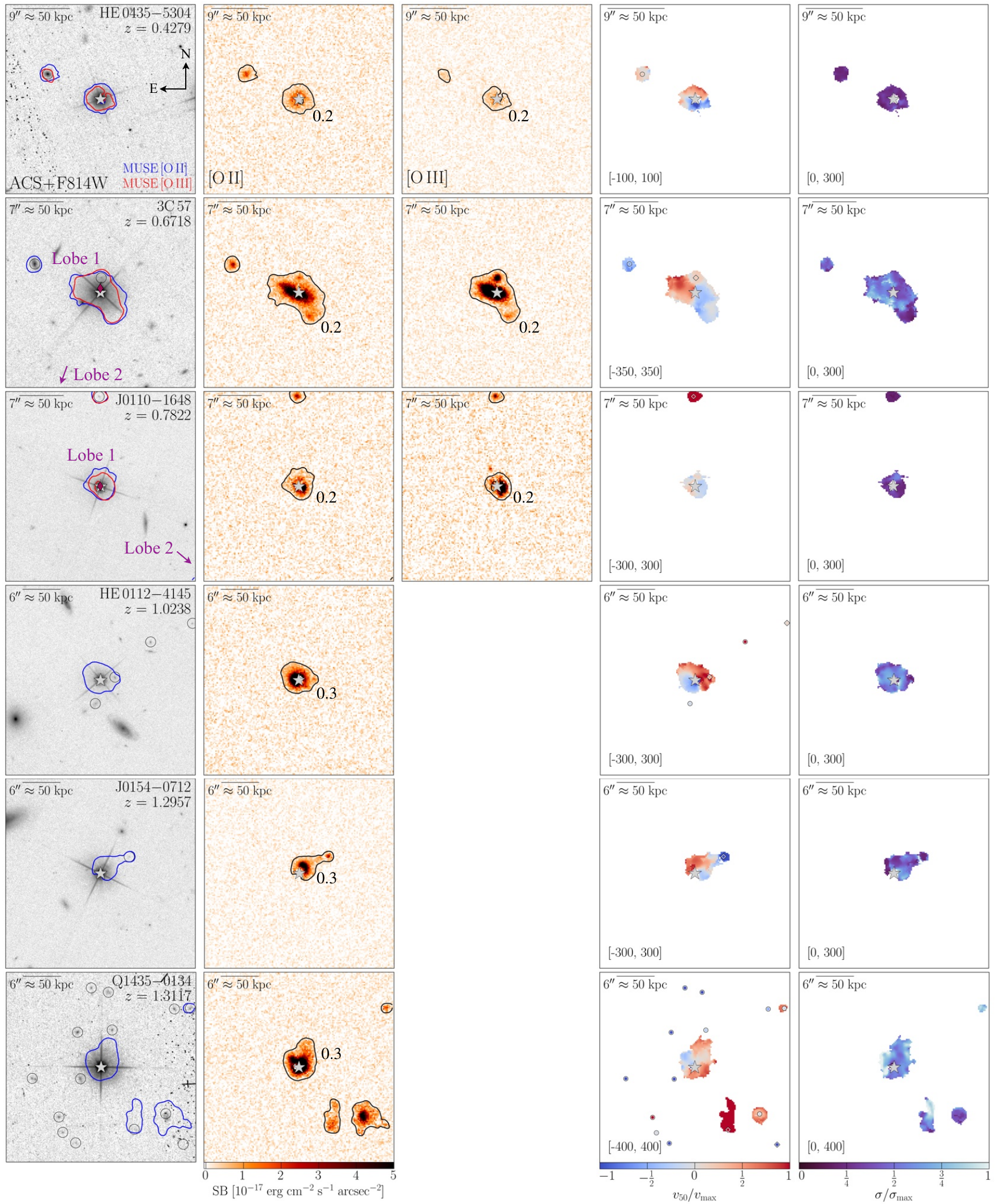}
    \caption{Same as Figure \ref{fig:L_1}, but for host-galaxy–scale nebulae (S) with blueshifted-redshifted kinematics pattern and patternless kinematics. The two subgroups are separated by a black line.}
    \label{fig:S_1}
\end{figure*}

\begin{figure*}
    \centering
    \addtocounter{figure}{-1} 
    \renewcommand{\thefigure}{\arabic{figure}} 
    \includegraphics[scale=0.6]{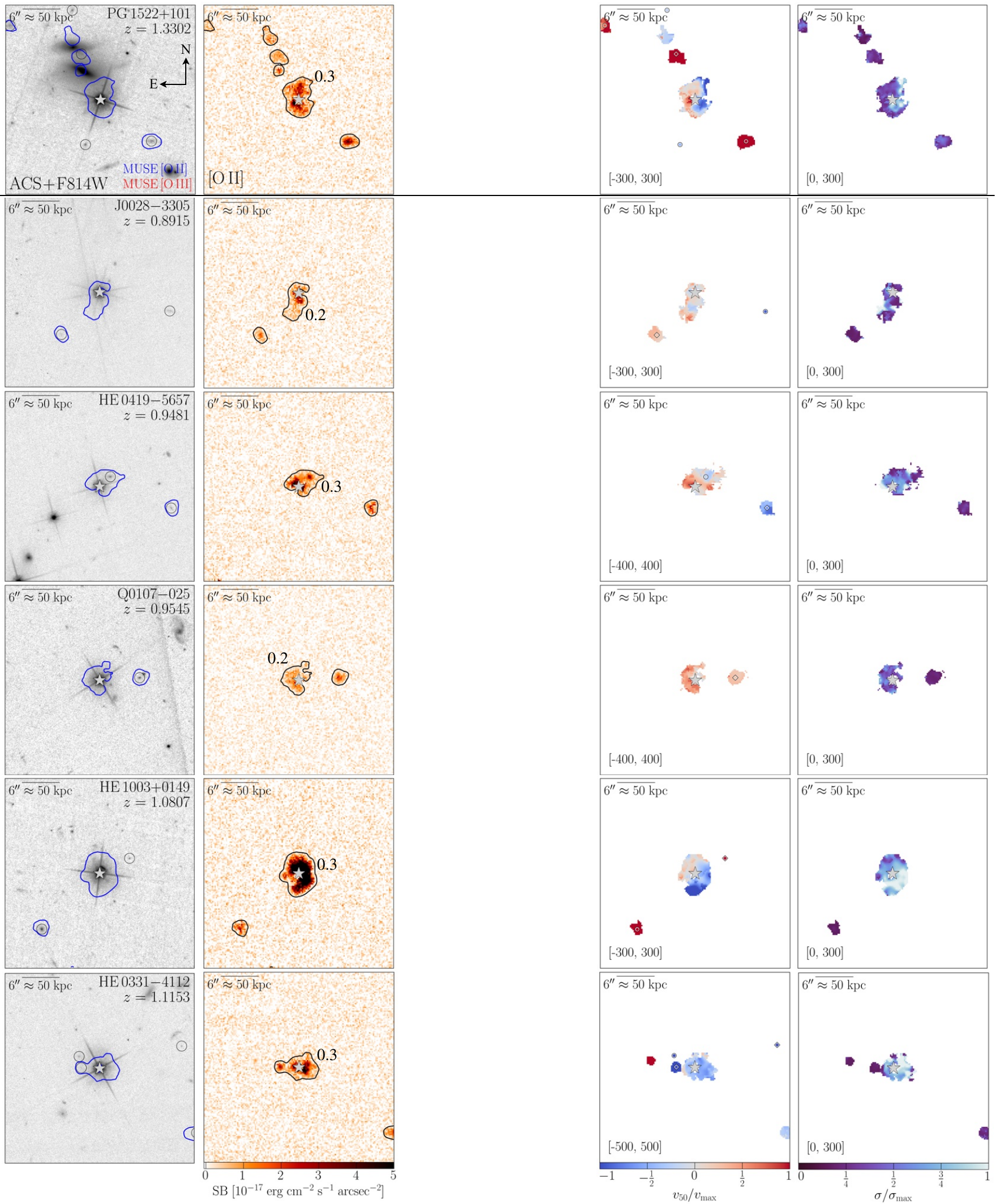}
    \caption{Continued for host-galaxy–scale nebulae (S) with blueshifted-redshifted kinematics pattern and patternless kinematics. The two subgroups are separated by a black line.}
    \label{fig:S_2}
    \renewcommand{\thefigure}{\arabic{figure}} 
\end{figure*}

\begin{figure*}
    \centering
    \includegraphics[scale=0.6]{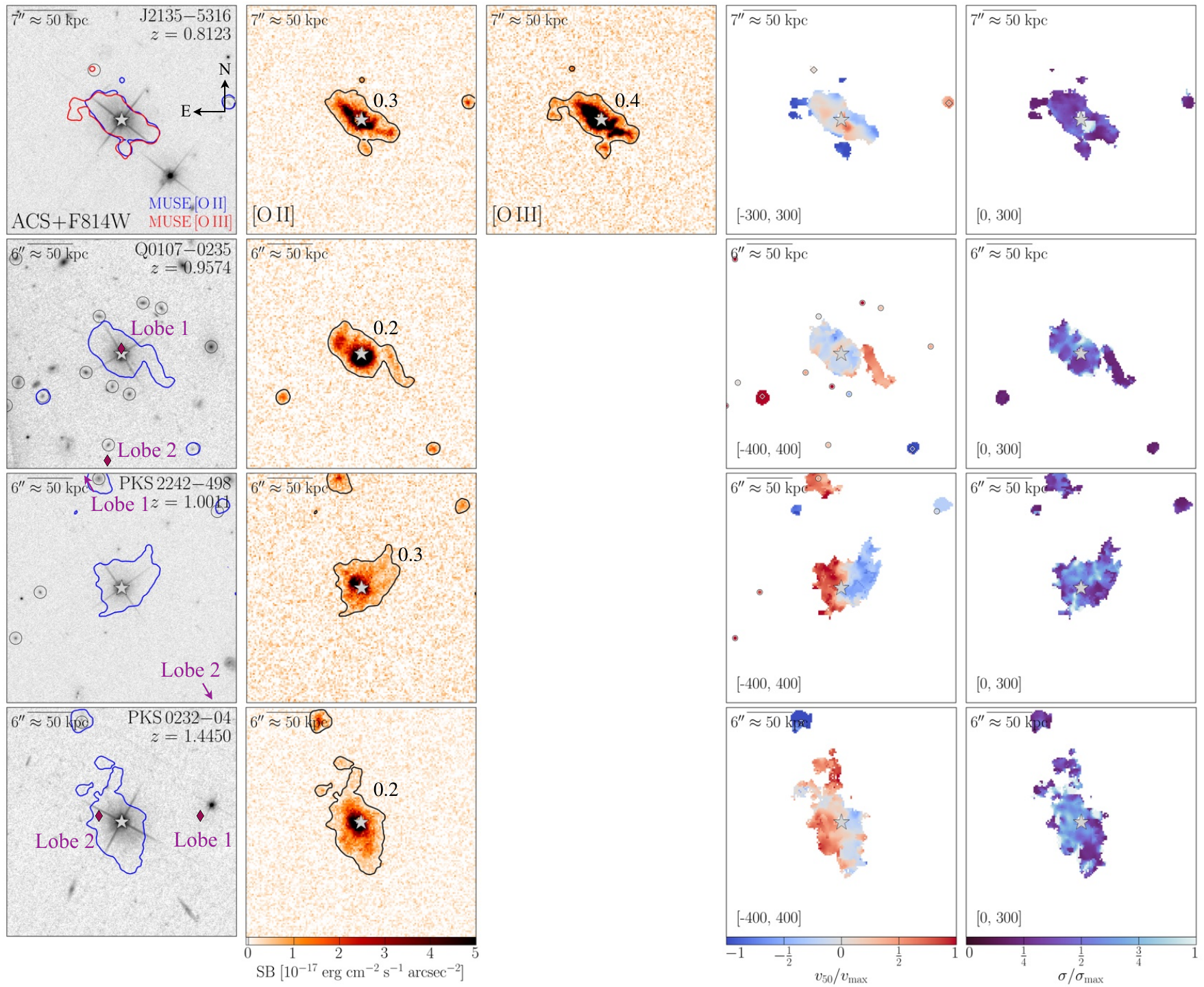}
    \caption{Same as Figure \ref{fig:L_1}, but for nebulae with complex morphology and kinematics nebulae (A).}
    \label{fig:A_1}
\end{figure*}

\begin{figure*}
    \centering
    \includegraphics[scale=0.6]{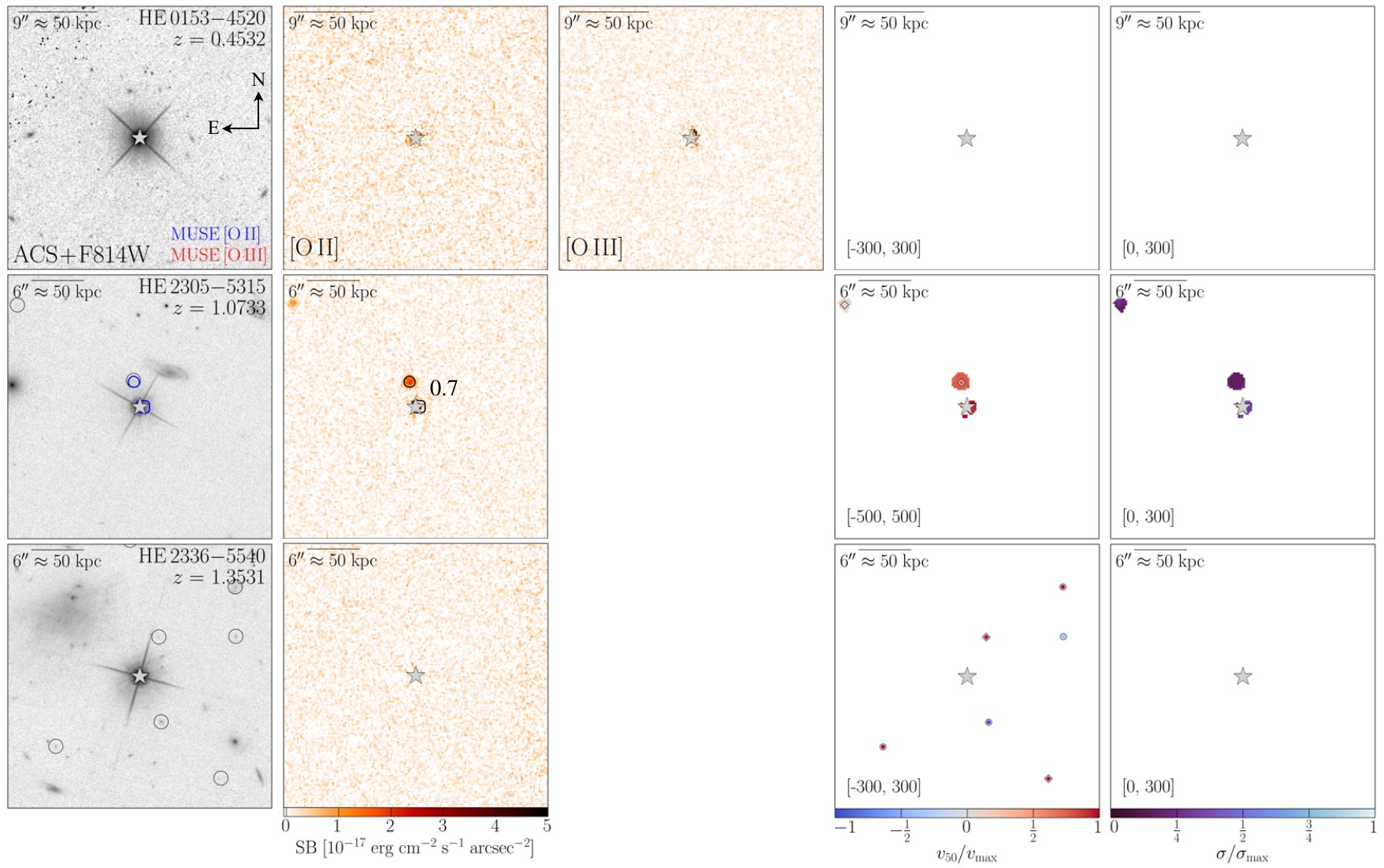}
    \caption{Same as Figure \ref{fig:L_1}, but for systems without detected nebulae (N).}
    \label{fig:N_1}
\end{figure*}

We identify 11 irregular, large-scale nebulae in the CUBS+MUSE survey. These systems often display asymmetric morphologies disordered kinematics, while encompassing multiple group galaxies within their extents. They also often exhibit velocity coherence with nearby group members. We will further investigate the velocity coherence in Section \ref{sec:D-CBGN}. The irregular, large-scale nebulae identified in our sample, including those around quasars, HE\,0226$-$4110, PKS\,0405$-$123, HE\,0238$-$1904, PKS\,0552$-$640, J0454$-$6116, J0119$-$2010, HE\,0246$-$4101, PKS\,0355$-$483, HE\,0439$-$5254, TXS\,0206$-$048, and Q1354$+$048, are shown in Figure \ref{fig:L_1}. Among them, PKS\,0405$-$123, HE\,0238$-$1904, J0454$-$6116, PKS\,0552$-$640, and TXS\,0206$-$048 have been previously characterized \citep[See][]{2018ApJ...869L...1J, 2022ApJ...940L..40J, 2023MNRAS.518.2354C, 2024ApJ...962...98C, 2024MNRAS.527.5429L}. For the remaining systems, this paper presents the first detailed characterization of their properties including morphology, kinematics, and environment.

The HE\,0226$-$4110 nebula shows a kinematic field lacking coherent structure, with blueshifted velocities in the east, redshifted velocities in the west, and mixing between the two, spanning $v_{50}\approx-300$ to $+300\,\rm km\,s^{-1}$ with dispersions up to $\sigma\approx300\,\rm km\,s^{-1}$. The system also shows clear evidence of a late-stage merger event, including prominent tidal tails and a post-starburst region with strong Balmer absorptions (see top row of Figure \ref{fig:L_1}). The nebulae around PKS\,0552$-$640 and Q1354$+$048 exhibit head-tail morphology, with the head centered on the quasar and the tail extending away from the quasar. The PKS\,0552$-$640 nebula shows a large blueshifted region north of the quasar, while the rest of the nebula appears redshifted. Q1354$+$048 is predominantly redshifted, with a region on the eastern side lying close to the quasar’s systemic velocity. Both PKS\,0552$-$640 and Q1354$+$048 nebulae contain two group galaxies whose kinematics are consistent with those of the nebula. In Q1354$+$048, however, stellar absorption features are not clearly detected in these galaxies. The J0454$-$6116 nebula extends in all directions, with a more pronounced extension toward the south. This southern extension is associated with a nearby galaxy at a similar velocity. In contrast, the J0119$-$2010 nebula shows more pronounced extensions to both the south and east. A blueshifted galaxy, with a velocity consistent with the nebular emission, is detected along the eastern extension, while no galaxy is found near the southern extension. The HE\,0246$-$4101, PKS\,0355$-$483, and HE\,0439$-$5254 nebulae show no strong interaction-related tails, though each hosts galaxies with velocity coherence relative to the nebula. 

Small-scale nebulae are more compact ($<70\,\rm kpc$ in diameter), centered on the quasar, and generally lack overlapping group members within their extent. Based on their kinematic patterns, they can be further classified into two subtypes: blueshifted–redshifted velocity structures and patternless velocity fields. Blueshifted–redshifted refers to a nebula that is blueshifted on one side of the quasar and redshifted on the other. We show these nebulae in Figure \ref{fig:S_1} and separate the two subtypes with a black straight line.

Host-galaxy-scale nebulae with a blueshifted-redshifted pattern include HE\,0435$-$5304, 3C\,57, J0110$-$1648, HE\,0112$-$4145, J0154$-$0712, Q1435$-$0134, and PG\,1522$+$101. Their morphologies are frequently symmetric, with blueshifted and redshifted regions of comparable size in five out of seven systems, whereas the remaining two are dominated by one component. All blueshifted-redshifted nebulae, except those around 3C\,57, HE\,0112$-$4145, and J0154$-$0712, lack group galaxies within their extents. In 3C\,57, the group galaxy may be responsible for the positive velocity field in its vicinity, producing a region that is neither distinctly blueshifted nor redshifted \citep[see][for details]{2025ApJ...984..140L}. In HE\,0112$-$4145, the redshifted galaxy is clearly associated with the bright knot, while a blueshifted galaxy is also present on the opposite side $10 \, \rm kpc$ beyond the nebular extent. In J0154$-$0712, the associated galaxy is likely responsible for the extremely blueshifted component of the nebula and for the trailing structure that extends toward it. The velocity dispersion varies significantly across these systems. For example, the HE\,0435$-$5304 nebula shows a relatively low velocity dispersion of $50{-}100 \, \rm km\,s^{-1}$, whereas the HE\,0112$-$4145 nebula exhibits a much higher value of $\approx 200 \, \rm km\,s^{-1}$. In PG\,1522$+$101, the blueshifted side exhibits a significantly higher velocity dispersion than the redshifted side. We will further explore the variation in kinematics in Section \ref{sec:D-BR}. 

Host-galaxy-scale nebulae without a clear kinematic pattern include J0028$-$3305, HE\,0419$-$5657, Q0107$-$025, HE\,1003$+$0149, and HE\,0331$+$4112. These nebulae do not show consistent velocity structures or kinematic patterns. Notably, not all nebulae fully enclose their quasars; for instance, nebulae are detected only on one side of HE\,0419$-$5657 and Q0107$-$025. This asymmetry may reflect differences in the opening angles of the AGN emission cones or a real absence of cool gas at densities high enough to be detected in emission when photoionized by the quasar. The nebulae around J0028$-$3305, HE\,0419$-$5657, and Q0107$-$025 are predominantly redshifted, while those around HE\,1003$+$0149 and HE\,0331$+$4112 are primarily blueshifted. We also see a large variation in velocity dispersions across their extent, with $\sigma$ fluctuating between $\approx 100{-}300 \, \rm km \, s^{-1}$. 

Complex morphology and kinematics nebulae refers to nebulae that exhibit a combination of characteristics of features described earlier. These nebulae include J2135$-$5316, Q0107$-$0235, PKS\,2242$-$498, PKS\,0232$-$04, as shown in Figure \ref{fig:A_1}. The J2135$-$5316 nebula exhibits a cigar-like morphology, with a predominantly blueshifted boundary and two notably blueshifted components along its eastern and southern edges. The emission gradually transitions to redshifted velocities toward the major axis of the nebula. The Q0107$-$0235 nebula has a main body that resembles a host-galaxy-scale structure. It also contains a kinematically distinct, redshifted tail extending more than $50 \, \rm kpc$ from the western edge toward the southwest, despite the main nebular body being blueshifted. This redshifted tail resembles the filament observed in TXS\,0206$-$048 and may suggest filamentary accretion. The PKS\,2242$-$498 and PKS\,0232$-$04 nebulae exhibit blueshifted–redshifted patterns, but these differ from the rotation signatures typically observed in local early-type galaxies. In these galaxies, rotation usually appears along the major axis, resembling a disk. By contrast, the PKS\,2242$-$498 nebula has a more rectangular morphology, with velocity shear oriented along the minor axis. In PKS\,0232$-$04, the pattern is asymmetric, with a positional offset between the left and right halves and velocity shear aligned with the minor axis.

In addition to the classifications described above, we also identify a distinct feature--separate from the complex morphology and kinematics nebulae shown in Figure \ref{fig:A_1}--in which objects exhibit blended multi-component emission and elevated velocity dispersions across several nebulae, including quasars spanning multiple prior classifications. Because these regions are generally small compared to the dominant large-scale structures, we do not treat them as a separate category. In 3C\,57, multi-component emission is detected on both sides of the blueshifted–redshifted pattern \citep[see][]{2025ApJ...984..140L}. In HE\,0238$-$1904, comparable features appear to the east of the quasar, seen in both $\rm [O\,II]$ and $\rm [O\,III]$ \citep{2024MNRAS.527.5429L}. Beyond these known cases, we report two additional systems from the CUBS+MUSEQuBES survey—J2135$-$5316 and J0119$-$2010—that show comparable signatures \citep[see][]{2023MNRAS.518.2354C, 2024ApJ...962...98C}. In these systems, the multi-component regions coincide with localized large velocity dispersion knots. We also detect multiple kinematic components that are clearly separated in velocity space in some nebulae, which can arise from projection effects. These regions typically exhibit $\rm [O\,II]$ and $\rm [O\,III]$ emission lines at different velocities, or two distinct, non-overlapping $\rm [O\,II]$ components. 

No extended nebular emission is detected in three cases: HE\,0153$-$4520, HE\,2305$-$5315, and HE\,2336$-$5540. HE\,0153$-$4520 has no group members within the MUSE FoV, whereas HE\,2305$-$5315 hosts four group members, including one galaxy close to the quasar centroid. In contrast, HE\,2336$-$5540 hosts 16 group members, with two located near the quasar centroid. These three systems therefore reside in markedly different galactic environments, highlighting the diversity of cases without detectable extended nebular emission.

Giant quasar nebulae are often more complicated than suggested by our simplified classification scheme, which captures only the dominant feature of each system while additional structures are frequently present. For instance, while the nebulae around PKS\,0405$-$123 and HE\,0238$-$1904 are primarily irregular and extended, both also exhibit blueshifted–redshifted structures near the quasar centroid, and HE\,0238$-$1904 further shows evidence of multi-component emission. These additional features (blushifted-redshifted, multi-component emission features) are listed for each nebula in Table \ref{table:nebulae}. Additionally, nebular emission is detected around individual group galaxies in the field. In most cases, the emission is centered on the galaxy, while in some systems it appears more extended, as in the fields of J0454$-$6116, J0119$-$2010, TXS\,0206$-$048, Q1435$-$0134, and PG\,1522$+$101. Notably, the PG\,1522$+$101 nebula contains two distinct blobs to the northeast that overlap with two foreground galaxies at $z\approx 0.12$ in the field. We identify a nearby potential group galaxy for only one of them, while no continuum counterpart is detected for the other. We examined the emission-line spectra in these regions and confirmed that these features are real, rather than artifacts caused by imperfect continuum subtraction associated with the foreground galaxies. As the focus of this work is on large-scale structures, we do not discuss these features further. In summary, we detect extended nebulae around 27 of the 30 UV-luminous quasars in our sample, with the three non-detections shown in Figure \ref{fig:N_1}. Among the detected cases, 11 are irregular, large-scale systems, 12 are confined to host-galaxy scales, and 4 display complex morphology and kinematics. In addition, 11 nebulae exhibit a blueshifted–redshifted pattern across part of their extent, and 4 show multi-component emission with large velocity dispersions.

\section{Discussion} \label{sec:D}
In this section, we synthesize the morphological, kinematic, and environmental properties of all quasar nebulae in the CUBS+MUSEQuBES sample to investigate the physical origins of extended ionized gas around UV-luminous quasars. We then introduce a new statistical approach to quantify associations between group galaxies and nebulae. We further examine the role of radio jets, searching for correlations between jet activity and nebular presence. In addition, we conduct a quantitative analysis of nebular morphology and compare our results with both local and high-redshift analogs. Finally, we analyze the kinematic profiles of nebulae exhibiting blueshifted–redshifted patterns and compare them with rotating gas in local early-type galaxies.

\subsection{The Nebular Origin}
\label{TBO}
Giant quasar nebulae can arise from several primary sources: inflows, outflows, extended gas from the quasar host galaxy, galaxy interactions, and the surrounding IGM/CGM, each exhibiting diverse morphologies. Distinguishing these origins requires analyzing nebular morphology, gas kinematics, galactic environment. Several nebulae in our sample have already been examined in detail: PKS\,0405$-$123, TXS\,0206$-$048, HE\,0238$-$1904, and 3C\,57. The TXS\,0206$-$048 nebula has been attributed to filamentary accretion \citep{2022ApJ...940L..40J} with significant turbulence \citep{2023MNRAS.518.2354C}. The PKS\,0405$-$123 and HE\,0238$-$1904 nebulae appear to primarily arise from galaxy interactions involving tidal and ram-pressure stripping \citep{2018ApJ...869L...1J, 2024MNRAS.527.5429L}. The 3C\,57 nebula, in contrast, exhibits a blueshifted-redshifted pattern consistent with ordered kinematics, but with complex line profiles that also suggest the presence of AGN feedback \citep{2025ApJ...984..140L}. In this subsection, we qualitatively assess the origin of each nebula in the CUBS+MUSEQuBES sample based on its morphology and kinematics, which provide direct clues about their nature. 

Irregular large-scale nebulae are often linked to galaxy interactions. Their morphologies frequently retain signatures of these interactions, such as tidal tails or head–tail structures, resembling those seen in nebulae produced by ram-pressure stripping, including jellyfish galaxies \citep{2014ApJ...781L..40E}. We identify a late-stage merger event in 1 out of 30 systems (HE\,0226$-$4110), marked by two prominent tidal tails and a post-starburst region located $10 \rm \,kpc$ southwest of the quasar centroid (See Figure \ref{fig:L_1}).

Nebular head–tail structures encompassing multiple group galaxies are found in 4 out of 30 systems (PKS\,0405$-$123, HE\,0238$-$1904, PKS\,0552$-$640 and Q1354$+$048)\footnote{We also identify head--tail features associated with individual group galaxies, likely driven by ram-pressure stripping, for example in J0454$-$6116, TXS\,0206$-$048, and Q1435$-$0134. As our analysis focuses on large-scale head–tail structures, these smaller features are not included in the present discussion.}. In both PKS\,0552$-$640 and Q1354$+$048, the heads lie east of the quasars, with tails trailing westward. This structure is likely produced by group galaxies moving from west to east and experiencing ram-pressure stripping. In PKS\,0552$-$640, two galaxies east of the quasar are probably responsible for the nebula. They appear to have already crossed to the opposite side, leaving behind gas with redshifted emission along their path that contributes to most of the nebula. Similarly, in Q1354$+$048, two galaxies located to the west are likely responsible for the nebula.

Moreover, 5 other systems (J0454$-$6116, J0119$-$2010, HE\,0246$-$4101, PKS\,0355$-$483, and HE\,0439$-$5254) are also likely interaction-related nebulae, as multiple group galaxies are detected within their extents. Unlike the more striking cases above, these systems do not display clear morphological signatures, but their kinematics are consistent with those of the group galaxies enclosed by the nebula. The southern and southeastern regions of the J0454$-$6116 nebula are predominantly redshifted, matching the velocities of nearby galaxies. Similarly, the PKS\,0355$-$483 nebula encloses four redshifted galaxies that align in velocity along its northern and southern boundaries. In HE\,0439$-$5254, a galaxy embedded within the nebula coincides with a bright knot in the surface brightness map and matches the nebular kinematics. For HE\,0246$-$4101, four galaxies are detected within the nebula; two located closer to the quasar are spatially coincident with bright knots in the nebula and aligned with velocity though these two galaxies are only identified from emission line. The spatial coincidences and velocity coherence indicate that the nebulae most likely originate from interactions with these galaxies. In J0119$-$2010, we detect a galaxy east of the quasar that may drive the blueshifted emission, but this component is kinematically distinct from the main body. The $\rm [O\,III]$ morphology tentatively resembles a head–tail structure. Near the quasar centroid, we identify a galaxy with a large positive LOS velocity that may be responsible for the multi-component emission features and the localized enhancement in velocity dispersion. A bright emission knot appears in the eastern portion of the main body, likely associated with nearby galaxies. However, their presence cannot be confirmed due to overlap with the diffraction spike. The system also exhibits a southern extension with no detected associated galaxies. In total, 10/30 ($33\%$) quasars exhibit nebulae that, based on qualitative analysis, are likely associated with galaxy interactions within the quasar host group (one irregular large-scale system, TXS\,0206$-$048, is not included because it likely arises from filamentary accretion).

Host-galaxy-scale nebulae centered around the quasar and enclosing few or no group galaxies are likely associated with the quasar host galaxy. When the host galaxy’s ISM and/or CGM are ionized by the quasar, they show up as host-galaxy-scale nebulae. Their morphologies and kinematics fall into two categories: relatively symmetric blueshifted–redshifted patterns and patternless velocity fields. The former case likely trace rotating gas in the quasar host galaxy, which naturally produces a blueshifted–redshifted pattern along the major axis. In the latter case, angular momentum may have been lost through past mergers or interactions, leaving behind dynamically unsettled gas. Similar behaviors are observed in both spiral galaxies and local early-type galaxies traced by H\,I 21 cm emission \citep{2012MNRAS.422.1835S, 2014MNRAS.444.3388S}, where some systems display regular disk-like rotation while others show disturbed, irregular kinematics. Together, these two types of nebulae suggest that host-galaxy-scale systems can exhibit both ordered and dynamically disturbed kinematics. Even systems with blueshifted–redshifted velocity patterns often retain signatures of disturbance \citep[e.g.,][]{2025ApJ...984..140L}, underscoring the complex dynamical histories of quasar hosts and offering insight into their evolutionary pathways.

Nebulae classified as having complex morphology and kinematics are often ambiguous in origin. The cigar-shaped J2135$-$5316 nebula does not enclose any group galaxies, and its blue-to-red velocity transition from the outer edge to the inner regions is intriguing but remains unexplained. The Q0107$-$0235 nebula consists of a main component and a filament-like extension. While the main component may originate from the quasar host galaxy, the redshifted tail has an uncertain origin. No absorption features are detected at the tail velocity in the HST/COS spectrum along the quasar sightline. This non-detection may indicate that the tail either has a small covering factor, misses the sightline, or lies behind the quasar. Both the PKS\,2242$-$498 and PKS\,0232$-$04 nebulae exhibit blueshifted–redshifted patterns; however, in these cases the gradients are along the minor axis of the nebula rather than along the major axis of the nebula, differing from the rotation signatures typically seen in H\,I 21 cm disks. This geometry complicates an interpretation in terms of simple disk-like rotation, although at higher redshift the kinematic major and minor axes may be more difficult to distinguish \citep[e.g.,][]{2015ApJ...799..209W}.

In the three systems without detectable extended nebular emission, the non-detection may reflect either a lack of ambient gas in the temperature and ionization state needed to emit efficiently in [O\,II] or [O\,III] or limited illumination by the quasar, owing to its opening angle, covering factor, or the finite quasar lifetime. In one case, HE\,0153$-$4520 does not reside in a system with any group galaxies. In this system, stripped gas may simply be absent, while the ISM and/or CGM of the quasar host galaxy may not be illuminated. By contrast, the other two systems (HE\,2305$-$5315 and HE\,2336$-$5540) contain group galaxies close to the quasar centroid, making the presence of stripped gas from galaxy interactions more likely. In these cases, the absence of nebular emission may indicate either that little stripped gas is present near the quasar, or that any existing gas--including gas in the ISM and/or CGM of the quasar host galaxy--lies outside the quasar ionization cone.

\subsection{Quantifying Correlation Between Group Galaxies and Nebulae} \label{sec:D-CBGN}
Giant quasar nebulae are frequently found in galaxy-rich environments, suggesting a close connection between extended ionized gas and the surrounding galaxy population. Within the irregular, large-scale class, systems such as PKS\,0405$-$123 and HE\,0238$-$1904 extend beyond 100 kpc and coexist with more than 30 group galaxies, whereas host-galaxy-scale systems typically contain far fewer galaxies and exhibit little spatial overlap between galaxies and nebular emission. However, it remains unclear whether this trend simply reflects differences in nebular size or indicates a more physical association with galaxies in the quasar host group. These contrasting cases underscore the need for a quantitative framework that goes beyond simple galaxy counts or nebular size to assess the physical association between nebular gas and galaxies. In this section, we introduce a statistical approach that jointly incorporates spatial and kinematic information to quantify the association between nebular gas and galaxies in the surrounding environment.

Our statistical framework is designed to identify spatial and kinematic coincidences between galaxies and the surrounding nebula, and to compare the observed level of coincidence with that expected under a random galaxy distribution in each system. To characterize the level of coincidence, we use the Bhattacharyya coefficient \citep[BC; ][]{bhattacharyya1946measure}, which, in general, quantifies the level of overlap between two probability density functions (PDFs), $p(x)$ and $q(x)$, as ${\rm BC}=\int \sqrt{p(x)q(x)}\mathrm{d}x$.  In this context, we want to compute the BC for each galaxy--nebular spaxel pair, requiring an approximation of the 3D PDF (two dimensions for the projected spatial locations and one for LOS velocity) for each galaxy and each nebular spaxel.

We model the PDF of each galaxy and each nebular spaxel as independent three-dimensional Gaussian distributions. Specifically, for the $k$-th group galaxy ($k = 1, \dots, N_{\rm gal}$, where $N_{\rm gal}$ is the number of group members), we define a 3D Gaussian PDF, $p_k(\vec{x}) \sim \mathcal{N}(\boldsymbol{\mu}_k, \boldsymbol{C}_k)$, based on its projected position and LOS velocity. Here, $\boldsymbol{\mu}_k = (x_k, y_k, v_k)$ denotes the projected position in the $\hat{x}$ and $\hat{y}$ directions and the LOS velocity relative to the quasar, while $\boldsymbol{C}_k$ is the covariance matrix. We assume $\boldsymbol{C}_k$ is diagonal with no correlation between any pair of dimensions, adopting spatial and velocity standard deviations of $10\,\mathrm{kpc}$ and $20\,\mathrm{km\,s^{-1}}$, respectively, motivated by the typical size of a modest-mass galaxy disk and the redshift precision from MUSE (see Section \ref{sec:CGE}).

We adopt an analogous representation for $i$-th nebular spaxel ($i = 1, \dots, N_{\rm neb}$, where $N_{\rm neb}$ is the number of spaxels where nebular emission is detected; treating the 2D nebular map as flattened into a one-dimensional list), with a 3D Gaussian PDF $q_i(\vec{x}) \sim \mathcal{N}(\boldsymbol{\mu}_i, \boldsymbol{C}_i)$. In this case, $\boldsymbol{\mu}_i = (x_i, y_i, v_i)$ gives the spaxel position and LOS velocity relative to the quasar, and $\boldsymbol{C}_i$ is also taken to be diagonal. For the spatial dimensions in $\boldsymbol{C}_i$, we adopt a standard deviation of 1.5 pixels, corresponding to the radius of a typical seeing disk. For the velocity dimension, we use the local velocity dispersion from the $\sigma$ map at the location of the $i$-th spaxel. For each galaxy--nebular spaxel pair, $(k, i)$, the BC can be expressed as
\begin{equation}
\begin{aligned}
\mathrm{BC}_{k,i} &= \int \sqrt{p_k(\vec{x})\,q_i(\vec{x})}\, \mathrm{d}^3x, \\
&= \frac{\exp\!\left[ -\frac{1}{8} \Delta \boldsymbol{\mu}_{k,i}^{\top}
\boldsymbol{C}_{k,i}^{-1} \Delta \boldsymbol{\mu}_{k,i} \right]}
{\sqrt{\dfrac{\det \boldsymbol{C}_{k,i}}
{\sqrt{\det (\boldsymbol{C}_{k})\,\det (\boldsymbol{C}_{i})}}}},
\end{aligned}
\end{equation}
where $\Delta \boldsymbol{\mu}_{k,i} = \boldsymbol{\mu}_{k} - \boldsymbol{\mu}_{i}$, $\boldsymbol{C}_{k,i} = \frac{1}{2}(\boldsymbol{C}_{k} + \boldsymbol{C}_{i})$. Larger BC values indicate a greater degree of spatial and kinematic coincidence between a galaxy and a nebular spaxel. The quasar host galaxy is excluded from this analysis.

For each nebula, we evaluate the BC for every galaxy--spaxel pair across the field and sum over galaxies (indexed by $k$) at each nebular spaxel (indexed by $i$) to construct a spatially resolved Kinematic Association Factor (KAF) map,
\begin{equation}
    \mathrm{KAF}(x_i, y_i) = \sum_{k=1}^{N_{\rm gal}} \mathrm{BC}_{k,i},
\end{equation}
Because $0 < \mathrm{BC}_{k,i} < 1$, summing over galaxies can yield spaxels with $\mathrm{KAF}(x_i, y_i)>1$. The KAF map therefore quantifies the degree to which a given nebular spaxel is spatially and kinematically associated with one or more galaxies.

For the analysis of the quasar host nebula, we mask compact nebular emission that is clearly separated from the main structure and associated with individual galaxies in order to avoid artificially enhancing the KAF at the galaxy positions. The galaxies themselves remain in the analysis, but their local emission is excluded.

\begin{figure*}
    \centering
    \includegraphics[scale=0.6]{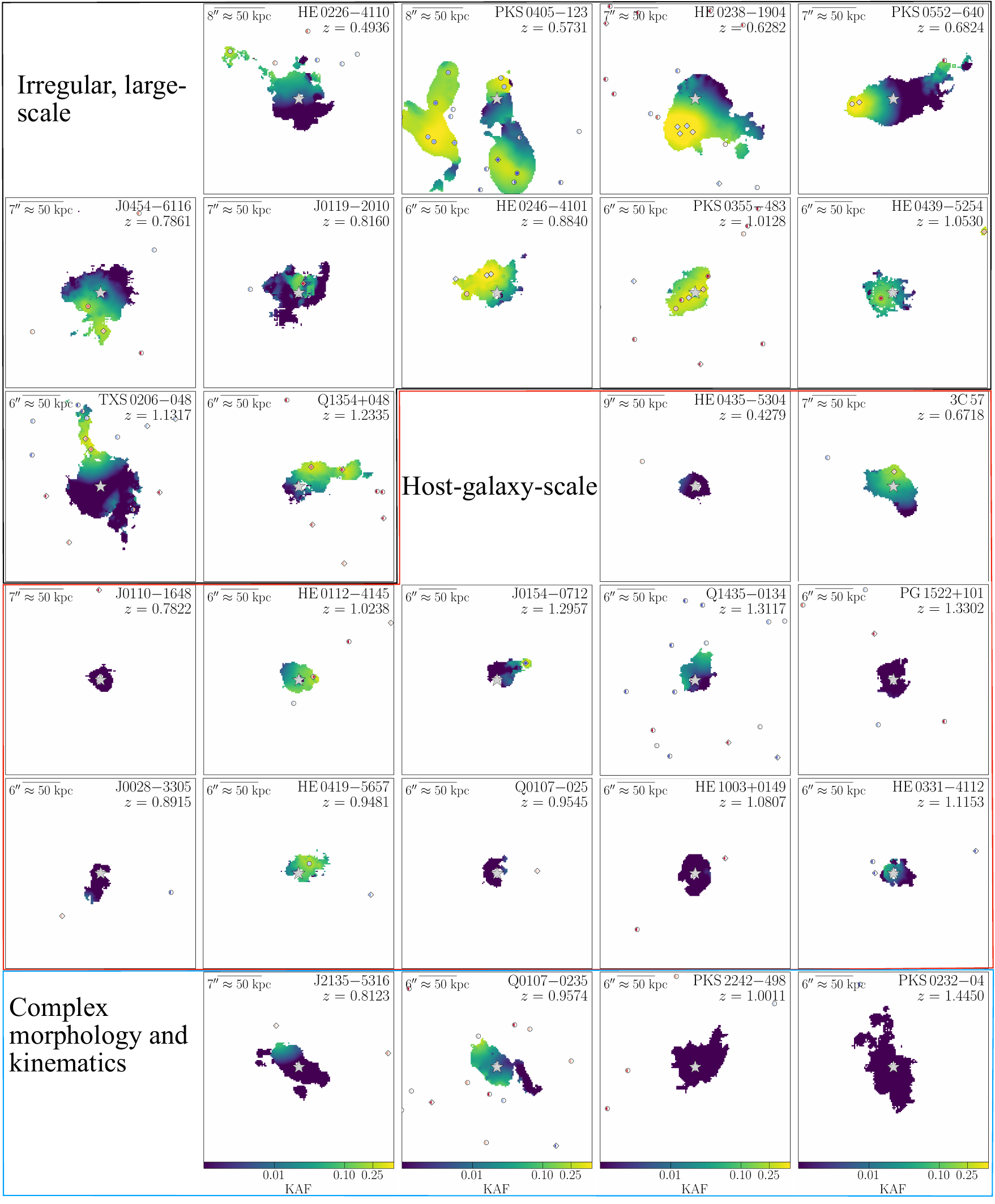}
    \caption{Visualization of the KAFs for all detected nebulae. The KAF quantifies the degree of spatial and kinematic overlap between nebular spaxels and galaxies in the quasar host group (See Section \ref{sec:D-CBGN}). The LOS velocities of the group galaxies are shown in the left half of each symbol and are color-coded using the same scale as the nebulae in Figures~\ref{fig:L_1}--\ref{fig:A_1}. For galaxies enclosed by the nebula, the nebular LOS velocity is shown in the right half. Small-scale nebular emission that is clearly separated from the main structure and associated with individual galaxies is masked in the analysis of the quasar host nebula, which leads to some differences compared to previous figures. All panels use the same colormap, with the color scale shown in the bottom row.}
    \label{fig:KAF}
\end{figure*}

\begin{figure}
    \centering
    \includegraphics[width=\linewidth]{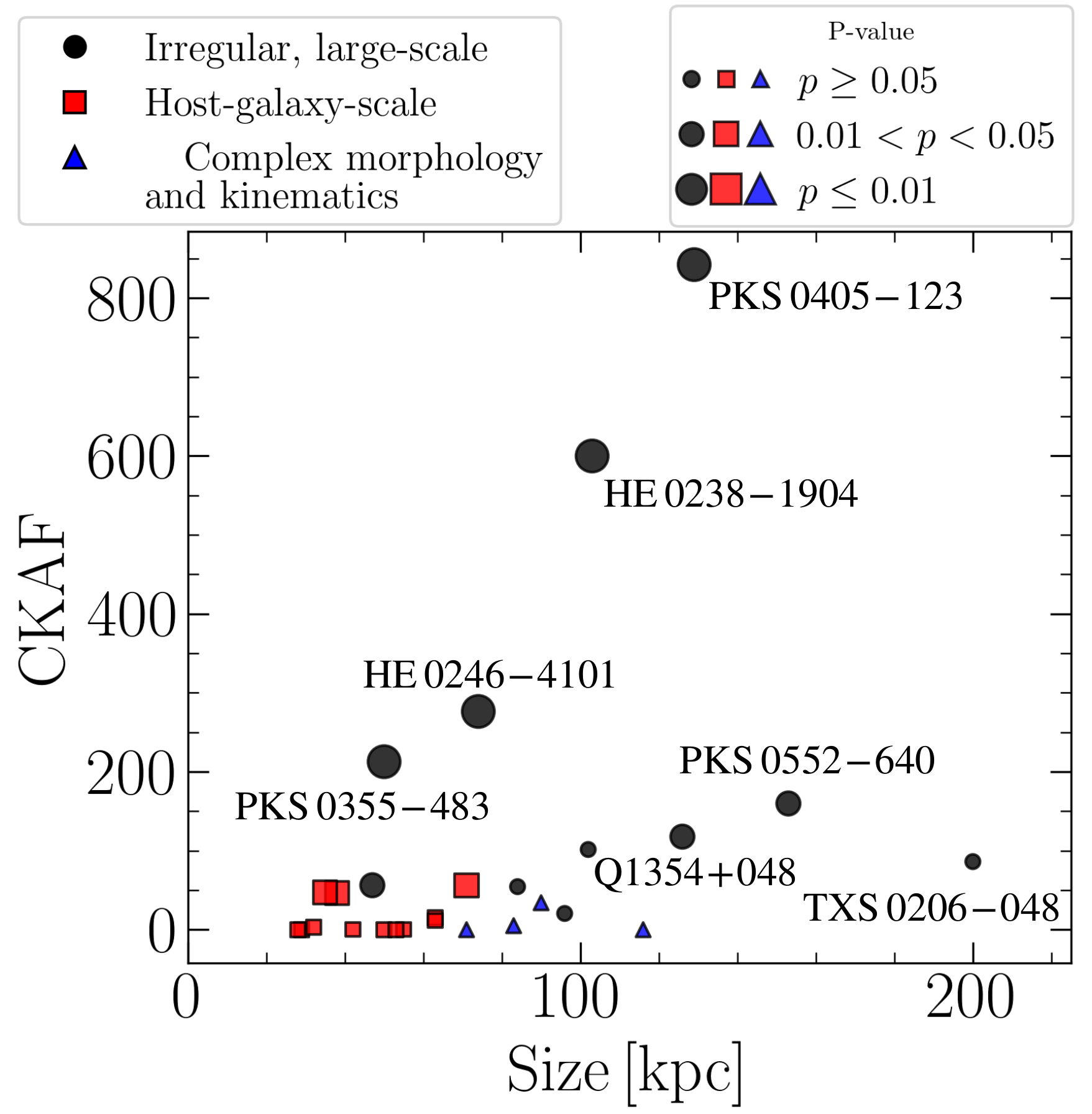}
    \caption{CKAF score versus nebular size for all detected nebulae. The CKAF quantifies the degree of spatial and kinematic overlap between a nebula as a whole and galaxies in the quasar host group (See Section \ref{sec:D-CBGN}). P-values are encoded by three scatter-point sizes. Most irregular, large-scale nebulae have p-values below 0.05, and in several cases below 0.01, indicating statistically significant associations that are highly unlikely to arise from a random galaxy distribution. Ten out of thirteen nebulae with $\rm CKAF \gtrsim 50$ have p-values below 0.05, demonstrating that CKAF robustly identifies genuine spatial–kinematic alignment. Of these 10 significant cases, 7 are irregular, large-scale, consistent with the qualitative analysis that most of these systems arise from galaxy interactions. }
    \label{fig:Ngal_size}
\end{figure}

To illustrate the application of this metric, we present KAF maps for each detected system in Figure \ref{fig:KAF}. Systems classified as irregular, large-scale structures are shown in the top rows, host-galaxy–scale systems in the middle rows, and systems with complex morphology and kinematics in the bottom rows. The KAF metric effectively captures the joint overlap in spatial position and kinematics between galaxies and nebulae, revealing clear differences among these classes. The irregular, large-scale systems exhibit the most prominent KAF values. Most nebulae show large KAF values across substantial regions. Notably, PKS\,0405$-$123 displays the most extended and pronounced KAF signal, with a prominent knot coincident with more than five nearby group galaxies. In contrast, HE\,0226$-$4110 and TXS\,0206$-$048 show weaker KAF signatures, consistent with scenarios in which the nebulae originate from late-stage mergers or filamentary accretion rather than from stripped galaxy gas. In the host-galaxy–scale class, nine nebulae exhibit uniformly low KAF values across their full extent or over most of their spatial coverage. Three systems—3C\,57, HE\,0112$-$4145, and HE\,0419$-$5657—show relatively elevated KAF values, driven by their proximity to one or two galaxies with velocities consistent with the nebular gas. In the complex morphology and kinematics class, three out of four systems do not display large, coherent regions of high KAF. The exception is Q0107$-$0235, which exhibits modestly elevated KAF values due to the presence of 22 galaxies in the field, many of which share velocities similar to that of the nebula.

To place individual systems in a broader context, we introduce a scalar score, Cumulative KAF (CKAF), for each nebula , defined as 
\begin{equation}
    \mathrm{CKAF} = \sum_{k=1}^{N_{\rm gal}}\sum_{i=1}^{N_{\rm neb}} \mathrm{BC}_{k, i},
\end{equation}
The CKAF metric quantifies the degree that a nebula, considered as a whole, is associated with group galaxies. The CKAF is intentionally not normalized by nebular size, because both the nebular extent and the number of group galaxies are intrinsic properties of each system; normalizing the statistic would therefore remove physically relevant information. We instead quantify the significance of the CKAF metric separately using a randomized resampling-based approach. We compute CKAF scores for all detected nebulae, with values listed in Table \ref{table:nebulae}.

Because larger nebulae are more likely to yield higher CKAF values purely by chance than smaller systems, we compare the measured CKAF values for each nebula with the distribution expected under a random galaxy population. For each quasar field, we assemble a parent catalog of candidate group galaxies drawn from all fields except the one being analyzed, and retain their projected positions and velocities. We convert these to coordinates relative to the quasar, expressed as physical distances and position angles. We then perform 10,000 Monte Carlo realizations, drawing the same number of galaxies as observed for the system (without replacement) in each trial. For each realization, we compute the CKAF with the observed nebula. The p-value is defined as the fraction of simulations in which the CKAF exceeds the observed value. Smaller p-values indicate that the observed CKAF values are less likely to arise from a random galaxy distribution. The resulting CKAF–nebular size relation is shown in Figure \ref{fig:Ngal_size}, with p-values encoded by the marker sizes.

Irregular, large-scale systems generally exhibit higher CKAF values, albeit with substantial scatter. Seven out of eleven systems have p-values below 0.05, and four out of eleven have p-values below 0.01, indicating statistically significant associations with group members that are unlikely to arise from a random galaxy distribution. The nebulae with the largest CKAF values are all found within this class. When considered jointly with nebular size, these systems tend to occupy regions of either large spatial extent or high CKAF, although three systems exhibit both sizes below 100 kpc and CKAF values below 100. The largest nebula by size, TXS\,0206$-$048, has a relatively modest CKAF value of $\sim$70, consistent with a scenario in which its nebular emission is dominated by filamentary accretion with minimal contribution from group galaxies. Host-galaxy–scale and complex-morphology nebulae typically have modest CKAF values, below 100, and are largely consistent with random galaxy distributions: only three out of eleven host-galaxy–scale systems have p-values below 0.05, and none of the systems in the complex morphology and kinematics class reach this significance threshold.  

Taken together, these results demonstrate that the KAF and CKAF metrics provide an effective quantitative framework for assessing the association between quasar nebulae and their surrounding galaxy populations. By requiring simultaneous spatial and kinematic coincidence, the metrics avoid bias toward large nebulae or galaxy-rich environments without true alignment. We find that $\approx\!75\%$ (10 out of 13) of the nebulae with $\rm CKAF \gtrsim 50$ have p-values below 0.05, demonstrating that CKAF robustly identifies genuine spatial–kinematic alignment. Of these 10 significant cases, 7 are irregular, large-scale nebulae, consistent with the qualitative analysis that most of these systems arise from galaxy interactions. \textcolor{black}{We verify that these conclusions are robust against differences in redshift and observational depth by recomputing the CKAF after applying a redshift-dependent surface brightness threshold to each nebula. We find that the statistical significance and overall trends remain unchanged, indicating that our results are not driven by variations in surface brightness sensitivity across the sample.}

\begin{figure*}
    \centering
    \includegraphics[scale=0.7]{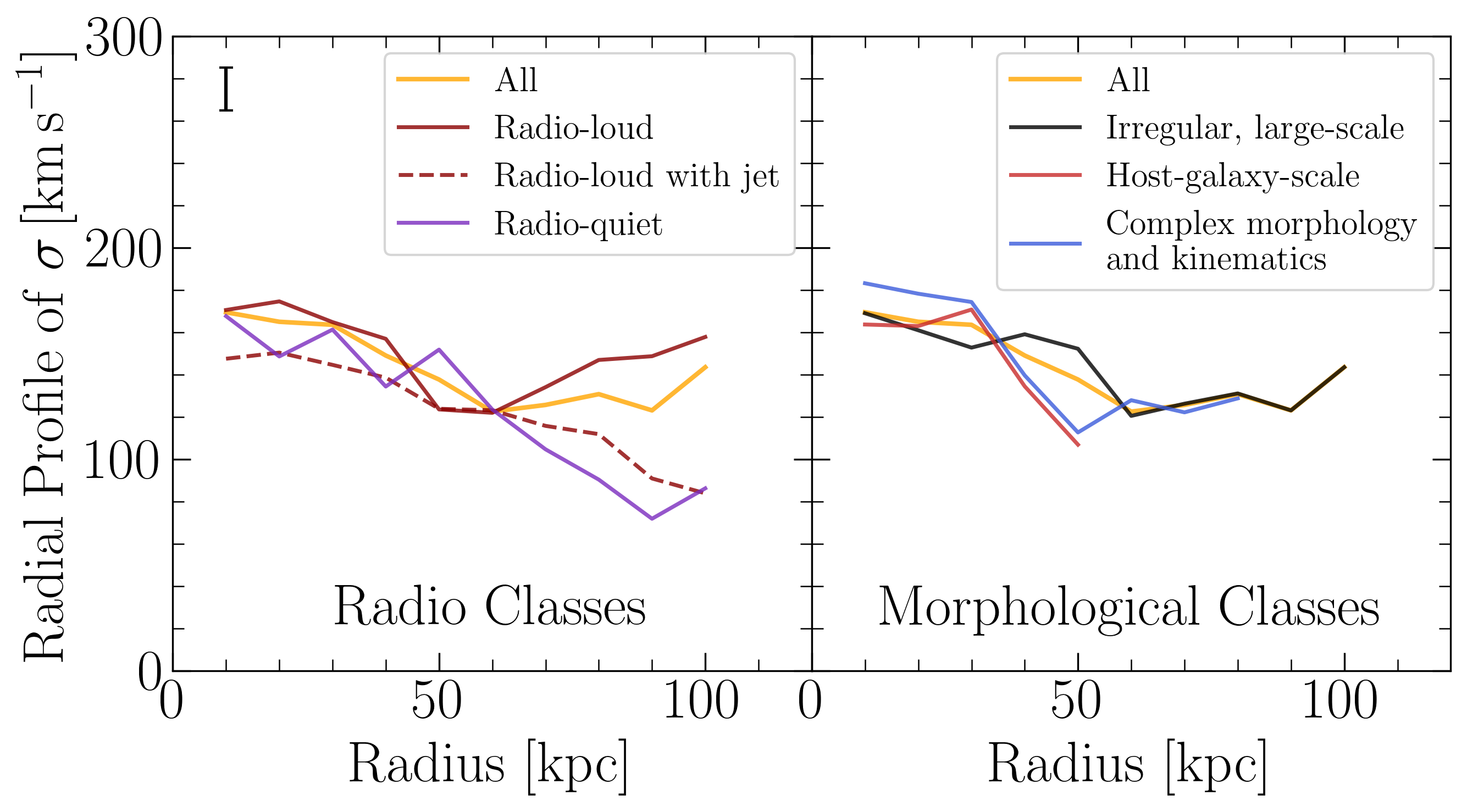}
    \caption{Ensemble radial profile of the nebular velocity dispersion. The ensemble nebular profile is shown in both panels as an orange line. Left: Ensemble nebular profiles grouped by radio class. The profile for all radio-loud systems is shown in dark red, with radio-loud systems hosting jets shown as a dark red dashed line. The profile for all radio-quiet systems is shown in purple. Right: Ensemble profiles for each morphological–kinematic class. The profiles for the irregular, large-scale class, the host-galaxy-scale class, and the complex morphology and kinematics class are shown as solid black, solid red, and solid blue lines, respectively. The characteristic uncertainty (error on the mean) in these profiles is indicated in the top left of the left panel. The ensemble velocity-dispersion profiles are broadly similar, with centrally peaked values that decline with radius, consistent with a similar trend recently reported in the Perseus Cluster \citep{2025arXiv251212754B}. This pattern suggests that AGN feedback is strongest in the inner nebulae and may be common throughout the sample, while manifesting in distinct forms such as radio-mode and quasar-mode feedback.}
    \label{fig:SigmaRadial}
\end{figure*}

\subsection{AGN Feedback}
AGN feedback is thought to play a crucial role in regulating galaxy formation and evolution. In our sample, we identify regions with multi-component emission and broad velocity dispersions, signatures that may reflect AGN feedback through either radio mode \citep[for a review, see][]{2007ARA&A..45..117M} or quasar mode \citep[for a review, see][]{2015ARA&A..53..115K}, though the underlying mechanisms remain uncertain. Additionally, previous studies suggested that radio emissivity correlates with the presence of extended nebular emission \citep{1984ApJ...281..535B, 1987ApJ...316..584S, 2009ApJ...696.1693F, 2017ApJ...841...93Y}. In this subsection, we investigate the presence of radio jets, search for connections between radio jets and quasar nebulae, and investigate the impact of AGN feedback from velocity dispersion.

We detect radio lobes in six quasars---PKS\,0405$-$123, 3C\,57, J0110$-$1648, Q0107$-$0235, PKS\,2242$-$498, and PKS\,0232$-$04---using archival radio data from the Very Large Array (VLA; \citealt{1999ApJS..124..285R}), the VLA Sky Survey (VLASS; \citealt{2020PASP..132c5001L}), and the Rapid ASKAP Continuum Survey (RACS; \citealt{2020PASA...37...48M}). The lobe centroids are presented in Figures \ref{fig:L_1}, \ref{fig:S_1}, and \ref{fig:A_1}. In most cases, the radio lobes lie outside the field of view and are therefore indicated with arrows marking their directions. In three systems, two lobes are detected and clearly separated from the central source, so we omit centroids for the core emission. In the remaining cases, only a single distant lobe is observed together with a central component that is typically offset from the quasar centroid; in these instances, we show the centroid of the central emission. In PKS\,0405$-$123, one radio lobe overlaps the southern extension of the nebula, but its morphology and kinematics favor an interaction-driven origin over radio-driven outflows \citep[see][]{2018ApJ...869L...1J}. For PKS\,2242$-$498, one lobe is coincident with a foreground spiral galaxy at $z \approx 0.04$. To test whether the lobe originates from this galaxy, we estimated its absolute magnitude using K-correction \citep{2002astro.ph.10394H} and found its luminosity to be far below that of typical radio galaxies. Moreover, in Q0107$-$0235 and PKS\,0232$-$04, the radio lobe centroids coincide with regions of elevated velocity dispersion, hinting AGN feedback may be redistributing gas, injecting energy, and driving turbulence in these cases \citep{2023MNRAS.518.2354C, 2024ApJ...962...98C, 2025ApJ...978L..18C}. To better constrain the physical drivers, these regions will require deeper spectroscopic observations and higher-resolution radio or X-ray data.

Overall, we find no strong correlation between the orientation of radio lobes and the extent of quasar nebulae. Most lobes lie well beyond the nebular emission. By connecting the two radio lobes and measuring their position angle, we compare it with the morphological major axis of the nebula. Five of the six systems show offsets greater than $40^\circ$ between the two angles, with four exhibiting offsets of $60^\circ{-}80^\circ$, indicating that the radio-lobe axis is often significantly misaligned with the nebular morphology. This result differs from several early studies that reported close spatial associations between radio jets and extended emission-line nebulae, often interpreting them as directly linked. However, the CUBS+MUSEQuBES sample targets UV-luminous quasars and may differ from radio-selected populations investigated in \citet{1984ApJ...281..535B} and \citet{2009ApJ...696.1693F}. Our result is instead more consistent with recent IFU studies of radio galaxies, such as MURALES \citep{2018A&A...619A..83B,2022A&A...662A..23B}, which analyzed 10 Fanaroﬀ-Riley I (FR I, \citealt{1974MNRAS.167P..31F}) and 26 FR II radio galaxies. In that sample, large-scale ($\gtrsim4$ kpc) ionized gas is detected in 24 of the 26 FR II sources, typically in elongated or filamentary structures extending $\approx10{-}30$ kpc and up to $\approx80$ kpc, with a broad range of offsets from the radio axis rather than a tight alignment. These results support the interpretation that the extended nebulae often trace pre-existing gaseous reservoirs that are subsequently illuminated and perhaps locally disturbed by the AGN, rather than being formed primarily by the jets. A more definitive test will require a larger and more representative quasar sample, as well as deeper, higher-resolution radio observations, since some quasars---including both radio-loud and radio-quiet systems---are only covered by RACS, whose angular resolution may be insufficient to detect compact or low-power jets.

To further assess the potential impact of AGN feedback on the nebulae, we extracted radial profiles of the velocity dispersion, $\sigma$, for each nebula using \texttt{Photutils} \citep{2016ascl.soft09011B}. Similar to the procedure adopted in Section \ref{sec:D-CBGN} for the KAF analysis, we masked and excluded emission and kinematic features clearly associated with individual galaxies in order to isolate the main nebular structure. To better illustrate differences among the various classes, we divided the profiles by both radio class and morphological--kinematic class, adopting the radio classifications reported in \citet{2024ApJ...966..218J}. We then constructed ensemble profiles for each subsample. The resulting profiles are shown in Figure \ref{fig:SigmaRadial}.

The ensemble profiles are broadly similar, typically peaking at small radii and declining with increasing distance from the quasar, with characteristic values of $\approx 100{-}200 \ \rm km \ s^{-1}$ in the inner regions and $\approx 80{-}150 \ \rm km \ s^{-1}$ in the outer regions. The uncertainties remain substantial, reflecting the complexity, large fluctuations, and considerable diversity seen in the radial profiles of individual nebulae. A qualitatively similar radial trend has recently been reported in the Perseus Cluster, where the velocity dispersion peaks at $\approx 175 \ \rm km \ s^{-1}$ in the central region and declines out to $\approx 100$ kpc \citep{2025arXiv251212754B}. Although the physical environments differ substantially, that study suggests that AGN-driven stirring can elevate the velocity dispersion at small radii, with its influence weakening farther from the center, while gas sloshing may become increasingly important on larger scales \citep{2025arXiv251212754B}. In this context, the centrally peaked velocity-dispersion profiles in our sample, together with the complex, blended, multi-component emission features identified in some systems, may likewise indicate that AGN feedback contributes most strongly in the inner nebulae, while additional processes likely shape the gas kinematics at larger radii. Moreover, the broadly consistent ensemble radial profiles across the different groups may suggest that AGN feedback is present throughout the sample, albeit manifesting in distinct forms, including both radio-mode and quasar-mode feedback.

\subsection{Characterizing Nebular Morphology}
Giant quasar nebulae exhibit a wide range of morphologies. Many previous studies have focused on the morphology of Ly$\alpha$ nebulae \citep{2019MNRAS.482.3162A, 2020MNRAS.495.1874D, 2024A&A...691A.210H, 2025arXiv250716898G}. \citet{2020MNRAS.495.1874D} investigated asymmetry between nebula around Type I and Type II AGN and found that nebulae around type II AGNs are more asymmetric than around type I, consistent with the unified model of AGN \citep{1993ARA&A..31..473A}. \citet{2024A&A...691A.210H} investigated Ly$\alpha$ nebulae in quasar pairs and found that extended emission around quasar pairs is morphologically diverse, but on average, it is more asymmetric than single-quasar nebulae with large oﬀsets between the quasar position and the flux-weighted nebula centroid. \citet{2025arXiv250716898G} found that Ly$\alpha$ nebulae around single quasars are overall circular but tend to be displaced somewhat towards one side of the quasar. There are no particularly evident trends with quasar luminosity, apart a tentative increase of more lopsided and asymmetric systems at low luminosity. In the local Universe, H\,I 21\,cm observations provide a complementary view of extended gas morphology. Large H\,I disks commonly trace ordered rotation on galactic scales \citep{2008AJ....136.2648D}, while asymmetries and lopsided structures are strong indicators of tidal interactions \citep{2011MNRAS.416.2401H}. Early-type galaxies further exhibit diverse and frequently disturbed H\,I morphologies, often interpreted as signatures of gas accretion and galaxy interactions \citep{2012MNRAS.422.1835S}. In this subsection, we perform a quantitative analysis of the extended emission’s spatial morphology in terms of geometrical asymmetry and compare with other nebulae, including high-redshift Ly$\alpha$ nebulae \citep{2021MNRAS.503.3044F} and 21 cm emission around local early-type galaxies \citep{2012MNRAS.422.1835S}. 

Our morphological analysis is inspired by \citet{2024ApJ...974..273M}, who adopted the methodology of \citet{2019MNRAS.483.4140R} to study the $\rm H\alpha$ emission morphology around galaxies. Similarly, we use the \texttt{statmorph} package \citep{2019MNRAS.483.4140R} to compute nonparametric morphological statistics for the quasar nebulae. Among the suite of metrics provided by \texttt{statmorph}, we focus on shape asymmetry \citep{2016MNRAS.456.3032P}, which is defined as
\begin{align}
    A_{\rm shape} = \frac{\sum_{i,j} | M_{ij} - M_{ij}^{180}|}{\sum_{ij} |M_{ij}|},
\end{align}
where $M_{ij}$ is the binary mask value at pixel $(i,j)$ in the original segmentation map, instead of recalculating the shape segmentation as in \citet{2019MNRAS.483.4140R}. $M_{ij}^{180}$ is the corresponding value after a $180^\circ$ rotation about the center. For example, perfectly circular or elliptical nebulae yield $A_{\rm shape} = 0$, whereas a lopsided or irregular structure produces a positive asymmetry. Shape asymmetry is conceptually similar to the conventional asymmetry measure introduced earlier \citep[e.g.,][]{1995ApJ...451L...1S, 1996MNRAS.279L..47A, 2000ApJ...529..886C}. The difference is that the conventional asymmetry is measured on the flux map, whereas shape asymmetry is measured on the binary mask. Because the measurement is performed on the binary segmentation map rather than the flux image, shape asymmetry is particularly sensitive to faint, extended features in the nebular outskirts. To ensure consistency across our sample, we fix the asymmetry center to the quasar centroid throughout the analysis—unlike the default \texttt{statmorph} implementation, which iteratively adjusts the centroid to minimize asymmetry. Using these definitions, we computed asymmetry statistics for each nebula in the CUBS+MUSEQuBES sample, and reported in Table \ref{table:nebulae}.

\begin{figure}
    \centering
    \includegraphics[scale=0.7]{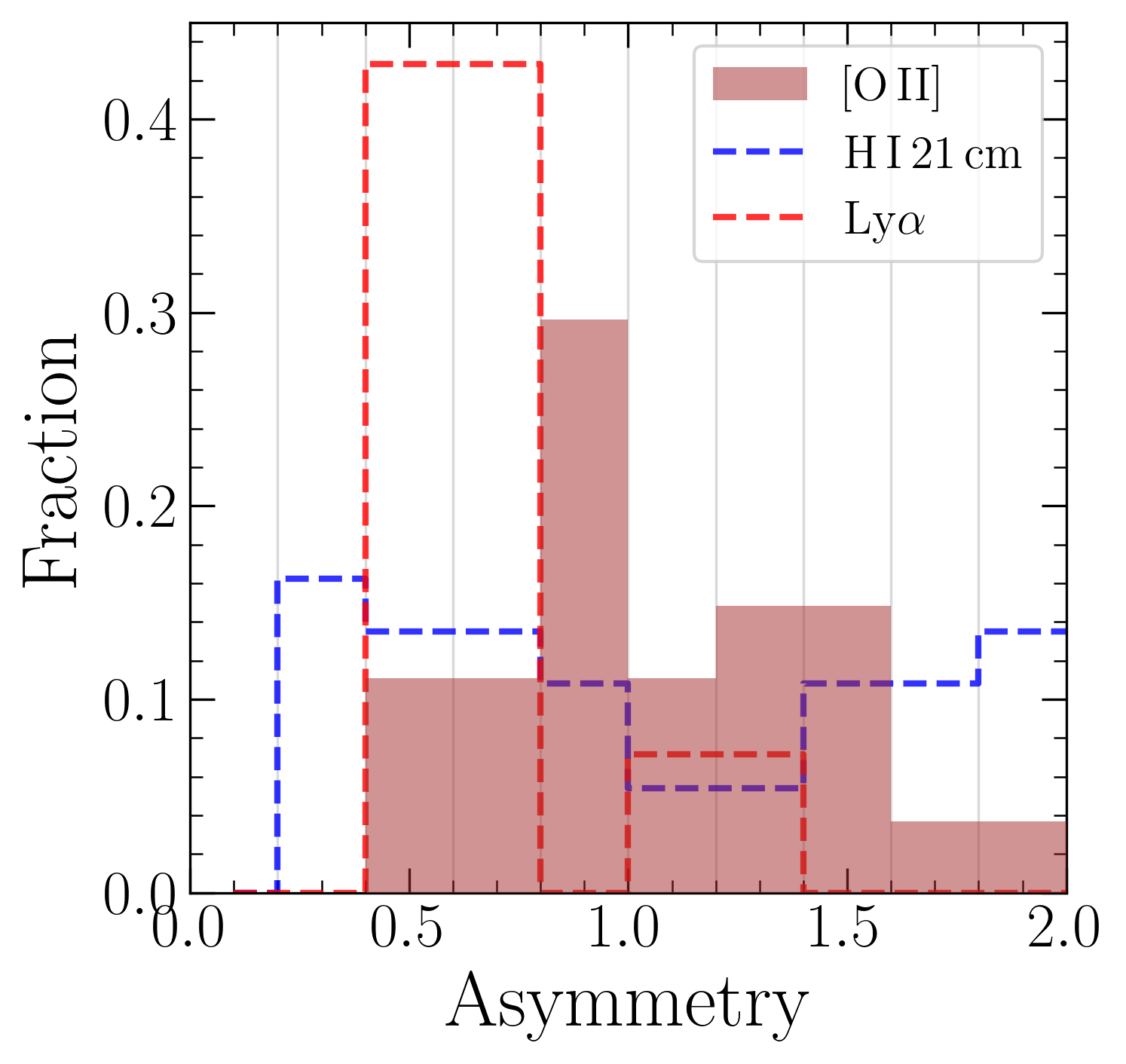}
    \caption{Distribution of nebular asymmetry for the $\rm [O\,II]$, H\,I 21 cm, and Ly$\alpha$ nebulae, normalized by the total number in each population. The light brown bars represent the $\rm [O\,II]$ nebulae presented in this study, while the Ly$\alpha$ nebulae and H\,I 21 cm emission are shown as blue and red dashed curves, respectively. Vertical lines mark the bin edges, corresponding to a bin width of 0.2. Both the $\rm [O\,II]$ nebulae and H\,I 21 cm emission show a broader asymmetry range, whereas the Ly$\alpha$ nebulae are generally more symmetric. This contrast likely stems from the larger spatial scales and resonant scattering of Ly$\alpha$ emission, which tend to smooth out small-scale structure.}
    \label{fig:AsymmetryGini}
\end{figure}

To place our analysis in a broader context, we apply the same morphological measurements to both Ly$\alpha$ nebulae and 21 cm emission around local early-type galaxies. We first examine 14 Ly$\alpha$ nebulae around high-redshift quasars at $z=3.2{-}4.2$, selected from the high-redshift subset of MUSEQuBES and the MUSE Analysis of Gas around Galaxies (MAGG) surveys \citep{2021MNRAS.503.3044F}. These nebulae are representative of their parent samples and additionally exhibit extended $\rm He\,II\,\lambda1640$ emission. A detailed presentation of this sample will be given in Travascio et al. (in prep.). For consistency, we align the asymmetry center with the continuum centroids and use nebula segmentation generated from \texttt{CUBEXTRACTOR} (CubEX; \citealt{2019MNRAS.483.5188C}), an optimal extraction pipeline comparable to the one used in this work (See Sec \ref{sec:DCN}).

For H\,I 21 cm emission, we use data from the $\rm ATLAS^{3D}$ H\,I survey \citep{2011MNRAS.413..813C, 2012MNRAS.422.1835S}, which observed 166 early-type galaxies, detecting H\,I in 53 of them. We adopt early-type galaxies rather than spirals because these UV-luminous quasars have a median black hole mass of $\log M_{\rm BH}/\rm M_{\odot} = 9.5$ \citep{2024ApJ...966..218J}. Through the $M_{\rm BH}$–$\sigma$ relation, such high black hole masses imply massive bulges and halos, indicating that the hosts are predominantly massive early-type systems and making the comparison sample appropriate. We exclude a few systems where the H\,I emission is significantly offset from the galaxy center and no H\,I is detected at the galaxy center, as such systems were interpreted as unbound gas in the intergalactic medium rather than material associated with the host \citep{2012MNRAS.422.1835S}. The final sample consists of 47 galaxies with centrally located H\,I. For consistency with other nebulae, we fix the asymmetry center at the galaxy centroid and construct segmentation maps by interpreting the H I velocity fields from the $\rm ATLAS^{3D}$ survey as binary masks \citep{2012MNRAS.422.1835S}.

We present the results in Figure \ref{fig:AsymmetryGini}, which highlight clear differences in asymmetry across these nebulae. The $\rm [O\,II]$ nebulae in our study exhibit a broad asymmetry range (0.4 to 2.0), reflecting diverse morphologies and spatial distributions. Among them, PKS\,0405$-$123 and J0154$-$0712 display the highest asymmetry: the former due to presence of multiple large nebulae to the east and south of the quasar, and the latter from a nebula concentrated toward the northwest. The 21\,cm H\,I nebulae likewise show substantial scatter in asymmetry, reflecting their wide range of ellipticities and consistent with the complex gas origins and dynamics of local early-type galaxies. In contrast, the high-redshift Ly$\alpha$ nebulae tend to be more symmetric, with only two objects exceeding an asymmetry value of 1. This relative symmetry likely arises from the resonant scattering nature of Ly$\alpha$ emission, which may smooth out smaller-scale irregularities. A K--S test shows that the $\rm [O\,II]$ asymmetry distribution differs significantly from that of Ly$\alpha$ ($p < 0.01$), but is broadly consistent with that of 21\,cm H\,I ($p \approx 0.1$). \textcolor{black}{We note that the K-S test is relatively insensitive to differences near the ends of the distribution \citep{2012msma.book.....F}. Therefore, although the K-S test suggests that the $\rm [O\,II]$ and 21\,cm H\,I distributions are not dramatically different, this result does not necessarily imply that they are drawn from the same parent distribution.} Although the shape asymmetry metric is designed to capture faint, low-surface-brightness features, the measurements may still be influenced by detection limits, observational depth, and differences in sensitivity among the datasets. As a result, part of the observed variation in shape asymmetry may reflect observational biases rather than purely intrinsic morphological differences.

\begin{figure*}
    \centering
    \includegraphics[width=\linewidth]{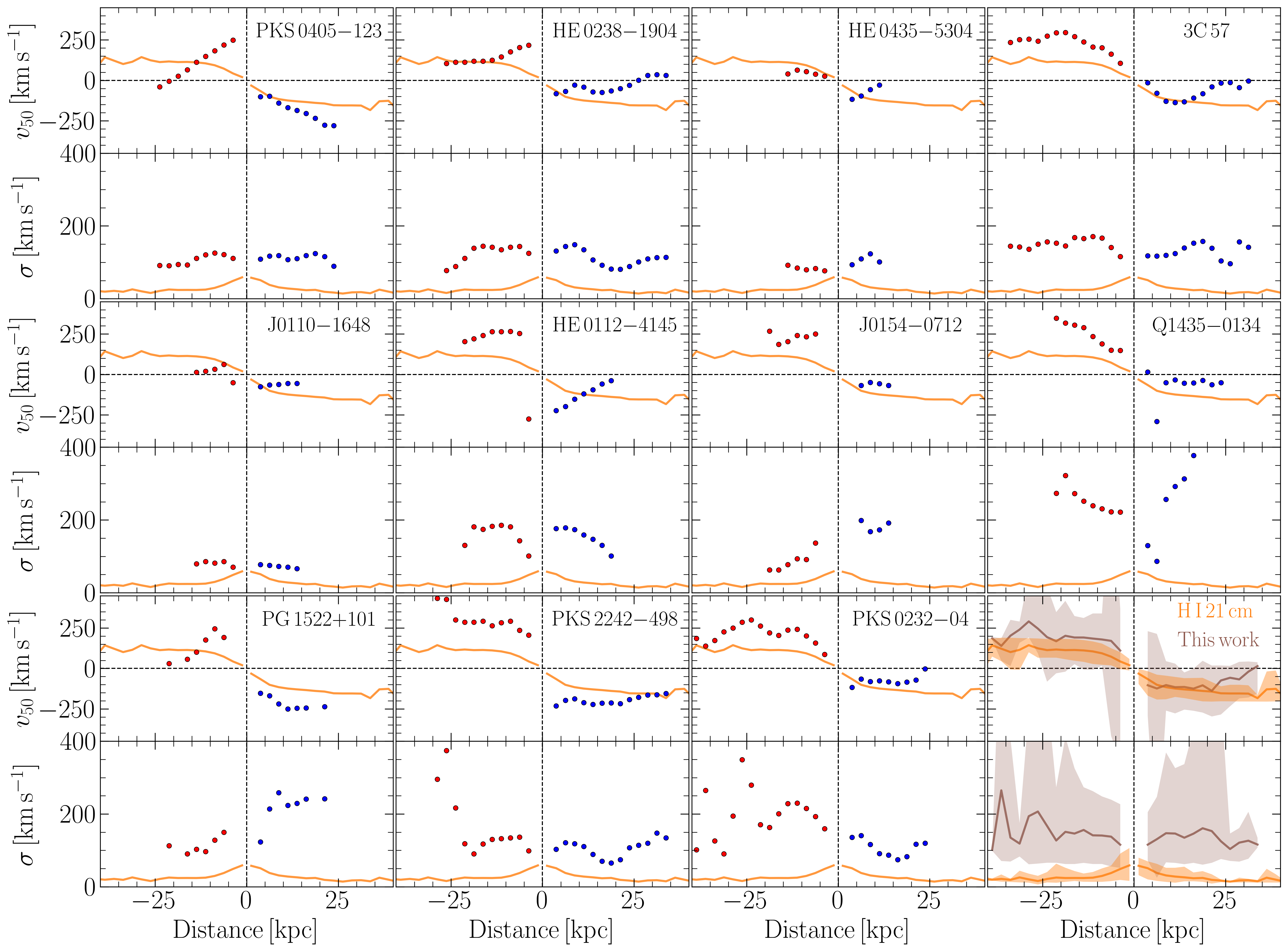}
    \caption{Position–velocity and velocity–dispersion diagrams are shown for each blueshifted-redshifted nebula in CUBS+MUSEQuBES, along with H\,I gas in early-type galaxies from \citet{2012MNRAS.422.1835S}. The radial profiles of $v_{50}$ and $\sigma$ are presented in the top and bottom panels, respectively. Binned data points for the blueshifted and redshifted sides are shown as red and blue circles. The mean radial profile of H\,I gas in early-type galaxies is shown as a solid line in each subpanel. The bottom-right panel compares the mean $\rm [O\,II]$ and H\,I profiles, with the shaded orange and brown regions indicating the full range (minimum to maximum) of values. Relative to the profile of H\,I gas, the profile of $\rm [O\,II]$ nebulae lack the smooth, symmetric trends characteristic and exhibit much higher velocity dispersions, highlighting the complexity of quasar nebula kinematics and possibly multi-origin nature of quasar nebulae.}
    \label{fig:vprofile}
\end{figure*}

\subsection{Investigating the Position-Velocity Profile of Blueshifted-Redshifted Nebulae} \label{sec:D-BR}
Early-type galaxies often exhibit extended rotating H\,I disks, traced by the 21\,cm line and provide a benchmark for studying large-scale gas dynamics in massive systems \citep{2014MNRAS.444.3388S}. Similarly, nebulae showing blueshifted-redshifted pattern (redshifted emission on one side of the quasar and blueshifted emission on the other side) in our sample may consist of rotating ISM and CGM gas associated with the quasar’s host galaxy. In this section, we examine the position–velocity profiles of nebulae exhibiting such patterns and compare them with H\,I gas observed in local early-type galaxies via 21\,cm emission. The H\,I 21\,cm line provides a useful analog for the gas seen in optical emission, as neutral gas in these systems would likely become ionized in the presence of the quasar’s intense radiation field. 

A blueshifted--redshifted velocity pattern can arise from any system with an ordered rotational component, but it can also be produced by purely radial motions such as inflow or outflow. In the discussion below, we therefore treat this pattern as suggestive of ordered kinematics rather than as definitive evidence for rotation.

We identified 11 out of 30 nebulae that exhibit either a prominent or partial blueshifted–redshifted velocity pattern. Although some of these systems show evidence of additional physical influences, our analysis focuses on the regions where the blueshifted–redshifted structure dominates. To further investigate this scenario, we extract the kinematic profiles of these nebulae and present them in position–velocity and position–velocity dispersion diagrams (Figure \ref{fig:vprofile}). We use a pseudo-slit $7\,\rm kpc$ wide and $100\,\rm kpc$ long. The pseudo-slit is placed along the major axis \citep[see Figure 3 of][]{2025ApJ...984..140L}, except for PKS\,2242$-$498 and PKS\,0232$-$04, where it is aligned with the minor axis to follow the observed velocity gradients. To enable the comparison, we follow the same methodology as \citet{2025ApJ...984..140L} by comparing with 21\,cm emission around local elliptical galaxies. 

We used data from the $\rm ATLAS^{3D}$ H\,I survey, selecting galaxies classified by \citet{2012MNRAS.422.1835S} as having large H I discs or rings---systems with regular rotation and gas distributions extending beyond the stellar body. To match the scale of our $\rm [O\,II]$ nebulae, we further restricted the sample to systems with rotating 21 cm gas extending over $40{-}100 \,\rm kpc$. We then derived corresponding $v_{50}$ and $\sigma$ maps by fitting single Gaussians to the H\,I datacubes, using the $\rm ATLAS^{3D}$ velocity fields as initial guesses. Kinematic profiles were extracted with a $7\,\rm kpc$–wide, $100\,\rm kpc$–long pseudo-slit along the major axis \citep[See Figure 3 in ][]{2025ApJ...984..140L}. To enable ensemble comparisons, we stacked the radial profiles to derive a mean trend, with the full range defined by the minimum and maximum values. The mean profiles are shown in Figure \ref{fig:vprofile}, with the bounds displayed in the lower-right panel for both $\rm [O\,II]$ and H\,I nebulae.

The $\rm [O\,II]$ nebulae show some similarities with the 21\,cm emission in the $v_{50}$ maps, though several key differences stand out. First, the $v_{50}$ profiles of the $\rm [O\,II]$ nebulae exhibit much stronger asymmetry between the blueshifted and redshifted sides. In contrast, the neutral hydrogen gas tends to have more symmetric profiles, with comparable maximum velocities on both sides. Second, the $\rm [O\,II]$ velocity profiles display more irregular behavior, often deviating from the typical pattern of rising from zero and flattening near the maximum. For example, the redshifted sides of PKS\,0405$-$123, J0110$-$1648, HE\,0112$-$4145, and PG\,1522$+$101 show asymptotically decreasing velocities, a trend not observed in the H\,I profiles, including both the mean profile and the individual galaxy profiles (not shown, but highly similar to the mean profile). These irregularities may reflect a more complex origin for the nebulae, involving a combination of processes beyond simple rotation, such as galaxy interactions and mergers, quasar-driven outflows, and cold gas accretion from the surrounding cosmic web. 

The velocity dispersion in the $\rm [O\,II]$ nebulae is generally higher and shows greater variation with distance from the quasar compared to the neutral gas. This difference may reflect intrinsic kinematic differences between ionized and neutral gas, suggesting that some gas observed in ionized lines around quasars would not remain neutral in the absence of quasar photoionization, as well as additional feedback and interactions that can enhance velocity dispersion in ionized regions through turbulent cascades. The elevated velocity dispersions may also be attributed to the significantly higher masses of the quasar host systems compared to local early-type galaxies. The stellar masses of the local early-type galaxies range from $\log(M_{*}/\rm M_{\odot}) \approx 10.3$ to 11.0, based on detailed axisymmetric dynamical modeling \citep{2013MNRAS.432.1709C}. Using stellar mass–halo mass relation from \citet{2018AstL...44....8K}, these stellar masses correspond to halo masses of $\log(M_{\rm 200}/\rm M_{\odot}) \approx 11.7$ to $12.4$. In contrast, the quasar host systems have halo masses of $\log(M_{\rm 200}/\rm M_{\odot}) \approx 12.1$ to $14.6$, inferred from their LOS velocity dispersions \citep{2024ApJ...965..143L}.

A blueshifted-redshifted velocity pattern can, in principle, arise from either ordered rotation or purely radial motions such as inflow or outflow. However, we do not favor a purely radial interpretation as the dominant explanation for these systems. In the detailed analysis of the 3C\,57 nebula \citep{2025ApJ...984..140L}, biconical outflow models can qualitatively reproduce the observed position-velocity profile, but the outflow scenario requires a small inclination angle ($\approx 10^\circ$--$20^\circ$) that is in tension with the unobscured Type I quasar geometry. All quasars in this work, drawn from the CUBS+MUSEQuBES survey, are unobscured, making a biconical outflow interpretation unlikely. Together with the extended and coherent velocity gradients observed here, this motivates an interpretation in which the large-scale gas kinematics include an ordered rotational component. We do not, however, claim that these systems are simple rotating disks, and AGN feedback, galaxy interactions, and accretion may also contribute significantly to the observed kinematics.

\section{Summary and Conclusions} \label{sec:SC}
In this work, we presented a comprehensive study of giant, optical-emitting, [O\,II] and [O\,III] nebulae around UV-luminous quasars at $z\approx0.4{-}1.4$ from the CUBS+MUSEQuBES survey, combining MUSE spectroscopy with HST imaging. We generated nebular surface-brightness maps using optimal extraction, characterized their kinematics through emission-line fitting, and examined their galactic environments, as surrounding galaxies can supply gas, drive interactions, and shape the formation of giant quasar nebulae. Using this information, we investigated nebular morphology, kinematics, and physical origins both qualitatively and quantitatively. Our main findings are summarized as follows:

\begin{enumerate}
    \item Based on nebular morphology and kinematics, we classified each nebula into four categories: irregular, large-scale systems (11 out of 30), likely driven by galaxy interactions; compact, host-galaxy-scale nebulae (11 out of 30), likely tracing ISM/CGM gas in and around the host; complex cases (5 out of 30) with ambiguous morphology and kinematics; and no nebular detection (3 out of 30). 

    \item Using the Bhattacharyya coefficient, we quantified the spatial–kinematic association between nebular gas and surrounding galaxies. We introduced the Kinematic Association Factor (KAF), which measures the level of coincidence between the position and velocity of nebular emission at a spaxel and those of galaxies in the quasar host group. We further defined the Cumulative KAF (CKAF) as the sum of the KAF over all spaxels, providing a global summary statistic. We demonstrated that KAF and CKAF robustly capture genuine spatial–kinematic alignment and identified statistically significant associations in ten nebulae. Most of these systems are irregular, large-scale, consistent with our qualitative analysis.

    \item We find no strong evidence that radio jets are the primary driver of the extended nebulae in our sample. Radio lobes are detected in six systems but are generally misaligned with the nebular major axis and rarely overlap the ionized gas, suggesting that radio jets are not responsible for producing the nebulae. At the same time, the presence of complex, multi-component emission features and the broadly similar, centrally peaked ensemble velocity-dispersion profiles suggest that AGN feedback likely contributes across the sample, with its influence strongest in the inner nebulae and potentially manifesting in both radio and quasar modes. Because the CUBS+MUSEQuBES sample targets UV-luminous quasars, these results may not be representative of the broader quasar population.

    \item We characterized nebular asymmetry and conducted comparisons with high-redshift Ly$\alpha$ nebulae around quasars and H\,I gas traced by 21\,cm line around local early-type galaxies. We found that the $\rm[O\,II]$ nebulae are generally more asymmetric than Ly$\alpha$ systems and broadly consistent with local H\,I gas observed in 21\,cm line, though the latter exhibits a wider asymmetry range. This difference likely reflects the resonant scattering of Ly$\alpha$ photons, which can smooth small-scale irregularities.
    
    \item We found that 11 of 30 nebulae exhibit a partial or prominent blueshifted–redshifted pattern, suggestive of ordered kinematics. While the corresponding 2D $v_{50}$ maps resemble rotating H\,I disks around local early-type galaxies, the morphology, velocity and dispersion profiles differ markedly from those of neutral gas. In particular, these quasar nebulae show stronger asymmetries, more irregular velocity trends, and systematically higher and more spatially variable velocity dispersions. These differences point to a complex, multi-origin nature for quasar nebulae.
\end{enumerate}
Taken together, these results show that giant quasar nebulae should not be interpreted as a single class of objects with a common origin. Instead, they represent a heterogeneous population shaped by the combined influence of host-galaxy gas, galaxy interactions, group environments, and quasar-driven activity. This diversity makes giant quasar nebulae powerful laboratories for studying how accretion, feedback, and environment jointly govern the assembly and evolution of circumgalactic gas around quasars.

\section{Acknowledgments}
We thank the anonymous referee for their constructive and insightful comments, which significantly improved this paper. We thank ZQL's thesis committee members, Camille Avestruz, Joel Bregman, and Feige Wang for their valuable comments and discussions. ZQL ackowledges partial support from NASA NNH24ZDA001N-FINESST. SDJ acknowledges partial support from NSF AST-2508761. SC gratefully acknowledges support from the European Research Council (ERC) under the European Union’s Horizon 2020 Research and Innovation programme grant agreement No 864361. JIL was supported by the Eric and Wendy Schmidt AI in Science Postdoctoral Fellowship, a Schmidt Futures program. MCC is supported by the Brinson Foundation through the Brinson Prize Fellowship Program. S.L. acknowledges support by FONDECYT grant 1231187. Based on observations with the NASA/ESA Hubble Space Telescope obtained at the Space Telescope Science Institute, which is operated by the Association of Universities for Research in Astronomy, Incorporated, under NASA contract NAS5-26555. Support for Program number 17815 was provided through a grant from the STScI under NASA contract NAS5-26555.

This paper is based on observations from the European Organization for Astronomical Research in the Southern Hemisphere under ESO (PI: H.-W. Chen, PID: 0104.A-0147; PI: J. Schaye, PID: 094.A-0131(B) \& 096.A-0222(A)), and the NASA/ESA Hubble Space Telescope (PI: L. Straka, PID: 14660; PI: J. S. Mulchaey, PID: 13024; PI: N. Lehner, PID: 14269; PI: A. Beckett, PID: 17815). The MAST-reduced (or STScI-reduced) HST images can be found in MAST: \dataset[10.17909/p63h-5b24]{http://dx.doi.org/10.17909/p63h-5b24}. Additionally, this paper made use of the NASA/IPAC Extragalactic Database, the NASA Astrophysics Data System, \texttt{Astropy} \citep[][]{2022ApJ...935..167A}, \texttt{Aplpy} \citep{aplpy2012}, and \texttt{Photutils} \citep{larry_bradley_2023_7946442}.

\section*{Data Availability}
The data used in this paper are available from the ESO and HST data archives. The MAST-reduced (or STScI-reduced) HST images can be found in MAST: \dataset[10.17909/p63h-5b24]{http://dx.doi.org/10.17909/p63h-5b24}.

\bibliography{main}{}

@ARTICLE{2020MNRAS.497..498C,
       author = {{Chen}, Hsiao-Wen and {Zahedy}, Fakhri S. and {Boettcher}, Erin and {Cooper}, Thomas M. and {Johnson}, Sean D. and {Rudie}, Gwen C. and {Chen}, Mandy C. and {Walth}, Gregory L. and {Cantalupo}, Sebastiano and {Cooksey}, Kathy L. and {Faucher-Gigu{\`e}re}, Claude-Andr{\'e} and {Greene}, Jenny E. and {Lopez}, Sebastian and {Mulchaey}, John S. and {Penton}, Steven V. and {Petitjean}, Patrick and {Putman}, Mary E. and {Rafelski}, Marc and {Rauch}, Michael and {Schaye}, Joop and {Simcoe}, Robert A. and {Weiner}, Benjamin J.},
        title = "{The Cosmic Ultraviolet Baryon Survey (CUBS) - I. Overview and the diverse environments of Lyman limit systems at z < 1}",
      journal = {\mnras},
         year = 2020,
        month = sep,
       volume = {497},
       number = {1},
        pages = {498-520},
          doi = {10.1093/mnras/staa1773},
archivePrefix = {arXiv},
       eprint = {2005.02408},
 primaryClass = {astro-ph.GA},
       adsurl = {https://ui.adsabs.harvard.edu/abs/2020MNRAS.497..498C}
}

@ARTICLE{2023ApJ...943L..25Z,
       author = {{Zhao}, Qinyuan and {Wang}, Junfeng},
        title = "{Discovery of a Spatially and Kinematically Resolved 55 kpc Scale Superbubble Inflated by an Intermediate-redshift Non-BAL Quasar}",
      journal = {\apjl},
         year = 2023,
        month = feb,
       volume = {943},
       number = {2},
          eid = {L25},
        pages = {L25},
          doi = {10.3847/2041-8213/acb546},
archivePrefix = {arXiv},
       eprint = {2302.00300},
 primaryClass = {astro-ph.GA},
       adsurl = {https://ui.adsabs.harvard.edu/abs/2023ApJ...943L..25Z}
}

@ARTICLE{2024MNRAS.528.3745D,
       author = {{Dutta}, Sayak and {Muzahid}, Sowgat and {Schaye}, Joop and {Mishra}, Sapna and {Chen}, Hsiao-Wen and {Johnson}, Sean and {Wisotzki}, Lutz and {Cantalupo}, Sebastiano},
        title = "{MUSEQuBES: mapping the distribution of neutral hydrogen around low-redshift galaxies}",
      journal = {\mnras},
         year = 2024,
        month = feb,
       volume = {528},
       number = {2},
        pages = {3745-3766},
          doi = {10.1093/mnras/stae206},
archivePrefix = {arXiv},
       eprint = {2303.16933},
 primaryClass = {astro-ph.GA},
       adsurl = {https://ui.adsabs.harvard.edu/abs/2024MNRAS.528.3745D}
}

@ARTICLE{2020MNRAS.496.1013M,
       author = {{Muzahid}, Sowgat and {Schaye}, Joop and {Marino}, Raffaella Anna and {Cantalupo}, Sebastiano and {Brinchmann}, Jarle and {Contini}, Thierry and {Wendt}, Martin and {Wisotzki}, Lutz and {Zabl}, Johannes and {Bouch{\'e}}, Nicolas and {Akhlaghi}, Mohammad and {Chen}, Hsiao-Wen and {Claeyssens}, Ad{\'e}la{\^\i}de and {Johnson}, Sean and {Leclercq}, Floriane and {Maseda}, Michael and {Matthee}, Jorryt and {Richard}, Johan and {Urrutia}, Tanya and {Verhamme}, Anne},
        title = "{MUSEQuBES: calibrating the redshifts of Ly {\ensuremath{\alpha}} emitters using stacked circumgalactic medium absorption profiles}",
      journal = {\mnras},
         year = 2020,
        month = aug,
       volume = {496},
       number = {2},
        pages = {1013-1022},
          doi = {10.1093/mnras/staa1347},
archivePrefix = {arXiv},
       eprint = {1910.03593},
 primaryClass = {astro-ph.GA},
       adsurl = {https://ui.adsabs.harvard.edu/abs/2020MNRAS.496.1013M}
}

@INPROCEEDINGS{2012SPIE.8451E..0BW,
       author = {{Weilbacher}, Peter M. and {Streicher}, Ole and {Urrutia}, Tanya and {Jarno}, Aur{\'e}lien and {P{\'e}contal-Rousset}, Arlette and {Bacon}, Roland and {B{\"o}hm}, Petra},
        title = "{Design and capabilities of the MUSE data reduction software and pipeline}",
    booktitle = {Software and Cyberinfrastructure for Astronomy II},
         year = 2012,
       editor = {{Radziwill}, Nicole M. and {Chiozzi}, Gianluca},
       series = {Society of Photo-Optical Instrumentation Engineers (SPIE) Conference Series},
       volume = {8451},
        month = sep,
          eid = {84510B},
        pages = {84510B},
          doi = {10.1117/12.925114},
       adsurl = {https://ui.adsabs.harvard.edu/abs/2012SPIE.8451E..0BW}
}

@ARTICLE{2021MNRAS.505.5497H,
       author = {{Helton}, Jakob M. and {Johnson}, Sean D. and {Greene}, Jenny E. and {Chen}, Hsiao-Wen},
        title = "{Discovery and origins of giant optical nebulae surrounding quasar PKS 0454-22}",
      journal = {\mnras},
         year = 2021,
        month = aug,
       volume = {505},
       number = {4},
        pages = {5497-5513},
          doi = {10.1093/mnras/stab1647},
archivePrefix = {arXiv},
       eprint = {2106.03858},
 primaryClass = {astro-ph.GA},
       adsurl = {https://ui.adsabs.harvard.edu/abs/2021MNRAS.505.5497H}
}

@ARTICLE{2018ApJ...869L...1J,
       author = {{Johnson}, Sean D. and {Chen}, Hsiao-Wen and {Straka}, Lorrie A. and {Schaye}, Joop and {Cantalupo}, Sebastiano and {Wendt}, Martin and {Muzahid}, Sowgat and {Bouch{\'e}}, Nicolas and {Herenz}, Edmund Christian and {Kollatschny}, Wolfram and {Mulchaey}, John S. and {Marino}, Raffaella A. and {Maseda}, Michael V. and {Wisotzki}, Lutz},
        title = "{Galaxy and Quasar Fueling Caught in the Act from the Intragroup to the Interstellar Medium}",
      journal = {\apjl},
         year = 2018,
        month = dec,
       volume = {869},
       number = {1},
          eid = {L1},
        pages = {L1},
          doi = {10.3847/2041-8213/aaf1cf},
archivePrefix = {arXiv},
       eprint = {1811.10615},
 primaryClass = {astro-ph.GA},
       adsurl = {https://ui.adsabs.harvard.edu/abs/2018ApJ...869L...1J}
}

@ARTICLE{2022ApJ...940L..40J,
       author = {{Johnson}, Sean D. and {Schaye}, Joop and {Walth}, Gregory L. and {Li}, Jennifer I. -Hsiu and {Rudie}, Gwen C. and {Chen}, Hsiao-Wen and {Chen}, Mandy C. and {Epinat}, Beno{\^\i}t and {Gaspari}, Massimo and {Cantalupo}, Sebastiano and {Kollatschny}, Wolfram and {Liu}, Zhuoqi (Will) and {Muzahid}, Sowgat},
        title = "{Directly Tracing Cool Filamentary Accretion over >100 kpc into the Interstellar Medium of a Quasar Host at z = 1}",
      journal = {\apjl},
         year = 2022,
        month = dec,
       volume = {940},
       number = {2},
          eid = {L40},
        pages = {L40},
          doi = {10.3847/2041-8213/aca28e},
archivePrefix = {arXiv},
       eprint = {2209.04245},
 primaryClass = {astro-ph.GA},
       adsurl = {https://ui.adsabs.harvard.edu/abs/2022ApJ...940L..40J}
}

@ARTICLE{2023MNRAS.518.2354C,
       author = {{Chen}, Mandy C. and {Chen}, Hsiao-Wen and {Rauch}, Michael and {Qu}, Zhijie and {Johnson}, Sean D. and {Li}, Jennifer I. -Hsiu and {Schaye}, Joop and {Rudie}, Gwen C. and {Zahedy}, Fakhri S. and {Boettcher}, Erin and {Cooksey}, Kathy L. and {Cantalupo}, Sebastiano},
        title = "{Empirical constraints on the turbulence in QSO host nebulae from velocity structure function measurements}",
      journal = {\mnras},
         year = 2023,
        month = jan,
       volume = {518},
       number = {2},
        pages = {2354-2372},
          doi = {10.1093/mnras/stac3193},
archivePrefix = {arXiv},
       eprint = {2209.04344},
 primaryClass = {astro-ph.GA},
       adsurl = {https://ui.adsabs.harvard.edu/abs/2023MNRAS.518.2354C}
}

@ARTICLE{2019MNRAS.483.5188C,
       author = {{Cantalupo}, Sebastiano and {Pezzulli}, Gabriele and {Lilly}, Simon J. and {Marino}, Raffaella Anna and {Gallego}, Sofia G. and {Schaye}, Joop and {Bacon}, Roland and {Feltre}, Anna and {Kollatschny}, Wolfram and {Nanayakkara}, Themiya and {Richard}, Johan and {Wendt}, Martin and {Wisotzki}, Lutz and {Prochaska}, J. Xavier},
        title = "{The large- and small-scale properties of the intergalactic gas in the Slug Ly {\ensuremath{\alpha}} nebula revealed by MUSE He II emission observations}",
      journal = {\mnras},
         year = 2019,
        month = mar,
       volume = {483},
       number = {4},
        pages = {5188-5204},
          doi = {10.1093/mnras/sty3481},
archivePrefix = {arXiv},
       eprint = {1811.11783},
 primaryClass = {astro-ph.GA},
       adsurl = {https://ui.adsabs.harvard.edu/abs/2019MNRAS.483.5188C}
}

@INPROCEEDINGS{2014ASPC..485..451W,
       author = {{Weilbacher}, P.~M. and {Streicher}, O. and {Urrutia}, T. and {P{\'e}contal-Rousset}, A. and {Jarno}, A. and {Bacon}, R.},
        title = "{The MUSE Data Reduction Pipeline: Status after Preliminary Acceptance Europe}",
    booktitle = {Astronomical Data Analysis Software and Systems XXIII},
         year = 2014,
       editor = {{Manset}, N. and {Forshay}, P.},
       series = {Astronomical Society of the Pacific Conference Series},
       volume = {485},
        month = may,
        pages = {451},
          doi = {10.48550/arXiv.1507.00034},
archivePrefix = {arXiv},
       eprint = {1507.00034},
 primaryClass = {astro-ph.IM},
       adsurl = {https://ui.adsabs.harvard.edu/abs/2014ASPC..485..451W}
}

@ARTICLE{2012AJ....144..144B,
       author = {{Bolton}, Adam S. and {Schlegel}, David J. and {Aubourg}, {\'E}ric and {Bailey}, Stephen and {Bhardwaj}, Vaishali and {Brownstein}, Joel R. and {Burles}, Scott and {Chen}, Yan-Mei and {Dawson}, Kyle and {Eisenstein}, Daniel J. and {Gunn}, James E. and {Knapp}, G.~R. and {Loomis}, Craig P. and {Lupton}, Robert H. and {Maraston}, Claudia and {Muna}, Demitri and {Myers}, Adam D. and {Olmstead}, Matthew D. and {Padmanabhan}, Nikhil and {P{\^a}ris}, Isabelle and {Percival}, Will J. and {Petitjean}, Patrick and {Rockosi}, Constance M. and {Ross}, Nicholas P. and {Schneider}, Donald P. and {Shu}, Yiping and {Strauss}, Michael A. and {Thomas}, Daniel and {Tremonti}, Christy A. and {Wake}, David A. and {Weaver}, Benjamin A. and {Wood-Vasey}, W. Michael},
        title = "{Spectral Classification and Redshift Measurement for the SDSS-III Baryon Oscillation Spectroscopic Survey}",
      journal = {\aj},
         year = 2012,
        month = nov,
       volume = {144},
       number = {5},
          eid = {144},
        pages = {144},
          doi = {10.1088/0004-6256/144/5/144},
archivePrefix = {arXiv},
       eprint = {1207.7326},
 primaryClass = {astro-ph.CO},
       adsurl = {https://ui.adsabs.harvard.edu/abs/2012AJ....144..144B}
}

@ARTICLE{2013ApJ...768...74T,
       author = {{Tacconi}, L.~J. and {Neri}, R. and {Genzel}, R. and {Combes}, F. and {Bolatto}, A. and {Cooper}, M.~C. and {Wuyts}, S. and {Bournaud}, F. and {Burkert}, A. and {Comerford}, J. and {Cox}, P. and {Davis}, M. and {F{\"o}rster Schreiber}, N.~M. and {Garc{\'\i}a-Burillo}, S. and {Gracia-Carpio}, J. and {Lutz}, D. and {Naab}, T. and {Newman}, S. and {Omont}, A. and {Saintonge}, A. and {Shapiro Griffin}, K. and {Shapley}, A. and {Sternberg}, A. and {Weiner}, B.},
        title = "{Phibss: Molecular Gas Content and Scaling Relations in z \raisebox{-0.5ex}\textasciitilde 1-3 Massive, Main-sequence Star-forming Galaxies}",
      journal = {\apj},
         year = 2013,
        month = may,
       volume = {768},
       number = {1},
          eid = {74},
        pages = {74},
          doi = {10.1088/0004-637X/768/1/74},
archivePrefix = {arXiv},
       eprint = {1211.5743},
 primaryClass = {astro-ph.CO},
       adsurl = {https://ui.adsabs.harvard.edu/abs/2013ApJ...768...74T}
}

@ARTICLE{2014Natur.506...63C,
       author = {{Cantalupo}, Sebastiano and {Arrigoni-Battaia}, Fabrizio and {Prochaska}, J. Xavier and {Hennawi}, Joseph F. and {Madau}, Piero},
        title = "{A cosmic web filament revealed in Lyman-{\ensuremath{\alpha}} emission around a luminous high-redshift quasar}",
      journal = {\nat},
         year = 2014,
        month = feb,
       volume = {506},
       number = {7486},
        pages = {63-66},
          doi = {10.1038/nature12898},
archivePrefix = {arXiv},
       eprint = {1401.4469},
 primaryClass = {astro-ph.CO},
       adsurl = {https://ui.adsabs.harvard.edu/abs/2014Natur.506...63C}
}

@ARTICLE{2016ApJ...831...39B,
       author = {{Borisova}, Elena and {Cantalupo}, Sebastiano and {Lilly}, Simon J. and {Marino}, Raffaella A. and {Gallego}, Sofia G. and {Bacon}, Roland and {Blaizot}, Jeremy and {Bouch{\'e}}, Nicolas and {Brinchmann}, Jarle and {Carollo}, C. Marcella and {Caruana}, Joseph and {Finley}, Hayley and {Herenz}, Edmund C. and {Richard}, Johan and {Schaye}, Joop and {Straka}, Lorrie A. and {Turner}, Monica L. and {Urrutia}, Tanya and {Verhamme}, Anne and {Wisotzki}, Lutz},
        title = "{Ubiquitous Giant Ly{\ensuremath{\alpha}} Nebulae around the Brightest Quasars at z {\ensuremath{\sim}} 3.5 Revealed with MUSE}",
      journal = {\apj},
         year = 2016,
        month = nov,
       volume = {831},
       number = {1},
          eid = {39},
        pages = {39},
          doi = {10.3847/0004-637X/831/1/39},
archivePrefix = {arXiv},
       eprint = {1605.01422},
 primaryClass = {astro-ph.GA},
       adsurl = {https://ui.adsabs.harvard.edu/abs/2016ApJ...831...39B}
}

@ARTICLE{2019ApJS..245...23C,
       author = {{Cai}, Zheng and {Cantalupo}, Sebastiano and {Prochaska}, J. Xavier and {Arrigoni Battaia}, Fabrizio and {Burchett}, Joe and {Li}, Qiong and {Chisholm}, John and {Bundy}, Kevin and {Hennawi}, Joseph F.},
        title = "{Evolution of the Cool Gas in the Circumgalactic Medium of Massive Halos: A Keck Cosmic Web Imager Survey of Ly{\ensuremath{\alpha}} Emission around QSOs at z {\ensuremath{\approx}} 2}",
      journal = {\apjs},
         year = 2019,
        month = dec,
       volume = {245},
       number = {2},
          eid = {23},
        pages = {23},
          doi = {10.3847/1538-4365/ab4796},
archivePrefix = {arXiv},
       eprint = {1909.11098},
 primaryClass = {astro-ph.GA},
       adsurl = {https://ui.adsabs.harvard.edu/abs/2019ApJS..245...23C}
}

@ARTICLE{2020ApJ...894....3O,
       author = {{O'Sullivan}, Donal B. and {Martin}, Christopher and {Matuszewski}, Mateusz and {Hoadley}, Keri and {Hamden}, Erika and {Neill}, James D. and {Lin}, Zeren and {Parihar}, Prachi},
        title = "{The FLASHES Survey. I. Integral Field Spectroscopy of the CGM around 48 z ≃ 2.3-3.1 QSOs}",
      journal = {\apj},
         year = 2020,
        month = may,
       volume = {894},
       number = {1},
          eid = {3},
        pages = {3},
          doi = {10.3847/1538-4357/ab838c},
archivePrefix = {arXiv},
       eprint = {1911.10740},
 primaryClass = {astro-ph.GA},
       adsurl = {https://ui.adsabs.harvard.edu/abs/2020ApJ...894....3O}
}

@ARTICLE{2021MNRAS.503.3044F,
       author = {{Fossati}, M. and {Fumagalli}, M. and {Lofthouse}, E.~K. and {Dutta}, R. and {Cantalupo}, S. and {Arrigoni Battaia}, F. and {Fynbo}, J.~P.~U. and {Lusso}, E. and {Murphy}, M.~T. and {Prochaska}, J.~X. and {Theuns}, T. and {Cooke}, R.~J.},
        title = "{MUSE analysis of gas around galaxies (MAGG) - III. The gas and galaxy environment of z = 3-4.5 quasars}",
      journal = {\mnras},
         year = 2021,
        month = may,
       volume = {503},
       number = {2},
        pages = {3044-3064},
          doi = {10.1093/mnras/stab660},
archivePrefix = {arXiv},
       eprint = {2103.01960},
 primaryClass = {astro-ph.GA},
       adsurl = {https://ui.adsabs.harvard.edu/abs/2021MNRAS.503.3044F}
}

@INPROCEEDINGS{2010SPIE.7735E..08B,
       author = {{Bacon}, R. and {Accardo}, M. and {Adjali}, L. and {Anwand}, H. and {Bauer}, S. and {Biswas}, I. and {Blaizot}, J. and {Boudon}, D. and {Brau-Nogue}, S. and {Brinchmann}, J. and {Caillier}, P. and {Capoani}, L. and {Carollo}, C.~M. and {Contini}, T. and {Couderc}, P. and {Daguis{\'e}}, E. and {Deiries}, S. and {Delabre}, B. and {Dreizler}, S. and {Dubois}, J. and {Dupieux}, M. and {Dupuy}, C. and {Emsellem}, E. and {Fechner}, T. and {Fleischmann}, A. and {Fran{\c{c}}ois}, M. and {Gallou}, G. and {Gharsa}, T. and {Glindemann}, A. and {Gojak}, D. and {Guiderdoni}, B. and {Hansali}, G. and {Hahn}, T. and {Jarno}, A. and {Kelz}, A. and {Koehler}, C. and {Kosmalski}, J. and {Laurent}, F. and {Le Floch}, M. and {Lilly}, S.~J. and {Lizon}, J.-L. and {Loupias}, M. and {Manescau}, A. and {Monstein}, C. and {Nicklas}, H. and {Olaya}, J.-C. and {Pares}, L. and {Pasquini}, L. and {P{\'e}contal-Rousset}, A. and {Pell{\'o}}, R. and {Petit}, C. and {Popow}, E. and {Reiss}, R. and {Remillieux}, A. and {Renault}, E. and {Roth}, M. and {Rupprecht}, G. and {Serre}, D. and {Schaye}, J. and {Soucail}, G. and {Steinmetz}, M. and {Streicher}, O. and {Stuik}, R. and {Valentin}, H. and {Vernet}, J. and {Weilbacher}, P. and {Wisotzki}, L. and {Yerle}, N.},
        title = "{The MUSE second-generation VLT instrument}",
    booktitle = {Ground-based and Airborne Instrumentation for Astronomy III},
         year = 2010,
       editor = {{McLean}, Ian S. and {Ramsay}, Suzanne K. and {Takami}, Hideki},
       series = {Society of Photo-Optical Instrumentation Engineers (SPIE) Conference Series},
       volume = {7735},
        month = jul,
          eid = {773508},
        pages = {773508},
          doi = {10.1117/12.856027},
archivePrefix = {arXiv},
       eprint = {2211.16795},
 primaryClass = {astro-ph.IM},
       adsurl = {https://ui.adsabs.harvard.edu/abs/2010SPIE.7735E..08B}
}

@BOOK{2021pdaa.book.....N,
       author = {{National Academies of Sciences}},
        title = "{Pathways to Discovery in Astronomy and Astrophysics for the 2020s}",
         year = 2021,
          doi = {10.17226/26141},
       adsurl = {https://ui.adsabs.harvard.edu/abs/2021pdaa.book.....N}
}

@ARTICLE{2012ARA&A..50..455F,
       author = {{Fabian}, A.~C.},
        title = "{Observational Evidence of Active Galactic Nuclei Feedback}",
      journal = {\araa},
         year = 2012,
        month = sep,
       volume = {50},
        pages = {455-489},
          doi = {10.1146/annurev-astro-081811-125521},
archivePrefix = {arXiv},
       eprint = {1204.4114},
 primaryClass = {astro-ph.CO},
       adsurl = {https://ui.adsabs.harvard.edu/abs/2012ARA&A..50..455F}
}

@ARTICLE{2006ApJ...647..910H,
       author = {{Hester}, J.~A.},
        title = "{Ram Pressure Stripping in Clusters and Groups}",
      journal = {\apj},
         year = 2006,
        month = aug,
       volume = {647},
       number = {2},
        pages = {910-921},
          doi = {10.1086/505614},
archivePrefix = {arXiv},
       eprint = {astro-ph/0610088},
 primaryClass = {astro-ph},
       adsurl = {https://ui.adsabs.harvard.edu/abs/2006ApJ...647..910H}
}

@ARTICLE{2016MNRAS.461.2630M,
       author = {{Marasco}, Antonino and {Crain}, Robert A. and {Schaye}, Joop and {Bah{\'e}}, Yannick M. and {van der Hulst}, Thijs and {Theuns}, Tom and {Bower}, Richard G.},
        title = "{The environmental dependence of H I in galaxies in the EAGLE simulations}",
      journal = {\mnras},
         year = 2016,
        month = sep,
       volume = {461},
       number = {3},
        pages = {2630-2649},
          doi = {10.1093/mnras/stw1498},
archivePrefix = {arXiv},
       eprint = {1606.06288},
 primaryClass = {astro-ph.GA},
       adsurl = {https://ui.adsabs.harvard.edu/abs/2016MNRAS.461.2630M}
}

@ARTICLE{2019ApJ...878L..33C,
       author = {{Chen}, Hsiao-Wen and {Boettcher}, Erin and {Johnson}, Sean D. and {Zahedy}, Fakhri S. and {Rudie}, Gwen C. and {Cooksey}, Kathy L. and {Rauch}, Michael and {Mulchaey}, John S.},
        title = "{A Giant Intragroup Nebula Hosting a Damped \{Ly\}\textbackslashalpha  Absorber at z = 0.313}",
      journal = {\apjl},
         year = 2019,
        month = jun,
       volume = {878},
       number = {2},
          eid = {L33},
        pages = {L33},
          doi = {10.3847/2041-8213/ab25ec},
archivePrefix = {arXiv},
       eprint = {1906.00005},
 primaryClass = {astro-ph.GA},
       adsurl = {https://ui.adsabs.harvard.edu/abs/2019ApJ...878L..33C}
}

@ARTICLE{2021MNRAS.507.4294Z,
       author = {{Zabl}, Johannes and {Bouch{\'e}}, Nicolas F. and {Wisotzki}, Lutz and {Schaye}, Joop and {Leclercq}, Floriane and {Garel}, Thibault and {Wendt}, Martin and {Schroetter}, Ilane and {Muzahid}, Sowgat and {Cantalupo}, Sebastiano and {Contini}, Thierry and {Bacon}, Roland and {Brinchmann}, Jarle and {Richard}, Johan},
        title = "{MusE GAs FLOw and Wind (MEGAFLOW) VIII. Discovery of a MgII emission halo probed by a quasar sightline}",
      journal = {\mnras},
         year = 2021,
        month = nov,
       volume = {507},
       number = {3},
        pages = {4294-4315},
          doi = {10.1093/mnras/stab2165},
archivePrefix = {arXiv},
       eprint = {2105.14090},
 primaryClass = {astro-ph.GA},
       adsurl = {https://ui.adsabs.harvard.edu/abs/2021MNRAS.507.4294Z}
}

@ARTICLE{2022A&A...663A..11L,
       author = {{Leclercq}, Floriane and {Verhamme}, Anne and {Epinat}, Benoit and {Simmonds}, Charlotte and {Matthee}, Jorryt and {Bouch{\'e}}, Nicolas F. and {Garel}, Thibault and {Urrutia}, Tanya and {Wisotzki}, Lutz and {Zabl}, Johannes and {Bacon}, Roland and {Abril-Melgarejo}, Valentina and {Boogaard}, Leindert and {Brinchmann}, Jarle and {Cantalupo}, Sebastiano and {Contini}, Thierry and {Kerutt}, Josephine and {Kusakabe}, Haruka and {Maseda}, Michael and {Michel-Dansac}, L{\'e}o and {Muzahid}, Sowgat and {Nanayakkara}, Themiya and {Richard}, Johan and {Schaye}, Joop},
        title = "{The MUSE eXtremely deep field: first panoramic view of an Mg II emitting intragroup medium}",
      journal = {\aap},
         year = 2022,
        month = jul,
       volume = {663},
          eid = {A11},
        pages = {A11},
          doi = {10.1051/0004-6361/202142179},
archivePrefix = {arXiv},
       eprint = {2203.05614},
 primaryClass = {astro-ph.GA},
       adsurl = {https://ui.adsabs.harvard.edu/abs/2022A&A...663A..11L}
}

@ARTICLE{2023MNRAS.522..535D,
       author = {{Dutta}, Rajeshwari and {Fossati}, Matteo and {Fumagalli}, Michele and {Revalski}, Mitchell and {Lofthouse}, Emma K. and {Nelson}, Dylan and {Papini}, Giulia and {Rafelski}, Marc and {Cantalupo}, Sebastiano and {Arrigoni Battaia}, Fabrizio and {Dayal}, Pratika and {Longobardi}, Alessia and {P{\'e}roux}, Celine and {Prichard}, Laura J. and {Prochaska}, J. Xavier},
        title = "{Metal line emission from galaxy haloes at z {\ensuremath{\approx}} 1}",
      journal = {\mnras},
         year = 2023,
        month = jun,
       volume = {522},
       number = {1},
        pages = {535-558},
          doi = {10.1093/mnras/stad1002},
archivePrefix = {arXiv},
       eprint = {2302.09087},
 primaryClass = {astro-ph.GA},
       adsurl = {https://ui.adsabs.harvard.edu/abs/2023MNRAS.522..535D}
}

@ARTICLE{2018Galax...6..114Z,
       author = {{Zhang}, Dong},
        title = "{A Review of the Theory of Galactic Winds Driven by Stellar Feedback}",
      journal = {Galaxies},
         year = 2018,
        month = nov,
       volume = {6},
       number = {4},
        pages = {114},
          doi = {10.3390/galaxies6040114},
archivePrefix = {arXiv},
       eprint = {1811.00558},
 primaryClass = {astro-ph.GA},
       adsurl = {https://ui.adsabs.harvard.edu/abs/2018Galax...6..114Z}
}

@ARTICLE{2021MNRAS.502..494M,
       author = {{Mackenzie}, Ruari and {Pezzulli}, Gabriele and {Cantalupo}, Sebastiano and {Marino}, Raffaella A. and {Lilly}, Simon and {Muzahid}, Sowgat and {Matthee}, Jorryt and {Schaye}, Joop and {Wisotzki}, Lutz},
        title = "{Revealing the impact of quasar luminosity on giant Ly {\ensuremath{\alpha}} nebulae}",
      journal = {\mnras},
         year = 2021,
        month = mar,
       volume = {502},
       number = {1},
        pages = {494-509},
          doi = {10.1093/mnras/staa3277},
archivePrefix = {arXiv},
       eprint = {2010.12589},
 primaryClass = {astro-ph.GA},
       adsurl = {https://ui.adsabs.harvard.edu/abs/2021MNRAS.502..494M}
}

@ARTICLE{2022ApJ...935..167A,
       author = {{Astropy Collaboration} and {Price-Whelan}, Adrian M. and {Lim}, Pey Lian and {Earl}, Nicholas and {Starkman}, Nathaniel and {Bradley}, Larry and {Shupe}, David L. and {Patil}, Aarya A. and {Corrales}, Lia and {Brasseur}, C.~E. and {N{\"o}the}, Maximilian and {Donath}, Axel and {Tollerud}, Erik and {Morris}, Brett M. and {Ginsburg}, Adam and {Vaher}, Eero and {Weaver}, Benjamin A. and {Tocknell}, James and {Jamieson}, William and {van Kerkwijk}, Marten H. and {Robitaille}, Thomas P. and {Merry}, Bruce and {Bachetti}, Matteo and {G{\"u}nther}, H. Moritz and {Aldcroft}, Thomas L. and {Alvarado-Montes}, Jaime A. and {Archibald}, Anne M. and {B{\'o}di}, Attila and {Bapat}, Shreyas and {Barentsen}, Geert and {Baz{\'a}n}, Juanjo and {Biswas}, Manish and {Boquien}, M{\'e}d{\'e}ric and {Burke}, D.~J. and {Cara}, Daria and {Cara}, Mihai and {Conroy}, Kyle E. and {Conseil}, Simon and {Craig}, Matthew W. and {Cross}, Robert M. and {Cruz}, Kelle L. and {D'Eugenio}, Francesco and {Dencheva}, Nadia and {Devillepoix}, Hadrien A.~R. and {Dietrich}, J{\"o}rg P. and {Eigenbrot}, Arthur Davis and {Erben}, Thomas and {Ferreira}, Leonardo and {Foreman-Mackey}, Daniel and {Fox}, Ryan and {Freij}, Nabil and {Garg}, Suyog and {Geda}, Robel and {Glattly}, Lauren and {Gondhalekar}, Yash and {Gordon}, Karl D. and {Grant}, David and {Greenfield}, Perry and {Groener}, Austen M. and {Guest}, Steve and {Gurovich}, Sebastian and {Handberg}, Rasmus and {Hart}, Akeem and {Hatfield-Dodds}, Zac and {Homeier}, Derek and {Hosseinzadeh}, Griffin and {Jenness}, Tim and {Jones}, Craig K. and {Joseph}, Prajwel and {Kalmbach}, J. Bryce and {Karamehmetoglu}, Emir and {Ka{\l}uszy{\'n}ski}, Miko{\l}aj and {Kelley}, Michael S.~P. and {Kern}, Nicholas and {Kerzendorf}, Wolfgang E. and {Koch}, Eric W. and {Kulumani}, Shankar and {Lee}, Antony and {Ly}, Chun and {Ma}, Zhiyuan and {MacBride}, Conor and {Maljaars}, Jakob M. and {Muna}, Demitri and {Murphy}, N.~A. and {Norman}, Henrik and {O'Steen}, Richard and {Oman}, Kyle A. and {Pacifici}, Camilla and {Pascual}, Sergio and {Pascual-Granado}, J. and {Patil}, Rohit R. and {Perren}, Gabriel I. and {Pickering}, Timothy E. and {Rastogi}, Tanuj and {Roulston}, Benjamin R. and {Ryan}, Daniel F. and {Rykoff}, Eli S. and {Sabater}, Jose and {Sakurikar}, Parikshit and {Salgado}, Jes{\'u}s and {Sanghi}, Aniket and {Saunders}, Nicholas and {Savchenko}, Volodymyr and {Schwardt}, Ludwig and {Seifert-Eckert}, Michael and {Shih}, Albert Y. and {Jain}, Anany Shrey and {Shukla}, Gyanendra and {Sick}, Jonathan and {Simpson}, Chris and {Singanamalla}, Sudheesh and {Singer}, Leo P. and {Singhal}, Jaladh and {Sinha}, Manodeep and {Sip{\H{o}}cz}, Brigitta M. and {Spitler}, Lee R. and {Stansby}, David and {Streicher}, Ole and {{\v{S}}umak}, Jani and {Swinbank}, John D. and {Taranu}, Dan S. and {Tewary}, Nikita and {Tremblay}, Grant R. and {de Val-Borro}, Miguel and {Van Kooten}, Samuel J. and {Vasovi{\'c}}, Zlatan and {Verma}, Shresth and {de Miranda Cardoso}, Jos{\'e} Vin{\'\i}cius and {Williams}, Peter K.~G. and {Wilson}, Tom J. and {Winkel}, Benjamin and {Wood-Vasey}, W.~M. and {Xue}, Rui and {Yoachim}, Peter and {Zhang}, Chen and {Zonca}, Andrea and {Astropy Project Contributors}},
        title = "{The Astropy Project: Sustaining and Growing a Community-oriented Open-source Project and the Latest Major Release (v5.0) of the Core Package}",
      journal = {\apj},
         year = 2022,
        month = aug,
       volume = {935},
       number = {2},
          eid = {167},
        pages = {167},
          doi = {10.3847/1538-4357/ac7c74},
archivePrefix = {arXiv},
       eprint = {2206.14220},
 primaryClass = {astro-ph.IM},
       adsurl = {https://ui.adsabs.harvard.edu/abs/2022ApJ...935..167A}
}

@misc{aplpy2012,
  author        = {{Robitaille}, T. and {Bressert}, E.},
  title         = "{APLpy: Astronomical Plotting Library in Python}",
  howpublished  = {Astrophysics Source Code Library},
  year          = 2012,
  month         = aug,
  archivePrefix = "ascl",
  eprint        = {1208.017},
  adsurl        = {http://adsabs.harvard.edu/abs/2012ascl.soft08017R}
}

@ARTICLE{2020MNRAS.495.1874D,
       author = {{den Brok}, J.~S. and {Cantalupo}, S. and {Mackenzie}, R. and {Marino}, R.~A. and {Pezzulli}, G. and {Matthee}, J. and {Johnson}, S.~D. and {Krumpe}, M. and {Urrutia}, T. and {Kollatschny}, W.},
        title = "{Probing the AGN unification model at redshift z {\ensuremath{\sim}} 3 with MUSE observations of giant Ly {\ensuremath{\alpha}} nebulae}",
      journal = {\mnras},
         year = 2020,
        month = jun,
       volume = {495},
       number = {2},
        pages = {1874-1887},
          doi = {10.1093/mnras/staa1269},
archivePrefix = {arXiv},
       eprint = {2005.01732},
 primaryClass = {astro-ph.GA},
       adsurl = {https://ui.adsabs.harvard.edu/abs/2020MNRAS.495.1874D}
}

@software{larry_bradley_2023_7946442,
  author       = {Larry Bradley},
  title        = {astropy/photutils: 1.8.0},
  month        = may,
  year         = 2023,
  publisher    = {Zenodo},
  version      = {1.8.0},
  doi          = {10.5281/zenodo.7946442},
  url          = {https://doi.org/10.5281/zenodo.7946442}
}

@ARTICLE{2019A&A...631A..18A,
       author = {{Arrigoni Battaia}, F. and {Obreja}, A. and {Prochaska}, J.~X. and {Hennawi}, J.~F. and {Rahmani}, H. and {Ba{\~n}ados}, E. and {Farina}, E.~P. and {Cai}, Z. and {Man}, A.},
        title = "{Discovery of intergalactic bridges connecting two faint z {\ensuremath{\sim}} 3 quasars}",
      journal = {\aap},
         year = 2019,
        month = nov,
       volume = {631},
          eid = {A18},
        pages = {A18},
          doi = {10.1051/0004-6361/201936211},
archivePrefix = {arXiv},
       eprint = {1909.00829},
 primaryClass = {astro-ph.GA},
       adsurl = {https://ui.adsabs.harvard.edu/abs/2019A&A...631A..18A}
}

@ARTICLE{2021ApJ...923..252S,
       author = {{Sanderson}, Kelly N. and {Prescott}, Moire K.~M. and {Christensen}, Lise and {Fynbo}, Johan and {M{\o}ller}, Palle},
        title = "{Mapping the Morphology and Kinematics of a Ly{\ensuremath{\alpha}}-selected Nebula at z = 3.15 with MUSE}",
      journal = {\apj},
         year = 2021,
        month = dec,
       volume = {923},
       number = {2},
          eid = {252},
        pages = {252},
          doi = {10.3847/1538-4357/ac3077},
archivePrefix = {arXiv},
       eprint = {2110.10865},
 primaryClass = {astro-ph.GA},
       adsurl = {https://ui.adsabs.harvard.edu/abs/2021ApJ...923..252S}
}

@ARTICLE{2024MNRAS.527.5429L,
       author = {{Liu}, Zhuoqi and {Johnson}, Sean D. and {Li}, Jennifer I. -Hsiu and {Rudie}, Gwen C. and {Schaye}, Joop and {Chen}, Hsiao-Wen and {Brinchmann}, Jarle and {Cantalupo}, Sebastiano and {Chen}, Mandy C. and {Kollatschny}, Wolfram and {Maseda}, Michael V. and {Mishra}, Nishant and {Muzahid}, Sowgat},
        title = "{The first comprehensive study of a giant nebula around a radio-quiet quasar in the z < 1 Universe}",
      journal = {\mnras},
         year = 2024,
        month = jan,
       volume = {527},
       number = {3},
        pages = {5429-5447},
          doi = {10.1093/mnras/stad3364},
archivePrefix = {arXiv},
       eprint = {2309.00053},
 primaryClass = {astro-ph.GA},
       adsurl = {https://ui.adsabs.harvard.edu/abs/2024MNRAS.527.5429L}
}

@ARTICLE{2024ApJ...966..218J,
       author = {{Johnson}, Sean D. and {Liu}, Zhuoqi (Will) and {Li}, Jennifer I. -Hsiu and {Schaye}, Joop and {Greene}, Jenny E. and {Cantalupo}, Sebastiano and {Rudie}, Gwen C. and {Qu}, Zhijie and {Chen}, Hsiao-Wen and {Rafelski}, Marc and {Muzahid}, Sowgat and {Chen}, Mandy C. and {Contini}, Thierry and {Kollatschny}, Wolfram and {Mishra}, Nishant and {Petitjean}, Patrick and {Rauch}, Michael and {Zahedy}, Fakhri S.},
        title = "{Discovery of Optically Emitting Circumgalactic Nebulae around the Majority of UV-luminous Quasars at Intermediate Redshift}",
      journal = {\apj},
         year = 2024,
        month = may,
       volume = {966},
       number = {2},
          eid = {218},
        pages = {218},
          doi = {10.3847/1538-4357/ad3911},
archivePrefix = {arXiv},
       eprint = {2404.00088},
 primaryClass = {astro-ph.GA},
       adsurl = {https://ui.adsabs.harvard.edu/abs/2024ApJ...966..218J}
}

@software{2022zndo...6462441W,
       author = {{Wenzl}, Lukas},
        title = "{lukaswenzl/astrometry: Astrometry v1.5}",
         year = 2022,
        month = apr,
          eid = {10.5281/zenodo.6462441},
          doi = {10.5281/zenodo.6462441},
      version = {v1.5},
    publisher = {Zenodo},
       adsurl = {https://ui.adsabs.harvard.edu/abs/2022zndo...6462441W}
}

@ARTICLE{2012MNRAS.422.1835S,
       author = {{Serra}, Paolo and {Oosterloo}, Tom and {Morganti}, Raffaella and {Alatalo}, Katherine and {Blitz}, Leo and {Bois}, Maxime and {Bournaud}, Fr{\'e}d{\'e}ric and {Bureau}, Martin and {Cappellari}, Michele and {Crocker}, Alison F. and {Davies}, Roger L. and {Davis}, Timothy A. and {de Zeeuw}, P.~T. and {Duc}, Pierre-Alain and {Emsellem}, Eric and {Khochfar}, Sadegh and {Krajnovi{\'c}}, Davor and {Kuntschner}, Harald and {Lablanche}, Pierre-Yves and {McDermid}, Richard M. and {Naab}, Thorsten and {Sarzi}, Marc and {Scott}, Nicholas and {Trager}, Scott C. and {Weijmans}, Anne-Marie and {Young}, Lisa M.},
        title = "{The ATLAS$^{3D}$ project - XIII. Mass and morphology of H I in early-type galaxies as a function of environment}",
      journal = {\mnras},
         year = 2012,
        month = may,
       volume = {422},
       number = {3},
        pages = {1835-1862},
          doi = {10.1111/j.1365-2966.2012.20219.x},
archivePrefix = {arXiv},
       eprint = {1111.4241},
 primaryClass = {astro-ph.CO},
       adsurl = {https://ui.adsabs.harvard.edu/abs/2012MNRAS.422.1835S}
}

@ARTICLE{2011MNRAS.413..813C,
       author = {{Cappellari}, Michele and {Emsellem}, Eric and {Krajnovi{\'c}}, Davor and {McDermid}, Richard M. and {Scott}, Nicholas and {Verdoes Kleijn}, G.~A. and {Young}, Lisa M. and {Alatalo}, Katherine and {Bacon}, R. and {Blitz}, Leo and {Bois}, Maxime and {Bournaud}, Fr{\'e}d{\'e}ric and {Bureau}, M. and {Davies}, Roger L. and {Davis}, Timothy A. and {de Zeeuw}, P.~T. and {Duc}, Pierre-Alain and {Khochfar}, Sadegh and {Kuntschner}, Harald and {Lablanche}, Pierre-Yves and {Morganti}, Raffaella and {Naab}, Thorsten and {Oosterloo}, Tom and {Sarzi}, Marc and {Serra}, Paolo and {Weijmans}, Anne-Marie},
        title = "{The ATLAS$^{3D}$ project - I. A volume-limited sample of 260 nearby early-type galaxies: science goals and selection criteria}",
      journal = {\mnras},
         year = 2011,
        month = may,
       volume = {413},
       number = {2},
        pages = {813-836},
          doi = {10.1111/j.1365-2966.2010.18174.x},
archivePrefix = {arXiv},
       eprint = {1012.1551},
 primaryClass = {astro-ph.CO},
       adsurl = {https://ui.adsabs.harvard.edu/abs/2011MNRAS.413..813C}
}

@ARTICLE{2014MNRAS.444.3388S,
       author = {{Serra}, Paolo and {Oser}, Ludwig and {Krajnovi{\'c}}, Davor and {Naab}, Thorsten and {Oosterloo}, Tom and {Morganti}, Raffaella and {Cappellari}, Michele and {Emsellem}, Eric and {Young}, Lisa M. and {Blitz}, Leo and {Davis}, Timothy A. and {Duc}, Pierre-Alain and {Hirschmann}, Michaela and {Weijmans}, Anne-Marie and {Alatalo}, Katherine and {Bayet}, Estelle and {Bois}, Maxime and {Bournaud}, Fr{\'e}d{\'e}ric and {Bureau}, Martin and {Crocker}, Alison F. and {Davies}, Roger L. and {de Zeeuw}, P.~T. and {Khochfar}, Sadegh and {Kuntschner}, Harald and {Lablanche}, Pierre-Yves and {McDermid}, Richard M. and {Sarzi}, Marc and {Scott}, Nicholas},
        title = "{The ATLAS$^{3D}$ project - XXVI. H I discs in real and simulated fast and slow rotators}",
      journal = {\mnras},
         year = 2014,
        month = nov,
       volume = {444},
       number = {4},
        pages = {3388-3407},
          doi = {10.1093/mnras/stt2496},
archivePrefix = {arXiv},
       eprint = {1401.3180},
 primaryClass = {astro-ph.GA},
       adsurl = {https://ui.adsabs.harvard.edu/abs/2014MNRAS.444.3388S}
}

@ARTICLE{1997ApJ...477..765C,
       author = {{Chiappini}, C. and {Matteucci}, F. and {Gratton}, R.},
        title = "{The Chemical Evolution of the Galaxy: The Two-Infall Model}",
      journal = {\apj},
         year = 1997,
        month = mar,
       volume = {477},
       number = {2},
        pages = {765-780},
          doi = {10.1086/303726},
archivePrefix = {arXiv},
       eprint = {astro-ph/9609199},
 primaryClass = {astro-ph},
       adsurl = {https://ui.adsabs.harvard.edu/abs/1997ApJ...477..765C}
}

@ARTICLE{2024ApJ...962...98C,
       author = {{Chen}, Mandy C. and {Chen}, Hsiao-Wen and {Rauch}, Michael and {Qu}, Zhijie and {Johnson}, Sean D. and {Schaye}, Joop and {Rudie}, Gwen C. and {Li}, Jennifer I. -Hsiu and {Liu}, Zhuoqi (Will) and {Zahedy}, Fakhri S. and {Cantalupo}, Sebastiano and {Boettcher}, Erin},
        title = "{An Ensemble Study of Turbulence in Extended QSO Nebulae at z {\ensuremath{\approx}} 0.5{\textendash}1}",
      journal = {\apj},
         year = 2024,
        month = feb,
       volume = {962},
       number = {1},
          eid = {98},
        pages = {98},
          doi = {10.3847/1538-4357/ad1406},
archivePrefix = {arXiv},
       eprint = {2310.18406},
 primaryClass = {astro-ph.GA},
       adsurl = {https://ui.adsabs.harvard.edu/abs/2024ApJ...962...98C}
}

@ARTICLE{2016A&A...587A..98W,
       author = {{Wisotzki}, L. and {Bacon}, R. and {Blaizot}, J. and {Brinchmann}, J. and {Herenz}, E.~C. and {Schaye}, J. and {Bouch{\'e}}, N. and {Cantalupo}, S. and {Contini}, T. and {Carollo}, C.~M. and {Caruana}, J. and {Courbot}, J. -B. and {Emsellem}, E. and {Kamann}, S. and {Kerutt}, J. and {Leclercq}, F. and {Lilly}, S.~J. and {Patr{\'\i}cio}, V. and {Sandin}, C. and {Steinmetz}, M. and {Straka}, L.~A. and {Urrutia}, T. and {Verhamme}, A. and {Weilbacher}, P.~M. and {Wendt}, M.},
        title = "{Extended Lyman {\ensuremath{\alpha}} haloes around individual high-redshift galaxies revealed by MUSE}",
      journal = {\aap},
         year = 2016,
        month = mar,
       volume = {587},
          eid = {A98},
        pages = {A98},
          doi = {10.1051/0004-6361/201527384},
archivePrefix = {arXiv},
       eprint = {1509.05143},
 primaryClass = {astro-ph.GA},
       adsurl = {https://ui.adsabs.harvard.edu/abs/2016A&A...587A..98W}
}

@ARTICLE{2023Natur.624...53G,
       author = {{Guo}, Yucheng and {Bacon}, Roland and {Bouch{\'e}}, Nicolas F. and {Wisotzki}, Lutz and {Schaye}, Joop and {Blaizot}, J{\'e}r{\'e}my and {Verhamme}, Anne and {Cantalupo}, Sebastiano and {Boogaard}, Leindert A. and {Brinchmann}, Jarle and {Cherrey}, Maxime and {Kusakabe}, Haruka and {Langan}, Ivanna and {Leclercq}, Floriane and {Matthee}, Jorryt and {Michel-Dansac}, L{\'e}o and {Schroetter}, Ilane and {Wendt}, Martin},
        title = "{Bipolar outflows out to 10 kpc for massive galaxies at redshift z {\ensuremath{\approx}} 1}",
      journal = {\nat},
         year = 2023,
        month = dec,
       volume = {624},
       number = {7990},
        pages = {53-56},
          doi = {10.1038/s41586-023-06718-w},
archivePrefix = {arXiv},
       eprint = {2312.05167},
 primaryClass = {astro-ph.GA},
       adsurl = {https://ui.adsabs.harvard.edu/abs/2023Natur.624...53G}
}

@ARTICLE{2024A&A...688A..37G,
       author = {{Guo}, Yucheng and {Bacon}, Roland and {Wisotzki}, Lutz and {Garel}, Thibault and {Blaizot}, J{\'e}r{\'e}my and {Schaye}, Joop and {Richard}, Johan and {Herrero Alonso}, Yohana and {Leclercq}, Floriane and {Boogaard}, Leindert and {Kusakabe}, Haruka and {Pharo}, John and {Vitte}, Elo{\"\i}se},
        title = "{Median surface-brightness profiles of Lyman-{\ensuremath{\alpha}} haloes in the MUSE Extremely Deep Field}",
      journal = {\aap},
         year = 2024,
        month = aug,
       volume = {688},
          eid = {A37},
        pages = {A37},
          doi = {10.1051/0004-6361/202347658},
archivePrefix = {arXiv},
       eprint = {2309.05513},
 primaryClass = {astro-ph.GA},
       adsurl = {https://ui.adsabs.harvard.edu/abs/2024A&A...688A..37G}
}

@ARTICLE{2020MNRAS.491.2057L,
       author = {{Lofthouse}, Emma K. and {Fumagalli}, Michele and {Fossati}, Matteo and {O'Meara}, John M. and {Murphy}, Michael T. and {Christensen}, Lise and {Prochaska}, J. Xavier and {Cantalupo}, Sebastiano and {Bielby}, Richard M. and {Cooke}, Ryan J. and {Lusso}, Elisabeta and {Morris}, Simon L.},
        title = "{MUSE Analysis of Gas around Galaxies (MAGG) - I: Survey design and the environment of a near pristine gas cloud at z {\ensuremath{\approx}} 3.5}",
      journal = {\mnras},
         year = 2020,
        month = jan,
       volume = {491},
       number = {2},
        pages = {2057-2074},
          doi = {10.1093/mnras/stz3066},
archivePrefix = {arXiv},
       eprint = {1910.13458},
 primaryClass = {astro-ph.GA},
       adsurl = {https://ui.adsabs.harvard.edu/abs/2020MNRAS.491.2057L}
}

@ARTICLE{2024ApJ...965..143L,
       author = {{Li}, Jennifer I. -Hsiu and {Johnson}, Sean D. and {Boettcher}, Erin and {Cantalupo}, Sebastiano and {Chen}, Hsiao-Wen and {Chen}, Mandy C. and {DePalma}, David R. and {Liu}, Zhuoqi (Will) and {Mishra}, Nishant and {Petitjean}, Patrick and {Qu}, Zhijie and {Rudie}, Gwen C. and {Schaye}, Joop and {Zahedy}, Fakhri S.},
        title = "{The Cosmic Ultraviolet Baryon Survey (CUBS). VIII. Group Environment of the Most Luminous Quasars at z {\ensuremath{\approx}} 1}",
      journal = {\apj},
         year = 2024,
        month = apr,
       volume = {965},
       number = {2},
          eid = {143},
        pages = {143},
          doi = {10.3847/1538-4357/ad2fad},
archivePrefix = {arXiv},
       eprint = {2403.03983},
 primaryClass = {astro-ph.GA},
       adsurl = {https://ui.adsabs.harvard.edu/abs/2024ApJ...965..143L}
}

@ARTICLE{1999ApJS..124..285R,
       author = {{Reid}, R.~I. and {Kronberg}, P.~P. and {Perley}, R.~A.},
        title = "{VLA Images at 5 GHZ of 212 Southern Extragalactic Objects}",
      journal = {\apjs},
         year = 1999,
        month = oct,
       volume = {124},
       number = {2},
        pages = {285-381},
          doi = {10.1086/313259},
       adsurl = {https://ui.adsabs.harvard.edu/abs/1999ApJS..124..285R}
}

@ARTICLE{1993ARA&A..31..473A,
       author = {{Antonucci}, Robert},
        title = "{Unified models for active galactic nuclei and quasars.}",
      journal = {\araa},
         year = 1993,
        month = jan,
       volume = {31},
        pages = {473-521},
          doi = {10.1146/annurev.aa.31.090193.002353},
       adsurl = {https://ui.adsabs.harvard.edu/abs/1993ARA&A..31..473A}
}

@ARTICLE{2017ARA&A..55..389T,
       author = {{Tumlinson}, Jason and {Peeples}, Molly S. and {Werk}, Jessica K.},
        title = "{The Circumgalactic Medium}",
      journal = {\araa},
         year = 2017,
        month = aug,
       volume = {55},
       number = {1},
        pages = {389-432},
          doi = {10.1146/annurev-astro-091916-055240},
archivePrefix = {arXiv},
       eprint = {1709.09180},
 primaryClass = {astro-ph.GA},
       adsurl = {https://ui.adsabs.harvard.edu/abs/2017ARA&A..55..389T}
}

@ARTICLE{2013MNRAS.432.1709C,
       author = {{Cappellari}, Michele and {Scott}, Nicholas and {Alatalo}, Katherine and {Blitz}, Leo and {Bois}, Maxime and {Bournaud}, Fr{\'e}d{\'e}ric and {Bureau}, M. and {Crocker}, Alison F. and {Davies}, Roger L. and {Davis}, Timothy A. and {de Zeeuw}, P.~T. and {Duc}, Pierre-Alain and {Emsellem}, Eric and {Khochfar}, Sadegh and {Krajnovi{\'c}}, Davor and {Kuntschner}, Harald and {McDermid}, Richard M. and {Morganti}, Raffaella and {Naab}, Thorsten and {Oosterloo}, Tom and {Sarzi}, Marc and {Serra}, Paolo and {Weijmans}, Anne-Marie and {Young}, Lisa M.},
        title = "{The ATLAS$^{3D}$ project - XV. Benchmark for early-type galaxies scaling relations from 260 dynamical models: mass-to-light ratio, dark matter, Fundamental Plane and Mass Plane}",
      journal = {\mnras},
         year = 2013,
        month = jul,
       volume = {432},
       number = {3},
        pages = {1709-1741},
          doi = {10.1093/mnras/stt562},
archivePrefix = {arXiv},
       eprint = {1208.3522},
 primaryClass = {astro-ph.CO},
       adsurl = {https://ui.adsabs.harvard.edu/abs/2013MNRAS.432.1709C}
}

@ARTICLE{2020PASP..132c5001L,
       author = {{Lacy}, M. and {Baum}, S.~A. and {Chandler}, C.~J. and {Chatterjee}, S. and {Clarke}, T.~E. and {Deustua}, S. and {English}, J. and {Farnes}, J. and {Gaensler}, B.~M. and {Gugliucci}, N. and {Hallinan}, G. and {Kent}, B.~R. and {Kimball}, A. and {Law}, C.~J. and {Lazio}, T.~J.~W. and {Marvil}, J. and {Mao}, S.~A. and {Medlin}, D. and {Mooley}, K. and {Murphy}, E.~J. and {Myers}, S. and {Osten}, R. and {Richards}, G.~T. and {Rosolowsky}, E. and {Rudnick}, L. and {Schinzel}, F. and {Sivakoff}, G.~R. and {Sjouwerman}, L.~O. and {Taylor}, R. and {White}, R.~L. and {Wrobel}, J. and {Andernach}, H. and {Beasley}, A.~J. and {Berger}, E. and {Bhatnager}, S. and {Birkinshaw}, M. and {Bower}, G.~C. and {Brandt}, W.~N. and {Brown}, S. and {Burke-Spolaor}, S. and {Butler}, B.~J. and {Comerford}, J. and {Demorest}, P.~B. and {Fu}, H. and {Giacintucci}, S. and {Golap}, K. and {G{\"u}th}, T. and {Hales}, C.~A. and {Hiriart}, R. and {Hodge}, J. and {Horesh}, A. and {Ivezi{\'c}}, {\v{Z}}. and {Jarvis}, M.~J. and {Kamble}, A. and {Kassim}, N. and {Liu}, X. and {Loinard}, L. and {Lyons}, D.~K. and {Masters}, J. and {Mezcua}, M. and {Moellenbrock}, G.~A. and {Mroczkowski}, T. and {Nyland}, K. and {O'Dea}, C.~P. and {O'Sullivan}, S.~P. and {Peters}, W.~M. and {Radford}, K. and {Rao}, U. and {Robnett}, J. and {Salcido}, J. and {Shen}, Y. and {Sobotka}, A. and {Witz}, S. and {Vaccari}, M. and {van Weeren}, R.~J. and {Vargas}, A. and {Williams}, P.~K.~G. and {Yoon}, I.},
        title = "{The Karl G. Jansky Very Large Array Sky Survey (VLASS). Science Case and Survey Design}",
      journal = {\pasp},
         year = 2020,
        month = mar,
       volume = {132},
       number = {1009},
          eid = {035001},
        pages = {035001},
          doi = {10.1088/1538-3873/ab63eb},
archivePrefix = {arXiv},
       eprint = {1907.01981},
 primaryClass = {astro-ph.IM},
       adsurl = {https://ui.adsabs.harvard.edu/abs/2020PASP..132c5001L}
}

@ARTICLE{1983ApJ...266..713O,
       author = {{Oke}, J.~B. and {Gunn}, J.~E.},
        title = "{Secondary standard stars for absolute spectrophotometry.}",
      journal = {\apj},
         year = 1983,
        month = mar,
       volume = {266},
        pages = {713-717},
          doi = {10.1086/160817},
       adsurl = {https://ui.adsabs.harvard.edu/abs/1983ApJ...266..713O}
}

@ARTICLE{2014MNRAS.438.1435C,
       author = {{Chen}, Hsiao-Wen and {Gauthier}, Jean-Ren{\'e} and {Sharon}, Keren and {Johnson}, Sean D. and {Nair}, Preethi and {Liang}, Cameron J.},
        title = "{Spatially resolved velocity maps of halo gas around two intermediate-redshift galaxies}",
      journal = {\mnras},
         year = 2014,
        month = feb,
       volume = {438},
       number = {2},
        pages = {1435-1450},
          doi = {10.1093/mnras/stt2288},
archivePrefix = {arXiv},
       eprint = {1312.0016},
 primaryClass = {astro-ph.GA},
       adsurl = {https://ui.adsabs.harvard.edu/abs/2014MNRAS.438.1435C}
}

@ARTICLE{2018Natur.554..493L,
       author = {{Lopez}, Sebastian and {Tejos}, Nicolas and {Ledoux}, C{\'e}dric and {Barrientos}, L. Felipe and {Sharon}, Keren and {Rigby}, Jane R. and {Gladders}, Michael D. and {Bayliss}, Matthew B. and {Pessa}, Ismael},
        title = "{A clumpy and anisotropic galaxy halo at redshift 1 from gravitational-arc tomography}",
      journal = {\nat},
         year = 2018,
        month = feb,
       volume = {554},
       number = {7693},
        pages = {493-496},
          doi = {10.1038/nature25436},
archivePrefix = {arXiv},
       eprint = {1801.10175},
 primaryClass = {astro-ph.GA},
       adsurl = {https://ui.adsabs.harvard.edu/abs/2018Natur.554..493L}
}

@FINPROCEEDINGS{2016SPIE.9909E..2SK,
       author = {{Kolb}, Johann and {Madec}, Pierre-Yves and {Arsenault}, Robin and {Oberti}, Sylvain and {Paufique}, J{\'e}r{\^o}me and {La Penna}, Paolo and {Str{\"o}bele}, Stefan and {Donaldson}, Robert and {Soenke}, Christian and {Su{\'a}rez Valles}, Marcos and {Kiekebusch}, Mario and {Argomedo}, Javier and {Le Louarn}, Miska and {Vernet}, Elise and {Haguenauer}, Pierre and {Duhoux}, Philippe and {Aller-Carpentier}, Emmanuel and {Valenzuela}, Jose Javier and {Guerra}, Juan Carlos},
        title = "{Laboratory results of the AOF system testing}",
    booktitle = {Adaptive Optics Systems V},
         year = 2016,
       editor = {{Marchetti}, Enrico and {Close}, Laird M. and {V{\'e}ran}, Jean-Pierre},
       series = {Society of Photo-Optical Instrumentation Engineers (SPIE) Conference Series},
       volume = {9909},
        month = jul,
          eid = {99092S},
        pages = {99092S},
          doi = {10.1117/12.2232788},
       adsurl = {https://ui.adsabs.harvard.edu/abs/2016SPIE.9909E..2SK}
}

@INPROCEEDINGS{2018SPIE10703E..02M,
       author = {{Madec}, P. -Y. and {Arsenault}, R. and {Kuntschner}, H. and {Kolb}, J. and {Pirard}, J. -F. and {Paufique}, J. and {La Penna}, P. and {Hackenberg}, W. and {Vernet}, E. and {Su{\'a}rez Valles}, M. and {Hubin}, N.},
        title = "{Adaptive Optics Facility: from an amazing present to a brilliant future...}",
    booktitle = {Adaptive Optics Systems VI},
         year = 2018,
       editor = {{Close}, Laird M. and {Schreiber}, Laura and {Schmidt}, Dirk},
       series = {Society of Photo-Optical Instrumentation Engineers (SPIE) Conference Series},
       volume = {10703},
        month = jul,
          eid = {1070302},
        pages = {1070302},
          doi = {10.1117/12.2312428},
       adsurl = {https://ui.adsabs.harvard.edu/abs/2018SPIE10703E..02M}
}

@ARTICLE{2007MNRAS.377L..74L,
       author = {{Liddle}, Andrew R.},
        title = "{Information criteria for astrophysical model selection}",
      journal = {\mnras},
         year = 2007,
        month = may,
       volume = {377},
       number = {1},
        pages = {L74-L78},
          doi = {10.1111/j.1745-3933.2007.00306.x},
archivePrefix = {arXiv},
       eprint = {astro-ph/0701113},
 primaryClass = {astro-ph},
       adsurl = {https://ui.adsabs.harvard.edu/abs/2007MNRAS.377L..74L}
}

@ARTICLE{2025ApJ...984..140L,
       author = {{Liu}, Zhuoqi (Will) and {Johnson}, Sean D. and {Li}, Jennifer I. -Hsiu and {Epinat}, Beno{\^\i}t and {Rudie}, Gwen C. and {Monreal-Ibero}, Ana and {Cantalupo}, Sebastiano and {Qu}, Zhijie and {Chen}, Mandy C. and {Kollatschny}, Wolfram and {Muzahid}, Sowgat and {Zahedy}, Fakhri S. and {Kesler}, Elise and {Mishra}, Nishant},
        title = "{The Morphology and Kinematics of a Giant, Symmetric Nebula around a Radio-loud Quasar 3C 57: Extended Rotating Gas or Biconical Outflows?}",
      journal = {\apj},
         year = 2025,
        month = may,
       volume = {984},
       number = {2},
          eid = {140},
        pages = {140},
          doi = {10.3847/1538-4357/adc1bf},
archivePrefix = {arXiv},
       eprint = {2503.12597},
 primaryClass = {astro-ph.GA},
       adsurl = {https://ui.adsabs.harvard.edu/abs/2025ApJ...984..140L}
}

@ARTICLE{2024ApJ...974..273M,
       author = {{Mintz}, Abby and {Greene}, Jenny E. and {Kado-Fong}, Erin and {Danieli}, Shany and {Li}, Jiaxuan and {Luo}, Yifei and {Leauthaud}, Alexie and {Baldassare}, Vivienne and {Huang}, Song and {Peter}, Annika H.~G. and {Bhattacharyya}, Joy and {Li}, Mingyu and {Pan}, Yue},
        title = "{A Nonparametric Morphological Analysis of H{\ensuremath{\alpha}} Emission in Bright Dwarfs Using the Merian Survey}",
      journal = {\apj},
         year = 2024,
        month = oct,
       volume = {974},
       number = {2},
          eid = {273},
        pages = {273},
          doi = {10.3847/1538-4357/ad6861},
archivePrefix = {arXiv},
       eprint = {2410.01886},
 primaryClass = {astro-ph.GA},
       adsurl = {https://ui.adsabs.harvard.edu/abs/2024ApJ...974..273M}
}

@ARTICLE{2019MNRAS.483.4140R,
       author = {{Rodriguez-Gomez}, Vicente and {Snyder}, Gregory F. and {Lotz}, Jennifer M. and {Nelson}, Dylan and {Pillepich}, Annalisa and {Springel}, Volker and {Genel}, Shy and {Weinberger}, Rainer and {Tacchella}, Sandro and {Pakmor}, R{\"u}diger and {Torrey}, Paul and {Marinacci}, Federico and {Vogelsberger}, Mark and {Hernquist}, Lars and {Thilker}, David A.},
        title = "{The optical morphologies of galaxies in the IllustrisTNG simulation: a comparison to Pan-STARRS observations}",
      journal = {\mnras},
         year = 2019,
        month = mar,
       volume = {483},
       number = {3},
        pages = {4140-4159},
          doi = {10.1093/mnras/sty3345},
archivePrefix = {arXiv},
       eprint = {1809.08239},
 primaryClass = {astro-ph.GA},
       adsurl = {https://ui.adsabs.harvard.edu/abs/2019MNRAS.483.4140R}
}

@ARTICLE{2025arXiv250716898G,
       author = {{Gonz{\'a}lez Lobos}, Jay and {Arrigoni Battaia}, Fabrizio and {Obreja}, Aura and {Kauffmann}, Guinevere and {Farina}, Emanuele Paolo and {Costa}, Tiago},
        title = "{QSO MUSEUM III: the circumgalactic medium in Ly$α$ emission around 120 $z\sim3$ quasars covering the SDSS parameter space. Witnessing the instantaneous AGN feedback on halo scales}",
      journal = {arXiv e-prints},
         year = 2025,
        month = jul,
          eid = {arXiv:2507.16898},
        pages = {arXiv:2507.16898},
          doi = {10.48550/arXiv.2507.16898},
archivePrefix = {arXiv},
       eprint = {2507.16898},
 primaryClass = {astro-ph.GA},
       adsurl = {https://ui.adsabs.harvard.edu/abs/2025arXiv250716898G}
}

@ARTICLE{2019MNRAS.482.3162A,
       author = {{Arrigoni Battaia}, Fabrizio and {Hennawi}, Joseph F. and {Prochaska}, J. Xavier and {O{\~n}orbe}, Jose and {Farina}, Emanuele P. and {Cantalupo}, Sebastiano and {Lusso}, Elisabeta},
        title = "{QSO MUSEUM I: a sample of 61 extended Ly {\ensuremath{\alpha}}-emission nebulae surrounding z {\ensuremath{\sim}} 3 quasars}",
      journal = {\mnras},
         year = 2019,
        month = jan,
       volume = {482},
       number = {3},
        pages = {3162-3205},
          doi = {10.1093/mnras/sty2827},
archivePrefix = {arXiv},
       eprint = {1808.10857},
 primaryClass = {astro-ph.GA},
       adsurl = {https://ui.adsabs.harvard.edu/abs/2019MNRAS.482.3162A}
}

@ARTICLE{2024A&A...691A.210H,
       author = {{Herwig}, Eileen and {Arrigoni Battaia}, Fabrizio and {Gonz{\'a}lez Lobos}, Jay and {Farina}, Emanuele P. and {Man}, Allison W.~S. and {Ba{\~n}ados}, Eduardo and {Kauffmann}, Guinevere and {Cai}, Zheng and {Obreja}, Aura and {Prochaska}, J. Xavier},
        title = "{QSO MUSEUM: II. Search for extended Ly{\ensuremath{\alpha}} emission around eight z {\ensuremath{\sim}} 3 quasar pairs}",
      journal = {\aap},
         year = 2024,
        month = nov,
       volume = {691},
          eid = {A210},
        pages = {A210},
          doi = {10.1051/0004-6361/202450959},
archivePrefix = {arXiv},
       eprint = {2408.16826},
 primaryClass = {astro-ph.GA},
       adsurl = {https://ui.adsabs.harvard.edu/abs/2024A&A...691A.210H}
}

@ARTICLE{2020PASA...37...48M,
       author = {{McConnell}, D. and {Hale}, C.~L. and {Lenc}, E. and {Banfield}, J.~K. and {Heald}, George and {Hotan}, A.~W. and {Leung}, James K. and {Moss}, Vanessa A. and {Murphy}, Tara and {O'Brien}, Andrew and {Pritchard}, Joshua and {Raja}, Wasim and {Sadler}, Elaine M. and {Stewart}, Adam and {Thomson}, Alec J.~M. and {Whiting}, M. and {Allison}, James R. and {Amy}, S.~W. and {Anderson}, C. and {Ball}, Lewis and {Bannister}, Keith W. and {Bell}, Martin and {Bock}, Douglas C. -J. and {Bolton}, Russ and {Bunton}, J.~D. and {Chippendale}, A.~P. and {Collier}, J.~D. and {Cooray}, F.~R. and {Cornwell}, T.~J. and {Diamond}, P.~J. and {Edwards}, P.~G. and {Gupta}, N. and {Hayman}, Douglas B. and {Heywood}, Ian and {Jackson}, C.~A. and {Koribalski}, B{\"a}rbel S. and {Lee-Waddell}, Karen and {McClure-Griffiths}, N.~M. and {Ng}, Alan and {Norris}, Ray P. and {Phillips}, Chris and {Reynolds}, John E. and {Roxby}, Daniel N. and {Schinckel}, Antony E.~T. and {Shields}, Matt and {Tremblay}, Chenoa and {Tzioumis}, A. and {Voronkov}, M.~A. and {Westmeier}, Tobias},
        title = "{The Rapid ASKAP Continuum Survey I: Design and first results}",
      journal = {\pasa},
         year = 2020,
        month = nov,
       volume = {37},
          eid = {e048},
        pages = {e048},
          doi = {10.1017/pasa.2020.41},
archivePrefix = {arXiv},
       eprint = {2012.00747},
 primaryClass = {astro-ph.IM},
       adsurl = {https://ui.adsabs.harvard.edu/abs/2020PASA...37...48M}
}

@ARTICLE{1984ApJ...281..535B,
       author = {{Boroson}, T.~A. and {Oke}, J.~B.},
        title = "{Spectroscopy of the nebulosity around eight high-luminosity QSOs.}",
      journal = {\apj},
         year = 1984,
        month = jun,
       volume = {281},
        pages = {535-544},
          doi = {10.1086/162126},
       adsurl = {https://ui.adsabs.harvard.edu/abs/1984ApJ...281..535B}
}

@ARTICLE{1987ApJ...316..584S,
       author = {{Stockton}, Alan and {MacKenty}, John W.},
        title = "{Extended Emission-Line Regions around QSOs}",
      journal = {\apj},
         year = 1987,
        month = may,
       volume = {316},
        pages = {584},
          doi = {10.1086/165227},
       adsurl = {https://ui.adsabs.harvard.edu/abs/1987ApJ...316..584S}
}

@ARTICLE{2002astro.ph.10394H,
       author = {{Hogg}, David W. and {Baldry}, Ivan K. and {Blanton}, Michael R. and {Eisenstein}, Daniel J.},
        title = "{The K correction}",
      journal = {arXiv e-prints},
         year = 2002,
        month = oct,
          eid = {astro-ph/0210394},
        pages = {astro-ph/0210394},
          doi = {10.48550/arXiv.astro-ph/0210394},
archivePrefix = {arXiv},
       eprint = {astro-ph/0210394},
 primaryClass = {astro-ph},
       adsurl = {https://ui.adsabs.harvard.edu/abs/2002astro.ph.10394H}
}

@ARTICLE{2025ApJ...992..155B,
       author = {{Beckett}, Alexander and {Rafelski}, Marc and {Scarlata}, Claudia and {Hu}, Wanjia and {Kim}, Keunho and {Goovaerts}, Ilias and {Malkan}, Matthew A. and {Webb}, Wayne and {Teplitz}, Harry and {Hayes}, Matthew and {Mehta}, Vihang and {Alavi}, Anahita and {Bunker}, Andrew J. and {Citro}, Annalisa and {Hathi}, Nimish and {Henry}, Alaina and {Le Reste}, Alexandra and {Moretti}, Alessia and {Rutkowski}, Michael J. and {Trebitsch}, Maxime and {Zanella}, Anita},
        title = "{The Parallel Ionizing Emissivity Survey (PIE). I. Survey Design and Selection of Candidate Lyman Continuum Leakers at 3.1 < z < 3.5}",
      journal = {\apj},
         year = 2025,
        month = oct,
       volume = {992},
       number = {1},
          eid = {155},
        pages = {155},
          doi = {10.3847/1538-4357/ae0291},
archivePrefix = {arXiv},
       eprint = {2503.20878},
 primaryClass = {astro-ph.GA},
       adsurl = {https://ui.adsabs.harvard.edu/abs/2025ApJ...992..155B}
}

@ARTICLE{2022ApJ...924...14P,
       author = {{Prichard}, Laura J. and {Rafelski}, Marc and {Cooke}, Jeff and {Me{\v{s}}tri{\'c}}, Uros and {Bassett}, Robert and {Ryan-Weber}, Emma V. and {Sunnquist}, Ben and {Alavi}, Anahita and {Hathi}, Nimish and {Wang}, Xin and {Revalski}, Mitchell and {Bajaj}, Varun and {O'Meara}, John M. and {Spitler}, Lee},
        title = "{Lyman Continuum Galaxy Candidates in COSMOS}",
      journal = {\apj},
         year = 2022,
        month = jan,
       volume = {924},
       number = {1},
          eid = {14},
        pages = {14},
          doi = {10.3847/1538-4357/ac3004},
archivePrefix = {arXiv},
       eprint = {2110.06945},
 primaryClass = {astro-ph.GA},
       adsurl = {https://ui.adsabs.harvard.edu/abs/2022ApJ...924...14P}
}

@INCOLLECTION{2021drzp.book....2H,
       author = {{Hoffman}, S.~L. and {Mack}, J. and {Avila}, R.~J. and {Martlin}, C. and {Bajaj}, V. and {Cohen}, Y.},
        title = "{The DrizzlePac Handbook, v. 2}",
    booktitle = {The DrizzlePac Handbook},
         year = 2021,
        pages = {2},
       adsurl = {https://ui.adsabs.harvard.edu/abs/2021drzp.book....2H}
}

@BOOK{2012drzp.book.....G,
       author = {{Gonzaga}, S. and {Hack}, W. and {Fruchter}, A. and {Mack}, J.},
        title = "{The DrizzlePac Handbook}",
         year = 2012,
       adsurl = {https://ui.adsabs.harvard.edu/abs/2012drzp.book.....G}
}

@ARTICLE{2016MNRAS.458.2423Z,
       author = {{Zahedy}, Fakhri S. and {Chen}, Hsiao-Wen and {Rauch}, Michael and {Wilson}, Michelle L. and {Zabludoff}, Ann},
        title = "{Probing the cool interstellar and circumgalactic gas of three massive lensing galaxies at z = 0.4-0.7}",
      journal = {\mnras},
         year = 2016,
        month = may,
       volume = {458},
       number = {3},
        pages = {2423-2442},
          doi = {10.1093/mnras/stw484},
archivePrefix = {arXiv},
       eprint = {1510.04307},
 primaryClass = {astro-ph.GA},
       adsurl = {https://ui.adsabs.harvard.edu/abs/2016MNRAS.458.2423Z}
}

@ARTICLE{2024ApJ...971..134Z,
       author = {{Zhao}, Qinyuan and {Wang}, Junfeng and {Li}, Zhenzhen},
        title = "{Serendipitous Catch of a Giant Jellyfish: An Ionized Nebula around 3C 275.1 with 170 kpc Long Tails}",
      journal = {\apj},
         year = 2024,
        month = aug,
       volume = {971},
       number = {2},
          eid = {134},
        pages = {134},
          doi = {10.3847/1538-4357/ad58d6},
archivePrefix = {arXiv},
       eprint = {2406.11433},
 primaryClass = {astro-ph.GA},
       adsurl = {https://ui.adsabs.harvard.edu/abs/2024ApJ...971..134Z}
}

@ARTICLE{2017ApJ...850...40R,
       author = {{Rupke}, David S.~N. and {G{\"u}ltekin}, Kayhan and {Veilleux}, Sylvain},
        title = "{Quasar-mode Feedback in Nearby Type 1 Quasars: Ubiquitous Kiloparsec-scale Outflows and Correlations with Black Hole Properties}",
      journal = {\apj},
         year = 2017,
        month = nov,
       volume = {850},
       number = {1},
          eid = {40},
        pages = {40},
          doi = {10.3847/1538-4357/aa94d1},
archivePrefix = {arXiv},
       eprint = {1708.05139},
 primaryClass = {astro-ph.GA},
       adsurl = {https://ui.adsabs.harvard.edu/abs/2017ApJ...850...40R}
}

@ARTICLE{2025ApJ...993L..18D,
       author = {{DePalma}, David and {Gupta}, Neeraj and {Chen}, Hsiao-Wen and {Simcoe}, Robert A. and {Balashev}, Sergei and {Boettcher}, Erin and {Cantalupo}, Sebastiano and {Chen}, Mandy C. and {Combes}, Fran{\c{c}}oise and {Faucher-Gigu{\`e}re}, Claude-Andr{\'e} and {Johnson}, Sean D. and {Kl{\"o}ckner}, Hans-Rainer and {Krogager}, Jens-Kristian and {Li}, Jennifer I.-Hsiu and {L{\'o}pez}, Sebasti{\'a}n and {Noterdaeme}, Pasquier and {Petitjean}, Patrick and {Qu}, Zhijie and {Rudie}, Gwen C. and {Schaye}, Joop and {Zahedy}, Fakhri},
        title = "{H I Properties of Field Galaxies at z ≍ 0.2─0.6: Insights into Declining Cosmic Star Formation}",
      journal = {\apjl},
         year = 2025,
        month = nov,
       volume = {993},
       number = {1},
          eid = {L18},
        pages = {L18},
          doi = {10.3847/2041-8213/ae0d8b},
archivePrefix = {arXiv},
       eprint = {2510.03400},
 primaryClass = {astro-ph.GA},
       adsurl = {https://ui.adsabs.harvard.edu/abs/2025ApJ...993L..18D}
}

@ARTICLE{2014ApJ...781L..40E,
       author = {{Ebeling}, H. and {Stephenson}, L.~N. and {Edge}, A.~C.},
        title = "{Jellyfish: Evidence of Extreme Ram-pressure Stripping in Massive Galaxy Clusters}",
      journal = {\apjl},
         year = 2014,
        month = feb,
       volume = {781},
       number = {2},
          eid = {L40},
        pages = {L40},
          doi = {10.1088/2041-8205/781/2/L40},
archivePrefix = {arXiv},
       eprint = {1312.6135},
 primaryClass = {astro-ph.GA},
       adsurl = {https://ui.adsabs.harvard.edu/abs/2014ApJ...781L..40E}
}

@article{bhattacharyya1946measure,
  author = {Bhattacharyya, A.},
  journal = {Sankhyā: The Indian Journal of Statistics },
  month = {Juli},
  number = 4,
  pages = {401-406},
  title = {On a Measure of Divergence between Two Multinomial Populations},
  volume = 7,
  year = 1946
}

@ARTICLE{2007ARA&A..45..117M,
       author = {{McNamara}, B.~R. and {Nulsen}, P.~E.~J.},
        title = "{Heating Hot Atmospheres with Active Galactic Nuclei}",
      journal = {\araa},
         year = 2007,
        month = sep,
       volume = {45},
       number = {1},
        pages = {117-175},
          doi = {10.1146/annurev.astro.45.051806.110625},
archivePrefix = {arXiv},
       eprint = {0709.2152},
 primaryClass = {astro-ph},
       adsurl = {https://ui.adsabs.harvard.edu/abs/2007ARA&A..45..117M}
}

@ARTICLE{2015ARA&A..53..115K,
       author = {{King}, Andrew and {Pounds}, Ken},
        title = "{Powerful Outflows and Feedback from Active Galactic Nuclei}",
      journal = {\araa},
         year = 2015,
        month = aug,
       volume = {53},
        pages = {115-154},
          doi = {10.1146/annurev-astro-082214-122316},
archivePrefix = {arXiv},
       eprint = {1503.05206},
 primaryClass = {astro-ph.GA},
       adsurl = {https://ui.adsabs.harvard.edu/abs/2015ARA&A..53..115K}
}

@ARTICLE{2009ApJ...696.1693F,
       author = {{Fu}, Hai and {Stockton}, Alan},
        title = "{FR II Quasars: Infrared Properties, Star Formation Rates, and Extended Ionized Gas}",
      journal = {\apj},
         year = 2009,
        month = may,
       volume = {696},
       number = {2},
        pages = {1693-1699},
          doi = {10.1088/0004-637X/696/2/1693},
archivePrefix = {arXiv},
       eprint = {0902.1714},
 primaryClass = {astro-ph.CO},
       adsurl = {https://ui.adsabs.harvard.edu/abs/2009ApJ...696.1693F}
}

@ARTICLE{2017ApJ...841...93Y,
       author = {{Yuma}, Suraphong and {Ouchi}, Masami and {Drake}, Alyssa B. and {Fujimoto}, Seiji and {Kojima}, Takashi and {Sugahara}, Yuma},
        title = "{Systematic Survey for [O II], [O III], and H{\ensuremath{\alpha}} Blobs at z = 0.1-1.5: The Implication for Evolution of Galactic-scale Outflow}",
      journal = {\apj},
         year = 2017,
        month = jun,
       volume = {841},
       number = {2},
          eid = {93},
        pages = {93},
          doi = {10.3847/1538-4357/aa709f},
archivePrefix = {arXiv},
       eprint = {1702.05107},
 primaryClass = {astro-ph.GA},
       adsurl = {https://ui.adsabs.harvard.edu/abs/2017ApJ...841...93Y}
}

@ARTICLE{2018AstL...44....8K,
       author = {{Kravtsov}, A.~V. and {Vikhlinin}, A.~A. and {Meshcheryakov}, A.~V.},
        title = "{Stellar Mass{\textemdash}Halo Mass Relation and Star Formation Efficiency in High-Mass Halos}",
      journal = {Astronomy Letters},
         year = 2018,
        month = jan,
       volume = {44},
       number = {1},
        pages = {8-34},
          doi = {10.1134/S1063773717120015},
archivePrefix = {arXiv},
       eprint = {1401.7329},
 primaryClass = {astro-ph.CO},
       adsurl = {https://ui.adsabs.harvard.edu/abs/2018AstL...44....8K}
}

@ARTICLE{2008AJ....136.2648D,
       author = {{de Blok}, W.~J.~G. and {Walter}, F. and {Brinks}, E. and {Trachternach}, C. and {Oh}, S.-H. and {Kennicutt}, Jr., R.~C.},
        title = "{High-Resolution Rotation Curves and Galaxy Mass Models from THINGS}",
      journal = {\aj},
         year = 2008,
        month = dec,
       volume = {136},
       number = {6},
        pages = {2648-2719},
          doi = {10.1088/0004-6256/136/6/2648},
archivePrefix = {arXiv},
       eprint = {0810.2100},
 primaryClass = {astro-ph},
       adsurl = {https://ui.adsabs.harvard.edu/abs/2008AJ....136.2648D}
}

@ARTICLE{2011MNRAS.416.2401H,
       author = {{Holwerda}, B.~W. and {Pirzkal}, N. and {de Blok}, W.~J.~G. and {Bouchard}, A. and {Blyth}, S.-L. and {van der Heyden}, K.~J. and {Elson}, E.~C.},
        title = "{Quantified H I morphology - I. Multi-wavelength analysis of the THINGS galaxies}",
      journal = {\mnras},
         year = 2011,
        month = oct,
       volume = {416},
       number = {4},
        pages = {2401-2414},
          doi = {10.1111/j.1365-2966.2011.18938.x},
archivePrefix = {arXiv},
       eprint = {1104.3291},
 primaryClass = {astro-ph.CO},
       adsurl = {https://ui.adsabs.harvard.edu/abs/2011MNRAS.416.2401H}
}

@ARTICLE{2025NatAs...9..577T,
       author = {{Tornotti}, Davide and {Fumagalli}, Michele and {Fossati}, Matteo and {Benitez-Llambay}, Alejandro and {Izquierdo-Villalba}, David and {Travascio}, Andrea and {Arrigoni Battaia}, Fabrizio and {Cantalupo}, Sebastiano and {Beckett}, Alexander and {Bonoli}, Silvia and {Dayal}, Pratika and {D'Odorico}, Valentina and {Dutta}, Rajeshwari and {Lusso}, Elisabeta and {Peroux}, Celine and {Rafelski}, Marc and {Revalski}, Mitchell and {Spinoso}, Daniele and {Swinbank}, Mark},
        title = "{High-definition imaging of a filamentary connection between a close quasar pair at z = 3}",
      journal = {Nature Astronomy},
         year = 2025,
        month = apr,
       volume = {9},
        pages = {577-588},
          doi = {10.1038/s41550-024-02463-w},
archivePrefix = {arXiv},
       eprint = {2406.17035},
 primaryClass = {astro-ph.CO},
       adsurl = {https://ui.adsabs.harvard.edu/abs/2025NatAs...9..577T}
}

@ARTICLE{2019AJ....157..168D,
       author = {{Dey}, Arjun and {Schlegel}, David J. and {Lang}, Dustin and {Blum}, Robert and {Burleigh}, Kaylan and {Fan}, Xiaohui and {Findlay}, Joseph R. and {Finkbeiner}, Doug and {Herrera}, David and {Juneau}, St{\'e}phanie and {Landriau}, Martin and {Levi}, Michael and {McGreer}, Ian and {Meisner}, Aaron and {Myers}, Adam D. and {Moustakas}, John and {Nugent}, Peter and {Patej}, Anna and {Schlafly}, Edward F. and {Walker}, Alistair R. and {Valdes}, Francisco and {Weaver}, Benjamin A. and {Y{\`e}che}, Christophe and {Zou}, Hu and {Zhou}, Xu and {Abareshi}, Behzad and {Abbott}, T.~M.~C. and {Abolfathi}, Bela and {Aguilera}, C. and {Alam}, Shadab and {Allen}, Lori and {Alvarez}, A. and {Annis}, James and {Ansarinejad}, Behzad and {Aubert}, Marie and {Beechert}, Jacqueline and {Bell}, Eric F. and {BenZvi}, Segev Y. and {Beutler}, Florian and {Bielby}, Richard M. and {Bolton}, Adam S. and {Brice{\~n}o}, C{\'e}sar and {Buckley-Geer}, Elizabeth J. and {Butler}, Karen and {Calamida}, Annalisa and {Carlberg}, Raymond G. and {Carter}, Paul and {Casas}, Ricard and {Castander}, Francisco J. and {Choi}, Yumi and {Comparat}, Johan and {Cukanovaite}, Elena and {Delubac}, Timoth{\'e}e and {DeVries}, Kaitlin and {Dey}, Sharmila and {Dhungana}, Govinda and {Dickinson}, Mark and {Ding}, Zhejie and {Donaldson}, John B. and {Duan}, Yutong and {Duckworth}, Christopher J. and {Eftekharzadeh}, Sarah and {Eisenstein}, Daniel J. and {Etourneau}, Thomas and {Fagrelius}, Parker A. and {Farihi}, Jay and {Fitzpatrick}, Mike and {Font-Ribera}, Andreu and {Fulmer}, Leah and {G{\"a}nsicke}, Boris T. and {Gaztanaga}, Enrique and {George}, Koshy and {Gerdes}, David W. and {Gontcho}, Satya Gontcho A. and {Gorgoni}, Claudio and {Green}, Gregory and {Guy}, Julien and {Harmer}, Diane and {Hernandez}, M. and {Honscheid}, Klaus and {Huang}, Lijuan Wendy and {James}, David J. and {Jannuzi}, Buell T. and {Jiang}, Linhua and {Joyce}, Richard and {Karcher}, Armin and {Karkar}, Sonia and {Kehoe}, Robert and {Kneib}, Jean-Paul and {Kueter-Young}, Andrea and {Lan}, Ting-Wen and {Lauer}, Tod R. and {Le Guillou}, Laurent and {Le Van Suu}, Auguste and {Lee}, Jae Hyeon and {Lesser}, Michael and {Perreault Levasseur}, Laurence and {Li}, Ting S. and {Mann}, Justin L. and {Marshall}, Robert and {Mart{\'\i}nez-V{\'a}zquez}, C.~E. and {Martini}, Paul and {du Mas des Bourboux}, H{\'e}lion and {McManus}, Sean and {Meier}, Tobias Gabriel and {M{\'e}nard}, Brice and {Metcalfe}, Nigel and {Mu{\~n}oz-Guti{\'e}rrez}, Andrea and {Najita}, Joan and {Napier}, Kevin and {Narayan}, Gautham and {Newman}, Jeffrey A. and {Nie}, Jundan and {Nord}, Brian and {Norman}, Dara J. and {Olsen}, Knut A.~G. and {Paat}, Anthony and {Palanque-Delabrouille}, Nathalie and {Peng}, Xiyan and {Poppett}, Claire L. and {Poremba}, Megan R. and {Prakash}, Abhishek and {Rabinowitz}, David and {Raichoor}, Anand and {Rezaie}, Mehdi and {Robertson}, A.~N. and {Roe}, Natalie A. and {Ross}, Ashley J. and {Ross}, Nicholas P. and {Rudnick}, Gregory and {Safonova}, Sasha and {Saha}, Abhijit and {S{\'a}nchez}, F. Javier and {Savary}, Elodie and {Schweiker}, Heidi and {Scott}, Adam and {Seo}, Hee-Jong and {Shan}, Huanyuan and {Silva}, David R. and {Slepian}, Zachary and {Soto}, Christian and {Sprayberry}, David and {Staten}, Ryan and {Stillman}, Coley M. and {Stupak}, Robert J. and {Summers}, David L. and {Sien Tie}, Suk and {Tirado}, H. and {Vargas-Maga{\~n}a}, Mariana and {Vivas}, A. Katherina and {Wechsler}, Risa H. and {Williams}, Doug and {Yang}, Jinyi and {Yang}, Qian and {Yapici}, Tolga and {Zaritsky}, Dennis and {Zenteno}, A. and {Zhang}, Kai and {Zhang}, Tianmeng and {Zhou}, Rongpu and {Zhou}, Zhimin},
        title = "{Overview of the DESI Legacy Imaging Surveys}",
      journal = {\aj},
         year = 2019,
        month = may,
       volume = {157},
       number = {5},
          eid = {168},
        pages = {168},
          doi = {10.3847/1538-3881/ab089d},
archivePrefix = {arXiv},
       eprint = {1804.08657},
 primaryClass = {astro-ph.IM},
       adsurl = {https://ui.adsabs.harvard.edu/abs/2019AJ....157..168D}
}

@ARTICLE{2023A&A...674A...1G,
  author = {{Gaia Collaboration} and {Vallenari}, A. and {Brown}, A.~G.~A. and others},
  title = "{Gaia Data Release 3. Summary of the content and survey properties}",
  journal = {A\&A},
  volume = {674},
  pages = {A1},
  year = {2023},
  doi = {10.1051/0004-6361/202243940}
}

@ARTICLE{2023MNRAS.521.1113B,
       author = {{Beckett}, Alexander and {Morris}, Simon L. and {Fumagalli}, Michele and {Tejos}, Nicolas and {Jannuzi}, Buell and {Cantalupo}, Sebastiano},
        title = "{Modelling gas around galaxy pairs and groups using the Q0107 quasar triplet}",
      journal = {\mnras},
         year = 2023,
        month = may,
       volume = {521},
       number = {1},
        pages = {1113-1143},
          doi = {10.1093/mnras/stad596},
archivePrefix = {arXiv},
       eprint = {2302.11609},
 primaryClass = {astro-ph.GA},
       adsurl = {https://ui.adsabs.harvard.edu/abs/2023MNRAS.521.1113B}
}

@ARTICLE{2025ApJ...978L..18C,
       author = {{Chen}, Mandy C. and {Chen}, Hsiao-Wen and {Rauch}, Michael and {Vayner}, Andrey and {Liu}, Weizhe and {Rupke}, David S.~N. and {Greene}, Jenny E. and {Zakamska}, Nadia L. and {Wylezalek}, Dominika and {Liu}, Guilin and {Veilleux}, Sylvain and {Nesvadba}, Nicole P.~H. and {Bertemes}, Caroline},
        title = "{Resolving Turbulence Drivers in Two Luminous Obscured Quasars with JWST/NIRSpec Integral Field Unit}",
      journal = {\apjl},
         year = 2025,
        month = jan,
       volume = {978},
       number = {2},
          eid = {L18},
        pages = {L18},
          doi = {10.3847/2041-8213/ad9bac},
archivePrefix = {arXiv},
       eprint = {2410.14785},
 primaryClass = {astro-ph.GA},
       adsurl = {https://ui.adsabs.harvard.edu/abs/2025ApJ...978L..18C}
}

@ARTICLE{2016MNRAS.456.3032P,
       author = {{Pawlik}, M.~M. and {Wild}, V. and {Walcher}, C.~J. and {Johansson}, P.~H. and {Villforth}, C. and {Rowlands}, K. and {Mendez-Abreu}, J. and {Hewlett}, T.},
        title = "{Shape asymmetry: a morphological indicator for automatic detection of galaxies in the post-coalescence merger stages}",
      journal = {\mnras},
         year = 2016,
        month = mar,
       volume = {456},
       number = {3},
        pages = {3032-3052},
          doi = {10.1093/mnras/stv2878},
archivePrefix = {arXiv},
       eprint = {1512.02000},
 primaryClass = {astro-ph.GA},
       adsurl = {https://ui.adsabs.harvard.edu/abs/2016MNRAS.456.3032P}
}

@ARTICLE{1995ApJ...451L...1S,
       author = {{Schade}, David and {Lilly}, S.~J. and {Crampton}, David and {Hammer}, F. and {Le Fevre}, O. and {Tresse}, L.},
        title = "{Canada-France Redshift Survey: Hubble Space Telescope Imaging of High-Redshift Field Galaxies}",
      journal = {\apjl},
         year = 1995,
        month = sep,
       volume = {451},
        pages = {L1},
          doi = {10.1086/309677},
archivePrefix = {arXiv},
       eprint = {astro-ph/9507028},
 primaryClass = {astro-ph},
       adsurl = {https://ui.adsabs.harvard.edu/abs/1995ApJ...451L...1S}
}

@ARTICLE{1996MNRAS.279L..47A,
       author = {{Abraham}, R.~G. and {Tanvir}, N.~R. and {Santiago}, B.~X. and {Ellis}, R.~S. and {Glazebrook}, K. and {van den Bergh}, S.},
        title = "{Galaxy morphology to I=25 mag in the Hubble Deep Field}",
      journal = {\mnras},
         year = 1996,
        month = apr,
       volume = {279},
       number = {3},
        pages = {L47-L52},
          doi = {10.1093/mnras/279.3.L47},
archivePrefix = {arXiv},
       eprint = {astro-ph/9602044},
 primaryClass = {astro-ph},
       adsurl = {https://ui.adsabs.harvard.edu/abs/1996MNRAS.279L..47A}
}

@ARTICLE{2000ApJ...529..886C,
       author = {{Conselice}, Christopher J. and {Bershady}, Matthew A. and {Jangren}, Anna},
        title = "{The Asymmetry of Galaxies: Physical Morphology for Nearby and High-Redshift Galaxies}",
      journal = {\apj},
         year = 2000,
        month = feb,
       volume = {529},
       number = {2},
        pages = {886-910},
          doi = {10.1086/308300},
archivePrefix = {arXiv},
       eprint = {astro-ph/9907399},
 primaryClass = {astro-ph},
       adsurl = {https://ui.adsabs.harvard.edu/abs/2000ApJ...529..886C}
}

@ARTICLE{2012ARA&A..50..491P,
       author = {{Putman}, M.~E. and {Peek}, J.~E.~G. and {Joung}, M.~R.},
        title = "{Gaseous Galaxy Halos}",
      journal = {\araa},
         year = 2012,
        month = sep,
       volume = {50},
        pages = {491-529},
          doi = {10.1146/annurev-astro-081811-125612},
archivePrefix = {arXiv},
       eprint = {1207.4837},
 primaryClass = {astro-ph.GA},
       adsurl = {https://ui.adsabs.harvard.edu/abs/2012ARA&A..50..491P}
}

@ARTICLE{2024A&A...691A.255K,
       author = {{Kusakabe}, Haruka and {Mauerhofer}, Valentin and {Verhamme}, Anne and {Garel}, Thibault and {Blaizot}, J{\'e}r{\'e}my and {Wisotzki}, Lutz and {Richard}, Johan and {Boogaard}, Leindert A. and {Leclercq}, Floriane and {Guo}, Yucheng and {Claeyssens}, Ad{\'e}la{\"\i}de and {Contini}, Thierry and {Herenz}, Edmund Christian and {Kerutt}, Josephine and {Maseda}, Michael V. and {Michel-Dansac}, Leo and {Nanayakkara}, Themiya and {Ouchi}, Masami and {Pessa}, Ismael and {Schaye}, Joop},
        title = "{The MUSE eXtremely Deep Field: Detections of circumgalactic Si II* emission at z {\ensuremath{\gtrsim}} 2}",
      journal = {\aap},
         year = 2024,
        month = nov,
       volume = {691},
          eid = {A255},
        pages = {A255},
          doi = {10.1051/0004-6361/202451009},
archivePrefix = {arXiv},
       eprint = {2406.04399},
 primaryClass = {astro-ph.GA},
       adsurl = {https://ui.adsabs.harvard.edu/abs/2024A&A...691A.255K}
}

@ARTICLE{2024SciA...10P8629Z,
       author = {{Zhang}, Huanian and {Zaritsky}, Dennis},
        title = "{A MUSE source-blind survey for emission from the circumgalactic medium}",
      journal = {Science Advances},
         year = 2024,
        month = nov,
       volume = {10},
       number = {47},
          eid = {eadp8629},
        pages = {eadp8629},
          doi = {10.1126/sciadv.adp8629},
archivePrefix = {arXiv},
       eprint = {2410.05392},
 primaryClass = {astro-ph.GA},
       adsurl = {https://ui.adsabs.harvard.edu/abs/2024SciA...10P8629Z}
}

@ARTICLE{2024A&A...683A.205E,
       author = {{Epinat}, B. and {Contini}, T. and {Mercier}, W. and {Ciesla}, L. and {Lemaux}, B.~C. and {Johnson}, S.~D. and {Richard}, J. and {Brinchmann}, J. and {Boogaard}, L.~A. and {Carton}, D. and {Michel-Dansac}, L. and {Bacon}, R. and {Krajnovi{\'c}}, D. and {Finley}, H. and {Schroetter}, I. and {Ventou}, E. and {Abril-Melgarejo}, V. and {Boselli}, A. and {Bouch{\'e}}, N.~F. and {Kollatschny}, W. and {Kova{\v{c}}}, K. and {Paalvast}, M. and {Soucail}, G. and {Urrutia}, T. and {Weilbacher}, P.~M.},
        title = "{MAGIC: MUSE gAlaxy Groups In COSMOS - A survey to probe the impact of environment on galaxy evolution over the last 8 Gyr}",
      journal = {\aap},
         year = 2024,
        month = mar,
       volume = {683},
          eid = {A205},
        pages = {A205},
          doi = {10.1051/0004-6361/202348038},
archivePrefix = {arXiv},
       eprint = {2312.00924},
 primaryClass = {astro-ph.GA},
       adsurl = {https://ui.adsabs.harvard.edu/abs/2024A&A...683A.205E}
}

@ARTICLE{2025arXiv251212754B,
       author = {{Bellomi}, Elena and {ZuHone}, John A. and {Truong}, Nhut and {Zhuravleva}, Irina and {Weinberger}, Rainer and {Pfrommer}, Christoph and {Zhang}, Congyao and {Heinrich}, Annie and {Ruszkowski}, Mateusz and {McNamara}, Brian and {Hlavacek-Larrondo}, Julie and {Gendron-Marsolais}, Marie-Lou and {Vigneron}, Benjamin},
        title = "{Disentangling AGN Feedback and Sloshing in the Perseus Cluster with XRISM: Insights from Simulations}",
      journal = {arXiv e-prints},
         year = 2025,
        month = dec,
          eid = {arXiv:2512.12754},
        pages = {arXiv:2512.12754},
          doi = {10.48550/arXiv.2512.12754},
archivePrefix = {arXiv},
       eprint = {2512.12754},
 primaryClass = {astro-ph.HE},
       adsurl = {https://ui.adsabs.harvard.edu/abs/2025arXiv251212754B}
}

@ARTICLE{2022A&A...662A..23B,
       author = {{Balmaverde}, B. and {Capetti}, A. and {Baldi}, R.~D. and {Baum}, S. and {Chiaberge}, M. and {Gilli}, R. and {Jimenez-Gallardo}, A. and {Marconi}, A. and {Massaro}, F. and {Meyer}, E. and {O'Dea}, C. and {Speranza}, G. and {Torresi}, E. and {Venturi}, G.},
        title = "{The MURALES survey. VI. Properties and origin of the extended line emission structures in radio galaxies}",
      journal = {\aap},
         year = 2022,
        month = jun,
       volume = {662},
          eid = {A23},
        pages = {A23},
          doi = {10.1051/0004-6361/202142823},
archivePrefix = {arXiv},
       eprint = {2204.00528},
 primaryClass = {astro-ph.GA},
       adsurl = {https://ui.adsabs.harvard.edu/abs/2022A&A...662A..23B}
}

@ARTICLE{2018A&A...619A..83B,
       author = {{Balmaverde}, B. and {Capetti}, A. and {Marconi}, A. and {Venturi}, G. and {Chiaberge}, M. and {Baldi}, R.~D. and {Baum}, S. and {Gilli}, R. and {Grandi}, P. and {Meyer}, E. and {Miley}, G. and {O'Dea}, C. and {Sparks}, W. and {Torresi}, E. and {Tremblay}, G.},
        title = "{The MURALES survey. I. A dual AGN in the radio galaxy 3C 459?}",
      journal = {\aap},
         year = 2018,
        month = nov,
       volume = {619},
          eid = {A83},
        pages = {A83},
          doi = {10.1051/0004-6361/201833515},
archivePrefix = {arXiv},
       eprint = {1809.04083},
 primaryClass = {astro-ph.GA},
       adsurl = {https://ui.adsabs.harvard.edu/abs/2018A&A...619A..83B}
}

@ARTICLE{2015ApJ...799..209W,
       author = {{Wisnioski}, E. and {F{\"o}rster Schreiber}, N.~M. and {Wuyts}, S. and {Wuyts}, E. and {Bandara}, K. and {Wilman}, D. and {Genzel}, R. and {Bender}, R. and {Davies}, R. and {Fossati}, M. and {Lang}, P. and {Mendel}, J.~T. and {Beifiori}, A. and {Brammer}, G. and {Chan}, J. and {Fabricius}, M. and {Fudamoto}, Y. and {Kulkarni}, S. and {Kurk}, J. and {Lutz}, D. and {Nelson}, E.~J. and {Momcheva}, I. and {Rosario}, D. and {Saglia}, R. and {Seitz}, S. and {Tacconi}, L.~J. and {van Dokkum}, P.~G.},
        title = "{The KMOS$^{3D}$ Survey: Design, First Results, and the Evolution of Galaxy Kinematics from 0.7 <= z <= 2.7}",
      journal = {\apj},
         year = 2015,
        month = feb,
       volume = {799},
       number = {2},
          eid = {209},
        pages = {209},
          doi = {10.1088/0004-637X/799/2/209},
archivePrefix = {arXiv},
       eprint = {1409.6791},
 primaryClass = {astro-ph.GA},
       adsurl = {https://ui.adsabs.harvard.edu/abs/2015ApJ...799..209W}
}

@ARTICLE{2024MNRAS.528.1895D,
       author = {{Dutta}, Rajeshwari and {Acebron}, Ana and {Fumagalli}, Michele and {Grillo}, Claudio and {Caminha}, Gabriel B. and {Fossati}, Matteo},
        title = "{Probing coherence in metal absorption towards multiple images of strong gravitationally lensed quasars}",
      journal = {\mnras},
         year = 2024,
        month = feb,
       volume = {528},
       number = {2},
        pages = {1895-1905},
          doi = {10.1093/mnras/stae048},
archivePrefix = {arXiv},
       eprint = {2401.03024},
 primaryClass = {astro-ph.GA},
       adsurl = {https://ui.adsabs.harvard.edu/abs/2024MNRAS.528.1895D}
}

@ARTICLE{2024CmPhy...7..286B,
       author = {{Barone}, Tania M. and {Kacprzak}, Glenn G. and {Nightingale}, James W. and {Nielsen}, Nikole M. and {Glazebrook}, Karl and {Tran}, Kim-Vy H. and {Jones}, Tucker and {Nateghi}, Hasti and {Vasan Gopala Chandrasekaran}, Keerthi and {Sahu}, Nandini and {Nanayakkara}, Themiya and {Skobe}, Hannah and {van de Sande}, Jesse and {Lopez}, Sebastian and {Lewis}, Geraint F.},
        title = "{Gravitational lensing reveals cool gas within 10-20 kpc around a quiescent galaxy}",
      journal = {Communications Physics},
         year = 2024,
        month = dec,
       volume = {7},
       number = {1},
          eid = {286},
        pages = {286},
          doi = {10.1038/s42005-024-01778-4},
archivePrefix = {arXiv},
       eprint = {2408.07984},
 primaryClass = {astro-ph.GA},
       adsurl = {https://ui.adsabs.harvard.edu/abs/2024CmPhy...7..286B}
}

@ARTICLE{2023ApJ...947...16A,
       author = {{Amiri}, Mandana and {Bandura}, Kevin and {Chen}, Tianyue and {Deng}, Meiling and {Dobbs}, Matt and {Fandino}, Mateus and {Foreman}, Simon and {Halpern}, Mark and {Hill}, Alex S. and {Hinshaw}, Gary and {H{\"o}fer}, Carolin and {Kania}, Joseph and {Landecker}, T.~L. and {MacEachern}, Joshua and {Masui}, Kiyoshi and {Mena-Parra}, Juan and {Milutinovic}, Nikola and {Mirhosseini}, Arash and {Newburgh}, Laura and {Ordog}, Anna and {Pen}, Ue-Li and {Pinsonneault-Marotte}, Tristan and {Polzin}, Ava and {Reda}, Alex and {Renard}, Andre and {Shaw}, J. Richard and {Siegel}, Seth R. and {Singh}, Saurabh and {Vanderlinde}, Keith and {Wang}, Haochen and {Wiebe}, Donald V. and {Wulf}, Dallas and {CHIME Collaboration}},
        title = "{Detection of Cosmological 21 cm Emission with the Canadian Hydrogen Intensity Mapping Experiment}",
      journal = {\apj},
         year = 2023,
        month = apr,
       volume = {947},
       number = {1},
          eid = {16},
        pages = {16},
          doi = {10.3847/1538-4357/acb13f},
archivePrefix = {arXiv},
       eprint = {2202.01242},
 primaryClass = {astro-ph.CO},
       adsurl = {https://ui.adsabs.harvard.edu/abs/2023ApJ...947...16A}
}

@ARTICLE{2022ApJ...937..103C,
       author = {{Chowdhury}, Aditya and {Kanekar}, Nissim and {Chengalur}, Jayaram N.},
        title = "{The Giant Metrewave Radio Telescope Cold-HI AT z ≍ 1 Survey}",
      journal = {\apj},
         year = 2022,
        month = oct,
       volume = {937},
       number = {2},
          eid = {103},
        pages = {103},
          doi = {10.3847/1538-4357/ac7d52},
archivePrefix = {arXiv},
       eprint = {2207.00031},
 primaryClass = {astro-ph.GA},
       adsurl = {https://ui.adsabs.harvard.edu/abs/2022ApJ...937..103C}
}

@ARTICLE{2025ApJS..279...19C,
       author = {{Chen}, Zhaoting and {Cunnington}, Steven and {Pourtsidou}, Alkistis and {Wolz}, Laura and {Spinelli}, Marta and {Bernal}, Jos{\'e} Luis and {Barberi-Squarotti}, Matilde and {Camera}, Stefano and {Carucci}, Isabella P. and {Fonseca}, Jos{\'e} and {Grainge}, Keith and {Irfan}, Melis O. and {Santos}, Mario G. and {Wang}, Jingying and {Meerklass Collaboration}},
        title = "{Emission-line Stacking of 21 cm Intensity Maps with MeerKLASS: Inference Pipeline and Application to the L-band Deep-field Data}",
      journal = {\apjs},
         year = 2025,
        month = jul,
       volume = {279},
       number = {1},
          eid = {19},
        pages = {19},
          doi = {10.3847/1538-4365/add897},
archivePrefix = {arXiv},
       eprint = {2504.03908},
 primaryClass = {astro-ph.CO},
       adsurl = {https://ui.adsabs.harvard.edu/abs/2025ApJS..279...19C}
}

@ARTICLE{1974MNRAS.167P..31F,
       author = {{Fanaroff}, B.~L. and {Riley}, J.~M.},
        title = "{The morphology of extragalactic radio sources of high and low luminosity}",
      journal = {\mnras},
         year = 1974,
        month = may,
       volume = {167},
        pages = {31P-36P},
          doi = {10.1093/mnras/167.1.31P},
       adsurl = {https://ui.adsabs.harvard.edu/abs/1974MNRAS.167P..31F}
}

@software{2016ascl.soft09011B,
       author = {{Bradley}, Larry and {Sipocz}, Brigitta and {Robitaille}, Thomas and {Tollerud}, Erik and {Deil}, Christoph and {Vin{\'\i}cius}, Z{\`e} and {Barbary}, Kyle and {G{\"u}nther}, Hans Moritz and {Bostroem}, Azalee and {Droettboom}, Michael and {Bray}, Erik and {Bratholm}, Lars Andersen and {Pickering}, T.~E. and {Craig}, Matt and {Pascual}, Sergio and {Greco}, Johnny and {Donath}, Axel and {Kerzendorf}, Wolfgang and {Littlefair}, Stuart and {Barentsen}, Geert and {D'Eugenio}, Francesco and {Weaver}, Benjamin Alan},
        title = "{Photutils: Photometry tools}",
 howpublished = {Astrophysics Source Code Library, record ascl:1609.011},
         year = 2016,
        month = sep,
          eid = {ascl:1609.011},
archivePrefix = {ascl},
       eprint = {1609.011},
       adsurl = {https://ui.adsabs.harvard.edu/abs/2016ascl.soft09011B}
}

@INPROCEEDINGS{2026enap....4..370C,
       author = {{Chen}, Hsiao-Wen and {Zahedy}, Fakhri S.},
        title = "{The circumgalactic medium}",
    booktitle = {Encyclopedia of Astrophysics, Volume 4},
         year = 2026,
       volume = {4},
        month = jan,
        pages = {370-400},
          doi = {10.1016/B978-0-443-21439-4.00059-6},
archivePrefix = {arXiv},
       eprint = {2412.10579},
 primaryClass = {astro-ph.GA},
       adsurl = {https://ui.adsabs.harvard.edu/abs/2026enap....4..370C}
}

@ARTICLE{2026A&A...707A.380G,
       author = {{Gonz{\'a}lez Lobos}, Jay and {Arrigoni Battaia}, Fabrizio and {Obreja}, Aura and {Kauffmann}, Guinevere and {Paolo Farina}, Emanuele and {Costa}, Tiago},
        title = "{QSO MUSEUM: III. The circumgalactic medium in the Ly{\ensuremath{\alpha}} emission around 120 z {\ensuremath{\sim}} 3 quasars covering the SDSS parameter space. Witnessing the instantaneous active galactic nucleus feedback on halo scales}",
      journal = {\aap},
         year = 2026,
        month = mar,
       volume = {707},
          eid = {A380},
        pages = {A380},
          doi = {10.1051/0004-6361/202554054},
archivePrefix = {arXiv},
       eprint = {2507.16898},
 primaryClass = {astro-ph.GA},
       adsurl = {https://ui.adsabs.harvard.edu/abs/2026A&A...707A.380G}
}

@BOOK{2012msma.book.....F,
       author = {{Feigelson}, Eric D. and {Babu}, G. Jogesh},
        title = "{Modern Statistical Methods for Astronomy}",
         year = 2012,
          doi = {10.48550/arXiv.1205.2064},
       adsurl = {https://ui.adsabs.harvard.edu/abs/2012msma.book.....F}
}
\bibliographystyle{aasjournalv7}



\end{document}